\documentclass[10pt, a4paper, twocolumn]{article} % 10pt font size (11 and 12 also possible), A4 paper (letterpaper for US letter) and two column layout (remove for one column)

\usepackage[english]{babel} % English language hyphenation

\usepackage{microtype} % Better typography

\usepackage{amsmath,amsfonts,amsthm} % Math packages for equations

\usepackage[svgnames]{xcolor} % Enabling colors by their 'svgnames'

\usepackage[small, labelfont=bf, up]{caption} % Custom captions under/above tables and figures

\usepackage{booktabs} % Horizontal rules in tables

\usepackage{lastpage} % Used to determine the number of pages in the document (for "Page X of Total")

\usepackage{graphicx} % Required for adding images

\usepackage{enumitem} % Required for customising lists
\setlist{noitemsep} % Remove spacing between bullet/numbered list elements

\usepackage{sectsty} % Enables custom section titles
\allsectionsfont{\usefont{OT1}{phv}{b}{n}} % Change the font of all section commands (Helvetica)
\usepackage{geometry} % Required for adjusting page dimensions

\usepackage[T1]{fontenc} % Output font encoding for international characters
\usepackage[utf8]{inputenc} % Required for inputting international characters

\usepackage{XCharter} % Use the XCharter font

\usepackage{fancyhdr} % Needed to define custom headers/footers
\fancypagestyle{firstpage}{ % Page style for the first page with the title
	\fancyhf{}
}

\newcommand{\authorstyle}[1]{{\large\usefont{OT1}{phv}{b}{n}\color{DarkRed}#1}} % Authors style (Helvetica)

\newcommand{\institution}[1]{{\footnotesize\usefont{OT1}{phv}{m}{sl}\color{Black}#1}} % Institutions style (Helvetica)

\usepackage{titling} % Allows custom title configuration

\newcommand{\HorRule}{\color{DarkGoldenrod}\rule{\linewidth}{1pt}} % Defines the gold horizontal rule around the title

\pretitle{
	\vspace{-30pt} % Move the entire title section up
	\HorRule\vspace{10pt} % Horizontal rule before the title
	\fontsize{21}{23}\usefont{OT1}{phv}{b}{n}\selectfont % Helvetica
	\color{DarkRed} % Text colour for the title and author(s)
}

\posttitle{\par\vskip 15pt} % Whitespace under the title

\preauthor{} % Anything that will appear before \author is printed

\postauthor{ % Anything that will appear after \author is printed
	\vspace{10pt} % Space before the rule
	\par\HorRule % Horizontal rule after the title
	\vspace{20pt} % Space after the title section
}

\usepackage{lettrine} % Package to accentuate the first letter of the text (lettrine)
\usepackage{fix-cm}	% Fixes the height of the lettrine

\usepackage{xstring} % Required for string manipulation

\usepackage[autostyle=true]{csquotes} % Required to generate language-dependent quotes in the bibliography
\usepackage[T1]{fontenc}
\usepackage[utf8]{inputenc}

\DeclareUnicodeCharacter{0301}{*************************************}

\usepackage{graphics}
\usepackage[margin=1.5cm, font=small]{caption}
\usepackage{graphicx}
\usepackage{amsmath}
\usepackage{amssymb}
\usepackage{setspace}
\usepackage{nicefrac}
\usepackage{bm}
\usepackage{xr}
\usepackage{relsize}
\usepackage{caption}
\usepackage[scr]{rsfso}
\usepackage{ulem}

\usepackage[colorlinks = true, linkcolor = purple, urlcolor  = purple, citecolor = blue,
            anchorcolor = purple]{hyperref}
\usepackage{xcolor}
\usepackage{cuted}
\usepackage{xurl}
\usepackage{hyperref}
\PassOptionsToPackage{hyphens}{url}
\usepackage{siunitx}

\RequirePackage[utf8]{inputenc}
\RequirePackage{calc}
\RequirePackage{indentfirst}
\RequirePackage{fancyhdr}
\RequirePackage{graphicx,epstopdf}
\RequirePackage{lastpage}
\RequirePackage{ifthen}
\RequirePackage{lineno}
\RequirePackage{float}
\RequirePackage{amsmath}
\RequirePackage{setspace}
\RequirePackage{enumitem}
\RequirePackage{mathpazo}

\usepackage{ulem}
\usepackage{mathtools}

\newtagform{brackets}{[}{]}
\usetagform{brackets}

\renewcommand{\labelenumi}{\theenumi}
\renewcommand{\theenumi}{(\roman{enumi})}

\definecolor{ao}{rgb}{0.0, 0.5, 0.0}

\title{Practical and scalable simulations of non-Markovian stochastic processes and temporal networks with individual node properties} 

\author{\authorstyle{Aur\'elien P\'elissier$^{1,2,3}$, Miroslav Phan$^{1,2}$, Didier Le Bail$^{4}$, Niko Beerenwinkel$^{2}$ and Mar\'ia Rodr\'iguez Mart\'inez$^{1,5,\dagger}$}\newline\newline
\textsuperscript{1}\institution{IBM Research Europe, 8803 Rüschlikon, Switzerland}\\
\textsuperscript{2}\institution{Department of Biosystems Science and Engineering, ETH Zurich, 4058 Basel, Switzerland}\\
\textsuperscript{3}\institution{Institute of Computational Life Sciences, Zürich University of Applied Sciences (ZHAW), 8820, Wädenswil, Switzerland}\\
\textsuperscript{4}\institution{Centre de Physique Théorique (CPT), Aix-Marseille University, CNRS, 13009 Marseille, France}\\
\textsuperscript{5}\institution{Department of Biomedical Informatics \& Data Science, Yale School of Medicine, New Haven, CT, United States.}\\
\textsuperscript{$\dagger$} \institution{Corresponding author \href{maria.rodriguezmartinez@yale.edu}{maria.rodriguezmartinez@yale.edu}}}

\date{\vspace{-5ex}}

\begin{document}

\maketitle

\begin{strip}

%current abstract is 182 words.
\begin{center}
\begin{minipage}{14.6cm}
\vspace{-2cm}
\begin{center}
{\large \textbf{Abstract}}
\end{center}
\vspace{-0.2cm}
Discrete stochastic processes are widespread in natural systems with many applications across physics, biochemistry, epidemiology, sociology, and finance. While analytic solutions often cannot be derived, existing simulation frameworks can generate stochastic trajectories compatible with the dynamical laws underlying the random phenomena.
%, such as radioactive decay or Brownian motion, although 
%have been used to model processes of different complexity, such as radioactive decay and Brownian motion.
However, most simulation algorithms assume the system dynamics are memoryless (Markovian assumption), under which assumption,  future occurrences only depend on the system's present state. Mathematically, the Markovian assumption models inter-event times as exponentially distributed variables, which enables the exact simulation of stochastic trajectories using the seminal Gillespie algorithm. 
Unfortunately, the majority of stochastic systems exhibit properties of memory, an inherently non-Markovian attribute. Non-Markovian systems are notoriously difficult to investigate analytically, and existing numerical methods are computationally costly or only applicable under strong simplifying assumptions, often not compatible with empirical observations. To address these challenges, we develop the Rejection-based Gillespie algorithm for non-Markovian Reactions (REGIR), a general and scalable framework to simulate non-Markovian stochastic systems with arbitrary inter-event time distributions. REGIR can achieve arbitrary user-defined accuracy while maintaining the same asymptotic computational complexity as the Gillespie algorithm.
We establish a lower bound on REGIR's approximation accuracy, and illustrate its modeling capabilities in three different classes of non-Markovian systems, namely delayed reaction channels, stochastic processes with individual reactant properties, and temporal networks driven by node activity. In all three cases, REGIR  efficiently models the underlying stochastic processes and demonstrates its competitive performance over existing methods.

%In the context of RNA transcription, B-cell maturation and social interaction 

%In all three cases, REGIR  efficiently models the underlying stochastic processes and demonstrates its utility to accurately investigate complex non-Markovian systems over existing method. 

%In the context of temporal networks, REGIR provide a more exact and adaptable simulation environment then activity driven modeling by utilizing dynamic and stochastic time-steps at each iteration.
%
%The algorithm is implemented as a python library \texttt{REGIR}(\url{https://github.com/Aurelien-Pelissier/REGIR}).

\end{minipage}

\end{center}

\end{strip}

\vfill

\normalsize
\onecolumn 

Discrete stochastic processes~\cite{ross1996stochastic} are prevalent in the study of a wide range of random phenomena in physics~\cite{van1992stochastic}, biochemistry~\cite{cao2009discrete}, epidemiology~\cite{yan2008distribution}, finance~\cite{kijima2002stochastic} and meteorology~\cite{tseng2020forecasting}. To simplify their study, the Markovian assumption is usually made, meaning the system dynamics are assumed to be memoryless, and therefore, the probability of any future occurrence depends only on the present state of the system. 
Examples of Markovian systems include first-order reaction kinetics in biochemical networks, where the rate of reaction only depends on the present concentration of the reactant~\cite{martinez2010messenger}, and Brownian motion, where the displacement of the particle does not depend on its past displacements~\cite{hida1980brownian}.
However, as the only memoryless continuous probability distribution is the exponential distribution,  Markovian dynamics require inter-event times to be modeled as exponentially distributed random variables, which often oversimplifies real-world system dynamics and fails to capture complex temporal patterns.

From a computational point of view, an advantage of using the Markovian assumption is that stochastic systems can be simulated using exact algorithms, such as the seminal Gillespie stochastic  algorithm~\cite{gillespie1976general,gillespie1977exact}, which generates statistically correct trajectories for a stochastic system of equations with constant reaction rates. 
The Gillespie algorithm is computationally more efficient than alternative simulation methods such as agent-based models with constant time increments~\cite{figueredo2014comparing}, and therefore, has been extensively used~\cite{arkin1998stochastic,martinez2010messenger, thomas2019probabilistic,pelissier2020computational,vestergaard2015temporal,sun2020stochastic}. 
However, real-world systems, particularly those comprising non-elementary reaction events that encapsulate multiple intermediate reaction steps, often exhibit dynamics that appear to have memory when viewed through observable states alone. To better model such systems, algorithmic generalizations of the Gillespie algorithm have been proposed. One such generalization involves Hidden Markov Models (HMMs), which, while still Markovian in their full state space, introduce hidden states that help capture longer-range dependencies in the observed data~\cite{rabiner1986introduction}. At their core, HMMs are mixture models that encode both visible (observable) and hidden (latent) states, allowing the system to model more complex stochastic behavior. HMMs have been extensively used to model sequences, where the decomposition into observable and unobservable states captures dependencies between consecutive measurements in a sequence. Besides modeling biological sequences~\cite{yoon2009hidden},  HMMs have been used for speech recognition~\cite{rose1996hidden}, economics~\cite{nguyen2015hidden}, and climate modeling~\cite{bracken2014hidden}. 
%~\cite{rabiner1986introduction, ghahramani1997factorial}, 
%
A drawback, however, is that HMMs require more parameters than simple Markov models and necessitate computationally expensive learning and inference algorithms. Furthermore, the hidden intermediate states can be difficult to interpret phenomenologically. 
%
%, which limits their predictive performance.
%
%In addition, the possible distributions are still constrained to Erlang and hyper exponential distributions~\cite{chiarugi2015modelling}.   
%\mrm{I commented a reference you added to support the use of Erlang and hyperexponential on HMMs. I checked the paper, and these distributions are used on generalized Petri Nets and Beta Workbench, not HMMs. }

Despite their elegance, the Gillespie algorithm and HMMs
might not be good modeling choices for real-world systems exhibiting strong non-Markovian behavior, such as quantum devices~\cite{white2020demonstration}, polymer reactions~\cite{guerin2012non}, molecular dynamics~\cite{ayaz2021non, vroylandt2022likelihood, lickert2020modeling}, biochemical reactions in single cells~\cite{bratsun2005delay, stumpf2017stem}, RNA transcription~\cite{cao2020analytical},  neuronal firing~\cite{baddeley1997responses}, social interactions~\cite{jiang2013calling,zhao2011social}, human activity patterns~\cite{barabasi2005origin}, or earthquakes~\cite{corral2004long}, just to cite a few.
In these cases, non-Markovian frameworks are necessary. However, non-Markovian stochastic processes are notoriously difficult to investigate, and exact analytical solutions have only been found for very simple problems, such as computing 
%such as cell division~\cite{jafarpour2018bridging} or 
stationary levels of a one-gene system~\cite{wang2020analytical}. In most other cases, their study relies on expensive numerical simulations.

To facilitate the investigation of non-Markovian systems, several methods have been developed. A common approach is to introduce delayed reactions, in which the effects of a reaction are not immediate but occur after a certain time lag~\cite{barrio2006oscillatory, cai2007exact, anderson2007modified, ramaswamy2011partial}. These models extend the Gillespie algorithm by incorporating a delay queue or event list that schedules the future execution of delayed reaction effects. When a delayed reaction occurs, the time of its effect is either fixed or sampled from a delay distribution, and the reaction is placed into a queue to be executed at the appropriate future time. We refer to this family of methods as stochastic simulation algorithms with delays (DelaySSA)~\cite{fu2022delayssatoolkit}. 
These methods are considered exact because they accurately simulate the system's stochastic behavior by incorporating delays, which preserve the integrity of the underlying probabilistic model. The delays extend the Gillespie algorithm to model time-dependent processes while maintaining its memoryless property, as the delays are computed based on the current state. The effectiveness of these methods, though, depends on our ability to compute the delays correctly and infer the appropriate delay distribution.
DelaySSA approaches face additional challenges. First, the distributions generating the delays must remain constant until the annotated event occurs. Hence, these methods may not adapt well to real-time changes in the system, such as those that evolve unpredictably.
Second, delays are computed dynamically when a reaction with a delay is selected. Specifically, all computed delays must be stored and updated in a sorted array, which significantly increases the algorithm's computational cost and complexity, especially with numerous delayed processes. 
These limitations underscore the trade-offs of using DelaySSA, where the benefits of incorporating delays into stochastic simulations must be weighed against increased computation and memory requirements and the potential for reduced accuracy in dynamical systems.

The non-Markovian Gillespie algorithm (nMGA) offers an alternative approach by assuming that inter-event times are exponentially distributed, similar to the standard Gillespie algorithm, but the individual rates depend on the elapsed times of the processes~\cite{boguna2014simulating}. nMGA becomes exact in the limit of an infinite number of reactants ($N \rightarrow \infty$). However, the approximation error can become large for systems with few reactants. Furthermore, nMGA requires recalculating the instantaneous event rates of each process after each event occurrence, a computationally expensive procedure.
Alternatively, the Laplace Gillespie algorithm~\cite{masuda2018gillespie} provides an exact and computationally efficient algorithm to simulate systems where the inter-event times are a continuous mixture of exponentials. Still,  the framework is applicable exclusively to monotone long-tailed distributions. 

\begin{figure}[h!t]
    \centering
    \captionsetup{width=1\linewidth}
    \includegraphics[width=0.5\linewidth]{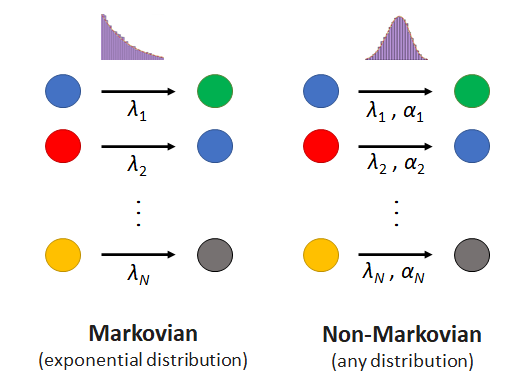}
    \caption{Diagram highlighting the differences between Markovian and non-Markovian stochastic models. \textbf{Left:} A set of reactants (blue, red, yellow,...) transforms into products through $N$ reaction channels. The reactions are memoryless, i.e. the inter-event times follow an exponential distribution with constant instantaneous rates $\lambda_i$. \textbf{Right:} In the non-Markovian case, the inter-event times follow more complex distributions characterized by a larger number of parameters, in this case, $\lambda_i, \alpha_i$. This results in instantaneous rates that depend on time.}
    \label{fig:intro}
\end{figure}

Another important class of stochastic systems is temporal networks, where interacting agents are depicted as nodes and their interactions as connecting edges, with links that change dynamically over time. These networks have been widely used to study mechanisms of information or signal propagation within populations, including the spread of opinions~\cite{li2006dynamics, zhan2019impact, wang2020public, liu2020homogeneity}, disease transmission~\cite{holme2016temporal, keeling2005implications, hazarie2021interplay}, brain dynamics~\cite{lynn2019physics, schmalzle2017brain}, ecosystem evolution~\cite{guimaraes2020structure, wang2020eco}, and the diffusion of innovations~\cite{cowan2004network, iacopini2018network}. However, the previously described methods -- DelaySSA, nMGA, Laplace Gillespie -- are primarily designed to model processes occurring on edges between nodes (i.e. pairwise interactions) rather than capturing the activity of individual nodes. While effective for certain dynamics, this edge-centric approach treats nodes as passive entities and cannot account for their intrinsic activity levels explicitly. Consequently, these methods struggle to represent node-driven phenomena, such as heterogeneous activation patterns, self-exciting processes~\cite{notarmuzi2021percolation}, or long inactivity periods, often critical in temporal networks.
For these processes, activity-driven (AD) models~\cite{perra2012activity, holme2012temporal} provide a more appropriate approach by offering a node-centric framework to capture temporal dynamics. These models assign an activity rate to each node, which dictates its probability of initiating interactions at each time step. In contrast to Gillespie’s process-oriented approach, which centers on interactions between pairs of nodes, AD models emphasize the activity of individual nodes, with interactions emerging based on predefined rules that govern these activity patterns. While AD models typically generate exponential inter-event time distributions (IEDs)~\cite{perra2012activity}, additional mechanisms—such as heterogeneous activity rates, self-exciting processes, or periodic resets with varying activity levels—can be used to capture more realistic bursty patterns and long-tailed event distributions~\cite{ubaldi2017burstiness, hiraoka2020modeling}. While these modifications enable the representation of diverse IEDs, most AD models lack rigorous mathematical guarantees and remain approximations of varying quality, where multiple events are typically aggregated within a single time step~\cite{ribeiro2013quantifying}. Recent advancements, such as the Spanning Tree method~\cite{sheng2023constructing}, partially address these challenges by precomputing activation times to enforce exact IEDs. Nevertheless, its reliance on conditional independence assumptions limits its ability to capture memory effects, such as first-order interdependencies. Thus, while AD models and their variants provide a flexible framework for exploring temporal networks, their computational constraints and reliance on approximations highlight the need for more robust methodologies to effectively model non-Markovian dynamics.

Here, we introduce REGIR (Rejection-based Gillespie for non-Markovian Reactions), a computationally efficient algorithm to simulate non-Markovian dynamics with arbitrary inter-event waiting time distributions (Figure~\ref{fig:intro}), while maintaining the same asymptotic computational complexity as the Gillespie algorithm. Our approach exploits rejection sampling, a process where propensities of unselected reactions are computed and subsequently discarded~\cite{thanh2014efficient,st2019efficient, thomas2019probabilistic,pelissier2020computational}. REGIR addresses many of the limitations of the DelaySSA~\cite{anderson2007modified, fu2022delayssatoolkit},  nMGA~\cite{boguna2014simulating}, Laplace Gillespie~\cite{masuda2018gillespie}, and AD modeling~\cite{perra2012activity, sheng2023constructing} methods. Namely, unlike DelaySSA, REGIR does not require storing all delays in a sorted array and can simulate arbitrary IEDs coupled to any instantaneous simulation rates, representing, for instance, individual reactant properties or macro-variables such as time or the number of reactants in each species.
Compared to nMGA, REGIR improves the approximation accuracy by allowing the reduction of the simulation's time step $\Delta t$ according to a user-defined threshold. Additionally, REGIR significantly reduces the computational complexity and running time by computing only the reaction rates of the sampled reactants instead of all reactants, unlike nMGA and Laplace Gillespie, which compute rates for all reactants. It also provides a versatile alternative to AD modeling~\cite{perra2012activity}, addressing its lack of mathematical guarantees in the context of node-driven non-Markovian temporal networks~\cite{holme2012temporal}, while preserving the ability to track intricate temporal neighborhood patterns that influence interactions.
Finally, REGIR's rejection sampling approach allows for a very high customization of stochastic simulations. Since each reactant can follow its own stochastic process, it is possible to efficiently simulate reactions where each reactant has \textit{individual properties}, such as individual reaction rates that continuously change according to time or external variables. This allows for the design of computationally efficient hybrid simulations, where the internal dynamics are coupled to macro-population variables.

This paper is structured as follows. We first introduce the REGIR algorithm and the standard parametrization of the distributions considered in the subsequent examples. Then, we demonstrate REGIR's advantages over other methods, such as the standard Gillespie, nMGA, DelaySSA, and AD methods using three different non-Markovian systems as examples. The first example describes RNA transcription~\cite{monk2003oscillatory} and exploits the classical framework of the Gillespie algorithm extended with delayed reactions and kinetic parameters that evolve over time. The second example models the maturation and competition of B cells during an immune response~\cite{pelissier2020computational}, differing from classical stochastic frameworks in that it allows reactants to have unique properties that influence their evolution. The third example simulates temporal social networks~\cite{le2023modeling} to study social interactions, where individuals are represented as nodes and their interactions are formalized as edges. Through these three examples, we demonstrate that REGIR
offers significant improvements in flexibility and computational over existing methods through a robust mathematical framework.

%\maria{Maybe you can move all prior examples to the supplementary? Unless you think they are adding noise.}

\section{Results}

\subsection{Rejection-based Gillespie for non-Markovian Reactions (REGIR).}
\noindent We consider a system of $N$ renewal processes. In the standard Gillespie (SG) algorithm, each process is represented by a reaction channel where identical reactants undergo reactions at the same rate. When the number of reaction channels increases, the simulations become increasingly computationally intense. This limits the applicability of SG to systems where reactants can be grouped into a few channels of identical particles. In contrast, REGIR can simulate systems where each reactant follows a distinct process with an individual reaction rate. For every single reactant, REGIR keeps track of the time elapsed since the last event, making the number of processes equal to the population size. In the SG, this would lead to a very large number of channels and infeasibly long running times, however, REGIR minimizes the computational cost using a rejection sampling approach described below.\\

\noindent \textbf{Rejection sampling}: We denote by $t_{j}$ the time elapsed since the last event of the $j$th process $(1 \leq j \leq N)$, and by $\lambda_{j}(t_j)$ the time-dependent reaction rate of the $j$th process. At each iteration, REGIR performs 4 steps, as described below:

\begin{enumerate}

    \item Set $\lambda_\text{max}$, the maximum reaction rate across all processes, such that:
    \begin{equation}
        \lambda_\text{max} \geq \max_{\{j \in [1,N]\}} \lambda_j (t_j).
    \end{equation}
    \item Compute the time increment to the next event as in SG using $\lambda_\text{max}$. Namely, a random variable $u$ is uniformly drawn from the interval $[0, 1]$, i.e. $u \in \mathcal{U}^{[0,1]}$. The time increment is computed as:  
    \begin{equation}
        \Delta t = \frac{\ln (1/u)}{N \cdot \lambda_\text{max}}.
        \label{eq:deltat}
    \end{equation}
    \item Select the process $j$ that produces the event. All processes have an equal probability of being drawn, and therefore, the probability of process $j$ is:
    \begin{equation}
        p_j = \frac{1}{N}.
        \label{eq:choice}    
    \end{equation}
    \item Accept the process with probability $p_\text{accept}$, given by:
    \begin{equation}
        p_\text{accept} = \frac{\lambda_j (t_j)}{\lambda_\text{max}}, 
        \label{eq:rejection}
    \end{equation}
    and update the reactants' population accordingly. If the process is rejected, the next event becomes an empty event, i.e. the reactant populations remain unchanged. Note that time advances regardless of if the event is accepted or rejected. 
    
\end{enumerate}

\noindent In the limit of $\Delta t \rightarrow 0$, which corresponds to $N\to\infty$ or $\lambda_\text{max}\to\infty$ according to Eq.~\ref{eq:deltat}, the probability  density function (PDF) of the IED $\psi_j (t_j)$ of the $j$th process becomes:
\begin{equation}
    \psi_j(t_j) = \lambda_j(t_j) \times \exp \left({-\int^{t_j}_{0} \lambda_j(\tau) d\tau}\right).
    \label{eq_PDF_main}
\end{equation}
The proof to Eq.~\ref{eq_PDF_main} can be found in the Supplementary Information (SI, Section~\ref{REGIR_proof}).

\paragraph{REGIR achieves arbitrarily high accuracy.} 

Strictly speaking, REGIR is not an exact algorithm, as rejection sampling introduces an approximation in simulating non-Markovian processes. However, REGIR becomes exact as $\Delta t \rightarrow 0$. In practice, arbitrarily high accuracy can be achieved by increasing $\lambda_\text{max}$, ensuring sufficiently small time steps at each iteration, and hence, preventing large time leaps that could introduce inaccuracies, as seen in other approximate methods such as nMGA.
We show here how an upper bound on the error can be established. Let us denote $\lambda_0$ the inverse of the mean IED of the process, which corresponds to the expected rate of the process observed at a random point in time (Supplementary section~\ref{mean_observed_rate}). We set $\lambda_\text{max}$ at each iteration such that
\begin{equation}
    \label{Eq:lambda_max}
    \lambda_\text{max} \geq \lambda_0 \ \ \ \text{where} \ \ \ \lambda_0 = \left(\int_{0}^{\infty} t \cdot \psi(t) \, dt\right)^{-1}.
\end{equation}
The time step increment is bounded by Eq.~\ref{eq:deltat}, which gives:  
\begin{equation}
    \label{eq:REGIR_increment_lambdamax}
    \Delta t  \leq \frac{\ln (1/u)}{N \cdot \lambda_0}.
\end{equation}
Analytically, the relative error in the simulated IED is bounded by the following approximation, as derived in Supplementary Section~\ref{quant_error_REGIR}:
\begin{equation}
    \label{Eq:REGIR_final_error}
 \mathbb{E} \left[ \mbox{error}_{ \mbox{\tiny REGIR} }\right] 
    \sim \frac{\langle \lambda' \rangle}{N \cdot \lambda_\text{max}^2} T \leq \frac{\langle \lambda' \rangle}{N \cdot \lambda_0^2} T
\end{equation}
where $\langle \lambda' \rangle = \int_{0}^{\infty} \lambda'(t) \cdot \psi(t) \, dt$ represents the average derivative of the process rates. Here, $\mathbb{E} \left[ \text{error}_{\text{\tiny REGIR}} \right]$  represents the expected total accumulated error over all time steps until a given process reacts, which occurs after approximately $\lambda_\text{max} N$ time steps. Since the system involves competition among multiple processes, a reaction from any particular one typically requires around $N$ simulation steps, where each step corresponds to an event occurring in the system. Additionally, due to the rejection sampling mechanism, the total number of time steps also scales proportionally with $\lambda_\text{max}$, reflecting the effect of discarded proposals in the simulation.

In practice, this error can be understood as the Earth Mover’s Distance (EMD)~\cite{rubner1998metric} between the ground truth and the simulated IED (Supplementary Section~\ref{arbitrary_accuracy}). IEDs are a fundamental component of non-Markovian simulations and are measurable in certain experimental systems such as bacterial inter-division times\cite{sauls2019control, si2019mechanistic, iyer2014scaling}. IEDs are a fundamental component of non-Markovian simulations and are measurable in certain experimental systems such as bacterial inter-division times\cite{sauls2019control, si2019mechanistic, iyer2014scaling}. Still, most non-Markovian studies report observables primarily regarding population dynamics, as this variable is generally more accessible experimentally~\cite{stumpf2017stem, england2010global, zeisel2011coupled}. However, establishing universal error bounds for REGIR in population dynamics is challenging, as the error is highly dependent on system design. While some systems are robust to variations in IEDs, others are highly sensitive to even minor perturbations. A detailed discussion on these aspects can be found in Supplementary Section~\ref{quant_error_REGIR}. 

%\maria{A minor revision: rearrange the order of the Supplementary Sections so they appear in numerical order in the main text, e.g., first B1, then B2, etc}
%The supplementary follows their own logical orders, so It's hard to do

\paragraph{Main differences between REGIR and SG.} 
There are several important differences between the REGIR and SG algorithms. First, the SG algorithm simulates systems in which \emph{each species has a single constant reaction rate}, leading to exponentially distributed inter-event times for its reactants (SI, Section 2). In contrast, REGIR allows reaction rates to vary dynamically for each process and \emph{reactant}.
Namely, for each process $j$, REGIR keeps track of the time since the last event $t_j$, enabling the simulation of non-Markovian processes under the assumption that $\Delta t \approx{0}$~\cite{boguna2014simulating}. 
Second, while SG tracks the rates of each reaction channel, REGIR only tracks the maximum rather than the individual rates of all processes at each time step. As we will show in Section~\ref{application1}, Figure~\ref{fig:hes1_model}E, this substantially reduces the computational cost of the algorithm~\cite{thanh2014efficient}, especially as the number of processes increases.  Finally, SG is exact for Markovian systems, while REGIR is an approximation algorithm. However, as we illustrate in Section~\ref{application1}, REGIR can achieve arbitrarily high accuracy for non-Markovian processes.

\paragraph{Distributions.}
In contrast to DelaySSA, which can sample time increments from any distribution, REGIR can only simulate IEDs with finite instantaneous rates. However, this is not a real limitation, as distributions that do not meet this condition can be simulated by setting the rate to an arbitrarily high value. Considering a renewal process with a probability density function (PDF) $\psi(t)$ and a
survival distribution function (SDF) $\Psi(t)= \int_{t}^{\infty} \psi(\tau)~d\tau$ , the instantaneous rate function is defined as:
\begin{equation}
    \lambda(t) = \frac{\psi(t)}{\Psi(t)}.
\end{equation}
%(i.e., the probability that the inter-event time is larger than ti) 
%
As an example, let us consider the Weibull distribution with PDF and SDF given by $\psi(t) =\beta t^{\alpha-1} \times \exp \left( - \frac{ \beta t^{\alpha}}{\alpha} \right)$ and $\Psi(t) =  \exp \left( - \frac{ \beta t^{\alpha}}{\alpha} \right)$ respectively~\cite{jiang2011study}. From these, we compute the instantaneous rate, $\lambda(t) = \beta t ^{\alpha-1}$. Setting $\alpha = 1$, we recover the  Markovian case with constant instantaneous rates. This is expected, as the Weibull distribution reduces to the exponential distribution when $\alpha = 1$. 
In the more general case, most distributions do not have 
instantaneous rates that can be written in simple analytical form (SI section~\ref{rate_proof}). However,  $\lambda(t)$ can be easily computed numerically with arbitrary precision. 

In this article, we illustrate REGIR's capability in simulating stochastic processes by employing gamma and Pareto distributions, both of which are representative of models commonly used to capture biological dynamics. For consistency with established practices in stochastic simulations, all processes are parameterized by a rate parameter $\lambda_0$ and a shape parameter $\alpha$. In Supplementary Section~\ref{REGIR_distributions}, we provide a detailed derivation of this alternative parameterization from the standard forms of these distributions, extending the framework to include other relevant distributions such as Weibull, Cauchy, normal, and log-normal. By unifying the parameterization across a diverse range of inter-event distributions, we facilitate REGIR's integration into various modeling workflows.

\subsection{Application I: Non-Markovian reactions with system-dependent inter-event times}

\label{application1}

\noindent We begin by introducing REGIR in the context of a general non-Markovian system. In the Gillespie kinetics framework, non-Markovian systems are typically represented through delayed reaction channels. These delays are either constant, allowing for easier analytical handling, or follow arbitrary probability distributions. Traditionally, such systems have been simulated using an annotated list of future events~\cite{fu2022delayssatoolkit, anderson2007modified}, a method distinct from the Gillespie approach. In contrast, frameworks that modify the Gillespie algorithm like REGIR, nMGA~\cite{boguna2014simulating}, and Laplace Gillespie~\cite{masuda2018gillespie}, introduce additional reaction channels to account for the hidden processes causing the observed delay. This allows for the simulation of complex IEDs by directly applying Gillespie's instantaneous kinetics approach to the underlying processes. We outline existing methods and their characteristics in Table~\ref{table:algorithm_comparison}. A notable distinction between DelaySSA and Gillespie-based approaches, such as nMGA and REGIR, is the way delays are handled. In DelaySSA, delays are calculated at the time of a reactant formation, instead of computing the instantaneous rate at each time step, as REGIR does. Hence, DelaySSA cannot adjust inter-event times once a delay is initiated, limiting its flexibility in modeling rapidly changing distributions. Another important difference is accuracy: DelaySSA is exact if the IED remains constant during the delay, whereas nMGA and REGIR are approximate, becoming exact only as $\Delta t \to 0$. However, REGIR can achieve any desired accuracy given a sufficiently high $\lambda_\text{max}$. We show in the  Supplementary Section~\ref{arbitrary_accuracy} that this can be accomplished at a reasonable computational cost, maintaining the same asymptotic complexity as SG—an advantage over nMGA, which suffers from higher computational overhead.

\begin{table*}[ht]
\renewcommand{\arraystretch}{1.3}
\centering
\captionsetup{width=1\linewidth}
\resizebox{\textwidth}{!}{
    \begin{tabular}{lcccll}
    \hline
    \textbf{Algorithm} & \textbf{\begin{tabular}[c]{@{}c@{}}Arbitrary inter-event\\ time distribution\end{tabular}} & \textbf{\begin{tabular}[c]{@{}c@{}}Dynamic IED\\params. adjustment\end{tabular}} & \textbf{Accuracy} & \textbf{\begin{tabular}[c]{@{}l@{}}Complexity/\\ simulation\end{tabular}} & \textbf{\begin{tabular}[c]{@{}l@{}}Complexity/\\ time step\end{tabular}} \\ 
    \hline
    Standard Gillespie & No & Yes & Exact & $O(N)$ & $O(1)$ \\ 
    Laplace Gillespie~\cite{masuda2018gillespie} & Limited* & Yes & Exact & $O(N \log N)$ & $O(\log N)$ \\ 
    DelaySSA~\cite{fu2022delayssatoolkit} & Yes & No & Exact & $O(N \log N)$ & $O(\log N)$ \\ 
    nMGA~\cite{boguna2014simulating} & Yes & Yes & Only when $N \rightarrow \infty$ & $O(N^2)$ & $O(N)$ \\ 
    REGIR  & Yes & Yes & Arbitrary high & 
    $O(rN) ^\dagger$ & $O(1)$\\
    %or $O(\lambda_\text{max})$
    \hline
    \end{tabular}}
\caption{Comparison of existing algorithms and REGIR for the simulation of a stochastic system with $N$ processes with the same inter-event time distribution. Here, we assume that executing the reaction and updating the population is $O(1)$. $^\dagger$REGIR requires more time steps than other methods due to event rejections, introducing a scaling factor $r$, the ratio of attempted to accepted events. This factor is proportional to $\lambda_\text{max}$, i.e. $r \propto \lambda_\text{max} / \lambda_0$. *Laplace Gillespie can only simulate inter-event times with monotonically decreasing probability density functions, e.g. it can simulate exponential, certain power-law, and some Weibull distributions, but it cannot handle log-normal, Pareto, or gamma and Weibull distributions with a shape parameter greater than one. Details on the complexity derivation can be found in Supplementary Section~\ref{SI_complexity}.}
\label{table:algorithm_comparison}
\end{table*}
\renewcommand{\arraystretch}{1}
%\maria{in the header of the table, "Adaptive process IED adjustment ". do you mean "Dynamic IED adjustment"?}

\noindent \textbf{Computational complexity.} To compare the computational performance of REGIR with other existing methods, let us first consider a simple reaction channel with $N$ reactants, all following the same IED over a duration $T$. Independently of the method chosen, the simulation will always be at least $O(N)$ as the number of time steps grows linearly with the number of ongoing processes. Let us estimate the computational cost associated with each time step for each method. In the case of Gillespie, updating the population is a straightforward $O(1)$ operation.  In contrast, DelaySSA incurs a cost of $O(\log N)$ per iteration because it maintains a sorted list of future events (Supplementary section~\ref{SI_complexity}). Similarly, Laplace Gillespie operates at $O(\log N)$, as it relies on a binary tree to draw instantaneous rates at each iteration. On the other hand, nMGA has a complexity of $O(N)$ per iteration since all individual rates have to be recalculated after each time step. Conversely, REGIR evaluates only the rate of the reactant being processed and avoids recalculating all rates at each iteration, thus maintaining an iteration complexity of $O(1)$. However, due to the possibility of event rejections, REGIR incurs more time steps compared to other methods, leading to an overall complexity of $O(r N)$, where $r$ represents the ratio of attempted over accepted events (Table~\ref{table:algorithm_comparison}, proof provided in Supplementary Section~\ref{proof_complexity}). This ratio is proportional to $\lambda_\text{max}$, a variable that can be used to set a lower bound on accuracy (Supplementary Section~\ref{arbitrary_accuracy}). In summary, the rejection framework introduced with REGIR leads to a significant simulation speed-up compared to alternative methods, as it is the only method that scales linearly with the number of reactants, apart from the standard Gillespie. This advantage is significant as long as the fraction of rejected reactions remains low, a balance that ultimately depends on the specific system considered. In the rest of this article, we illustrate REGIR's advantage in several common biological scenarios.

\subsubsection*{Modeling RNA transcription and protein synthesis with negative feedback loop}

\begin{figure}[h!t]
    \centering
    \captionsetup{width=1\linewidth}
    \includegraphics[width=0.95\linewidth]{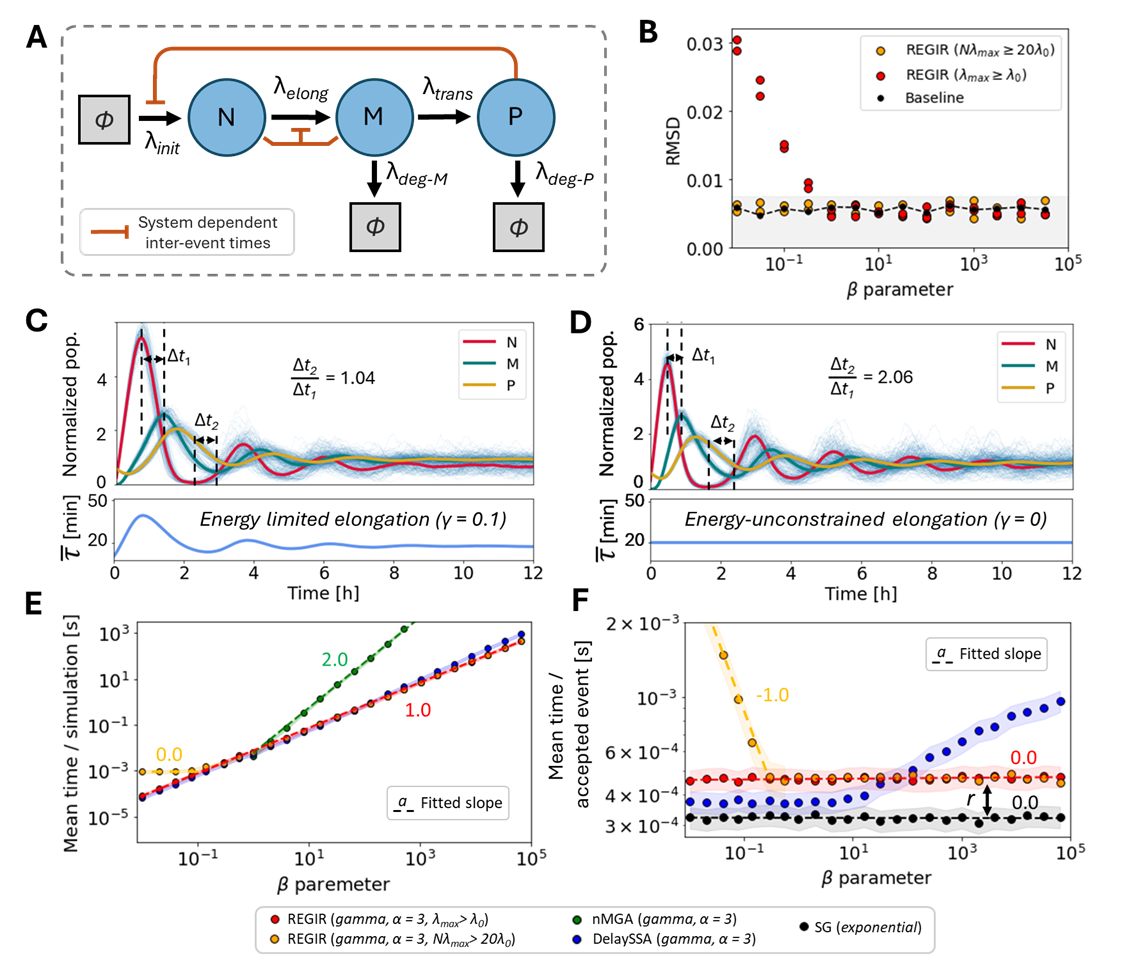}
    \caption{(A) Cartoon of the Hes1 model of RNA transcription and protein synthesis, where the reactions obey non-Markovian stochastic dynamics. A nascent RNA (N) forms and elongates until it matures into mature mRNA (M), which can be subsequently translated into proteins (P). Both mRNAs and proteins degrade at a constant rate. The brown arrow indicates that the instantaneous rates depend on the system's populations through a feedback mechanism. (B) Root mean square deviation (RMSD) between population trajectories simulated by REGIR and DelaySSA as a function of the scaling parameter $\beta$. Each RMSD value is averaged over $N_{\text{sim}} = 10^4 / \sqrt{\beta}$ independent simulations. Black dots indicate the RMSD between two DelaySSA averaged simulations, serving as a baseline, while red dots represent the RMSD between REGIR and DelaySSA. The shaded region marks the 95\% confidence interval for the RMSD expected between identical simulations. A parallel REGIR simulation, using the different condition of $N\lambda_\text{max} \geq 20 \lambda_0$ (Equation~\ref{Eq:lambda_max}) is also shown in orange. (C \& D) Observed oscillations in RNA and protein populations for energy-limited elongation (C) and energy-unconstrained elongation (D), averaged over 1000 simulations. Thin blue lines represent one single Gillespie simulation, while thick lines are the average over all simulations. Delays between nascent and mature mRNA's first peaks and first minima are depicted as $\Delta t_1$ and $\Delta t_2$, respectively. Changes in the ratio $\Delta t_2 / \Delta t_1$ illustrate the impact of changing elongation times during hes1 transcription. The corresponding mean elongation delay ($\bar{\tau} = 1/\lambda_0$) for both scenarios is also shown in blue. (E) The computational complexity of different methods per simulation as the scale parameter $\beta$ increases. Run times are averaged over 100 runs and slopes are obtained with a linear fit. (F) The same graph is shown after normalizing the computational time by the number of accepted events in each simulation. The standard deviation is highlighted by the shaded area. REGIR maintains a constant average runtime per accepted event while simulating non-Markovian processes. This runtime is $r$ times that of Gillespie, where $r$ represents the ratio of attempted over accepted events. In our model, $r = 1.6$.}
    \label{fig:hes1_model}
\end{figure}

\noindent To highlight the advantages of REGIR over other existing methods, we first consider a system describing hes1 RNA transcription. Extensive experimental evidence indicates that RNA transcriptions and protein synthesis often involve memory-dependent dynamics, underscoring the necessity for non-Markovian models to describe these biological processes~\cite{cao2020analytical, park2018chemical}. 
For instance, in the hes1 RNA system, the protein Hes1 represses the transcription of hes1 RNA, and this repression can be modeled using a delayed auto-inhibition loop (Figure~\ref{fig:hes1_model}A). This model was originally introduced to explain the oscillatory behavior observed in hes1 mRNA and protein expression~\cite{hirata2002oscillatory, monk2003oscillatory, barrio2006oscillatory}. 
In the following, we model the same system using REGIR. Generally, RNA transcription consists of three main steps. First, RNA polymerase catalyzes the elongation of a nascent nRNA molecule (denoted by N). Once elongation is complete, the RNA molecule becomes mature mRNA (denoted by M) and can be translated into a protein (P) by ribosomes. 
Focusing on the Hes1 system, we denote the number of nRNAs, mRNAs, and proteins as $N_N$, $N_M$ and $N_P$, respectively. 
For notation, reaction rates are labeled as $\xrightarrow{X}$, where $X$ indicates the inter-event time distribution. For reactions with a different IED, the specific details are explicitly stated. With this notation, the transcription kinetics for hes1 can be described as follows:
\begin{itemize}

    \item \textbf{RNA transcription initiation:} $\varnothing \xrightarrow{\beta  \cdot G(N_P)} \text{N}$. 
    The initiation of hes1 transcription is typically modeled with an instantaneous rate that decreases with the amount of Hes1 protein $N_P$~\cite{barrio2006oscillatory}. Here, we model $G(N_P)$ as a monotonically decreasing Hill function with coefficient $h$. $\beta$ is a scaling parameter representing the number of proteins necessary to inhibit the transcription rate by half, i.e.:
    \begin{equation}
        G(N_P) = \frac{1}{1 + \left(\frac{N_P}{\beta}\right)^h} \, . 
    \end{equation}
%
    %where $h$ the Hill coefficient and $N_P$ is normalized by $\beta$ to remain scale free.
    
    \item \textbf{nRNA elongation:} $\text{N} \xrightarrow{ \tau \sim \mathcal{G}(\lambda_0, \ \alpha_\text{elong})} \text{M}$. 

    The elongation of an RNA molecule typically takes 15 to 20 minutes to complete. In the context of DelaySSA, this elongation time is handled by adding a stochastic delay $\tau$, with mean $\tau_0$, to the RNA transcription initiation channel. In contrast,  REGIR and nMGA account for this delay by introducing an additional reaction channel with an arbitrary IED. For this example, we use a Gamma distribution $\mathcal{G}$ with a rate parameter $\lambda_\text{0} = 1/\tau_0$ and a shape parameter $\alpha_\text{elong} = 3$ (see Supplementary Eq.~\ref{Eq:gamma-param} for parametrization). The Gamma distribution is particularly well-suited to model biological delays, due to its flexibility in capturing a broad spectrum of stochastic processes, ranging from bursty dynamics to simple exponential behavior. Indeed,
    RNA polymerase elongation times often follow heavy-tailed distributions, characterized by frequent pauses during the process~\cite{qian2021basic, rajala2010effects}, which a Gamma distribution can effectively model.
    The choice of $\alpha_\text{elong} = 3$ reflects a scenario where the elongation process can be viewed as consisting of three sequential stages (e.g., initiation, elongation, and termination). While we use here $\alpha_\text{elong} = 3$  for illustrative purposes, we remind the reader the REGIR  is capable of accommodating any IED.

    Usually, the delay $\tau$ follows a probability distribution that does not change over time~\cite{brett2013stochastic}. However, it may be influenced by the total amount of mRNA being transcribed ($N_N$) or already present ($N_M$), a biologically plausible assumption given that resources needed for mRNA transcription, such as nucleotides or ATP energy, are finite~\cite{chen2016energy}. Thus, we consider  a more general form of the elongation rate parameter here~\cite{boguna2014simulating}:
%
    %\begin{equation}
    %    \lambda_0 = \tau_0^{-1} \cdot \exp \left(- \gamma \cdot \frac{N_N + N_M}{\beta}\right)
    %\end{equation}
%
    \begin{equation}
        \lambda_0 = \tau_0^{-1} \left(1 + \gamma \cdot \frac{N_N + N_M}{\beta}\right)^{-1} , 
    \end{equation}
    where $\gamma$ is a factor controlling the effect of limited energy on the elongation rate and mRNA populations are normalized by $\beta$ to ensure a scale-free behavior. Setting $\gamma = 0$ assumes no limiting effect. 

    \item \textbf{Protein synthesis:} $\text{M} \xrightarrow{\lambda_\text{trans}} \text{M} + \text{P}$. Mature mRNA molecules bind to ribosomes and translate into proteins. Since hes1 translation, once initiated, occurs within approximately one minute,  this delay is typically neglected~\cite{barrio2006oscillatory}. Note that a single mRNA molecule can be translated multiple times by ribosomes before being degraded.

    \item \textbf{RNA degradation:} $\text{M} \xrightarrow{\lambda_\text{deg-M}} \varnothing$.  mRNA molecules degrade over time. Experiments show that mRNA decay typically follows first-order kinetics~\cite{wang2002precision}, hence, we model this process as a Markovian process using an exponential IED.
    
    \item \textbf{Protein degradation} $\text{P} \xrightarrow{\lambda_\text{deg-P}} \varnothing$. Similarly, protein decay can also be described as a Markovian process using an exponential IED.

\end{itemize}

%This kind of stochatic feedback mechanisms is also useful to model the p53-Mdm2 feedback loop, NF-kB signaling, circadian rhythms involving CLOCK and BMAL1, and the Notch signaling pathway mechanism~\cite{a}. These systems have been pivotal in benchmarking stochastic simulation algorithms due to its characteristic delayed reactions~\cite{bayati2009d}. Additionally, complex interactions involving multiple proteins and mRNA productions further exemplify the broad applicability of these models in biological research~\cite{koshkin2024stochastic}.

Building upon the work of Monk~\cite{monk2003oscillatory}, where parameters were fit from  experimental data, we set $\lambda_\text{deg-M} = \SI{0.029}{\text{min}^{-1}}$, $\lambda_\text{deg-P} = \SI{0.031}{\text{min}^{-1}}$, $\lambda_\text{trans} = \SI{0.01}{\text{min}^{-1}}$, $\tau_0 = \SI{20}{min}$ and $h = 4.1$ to model the dynamics of RNA and protein populations in Hes1. 
%
%However, we opted to set the protein translation rate to $\lambda_\text{trans} = \SI{0.01}{\text{min}^{-1}}$ instead of $\lambda_\text{trans} = \SI{0.1}{\text{min}^{-1}}$ to prevent an overwhelming dominance of protein production in the simulation. A value of $\lambda_\text{trans} = \SI{0.1}{\text{min}^{-1}}$ resulted in over $90\%$ of reactions being involved in either protein degradation or translation, thereby impeding our ability to analyze the impact of non-Markovian mRNA transcription. 
As noted in earlier studies, methods relying on annotated lists of future events~\cite{fu2022delayssatoolkit} cannot be directly applied when there is a coupling between the distribution of time delays and the system's state. Consequently, we assess REGIR's accuracy under the specific condition of $\gamma = 0$, where the distribution of RNA elongation inter-event times remains unaffected by the RNA molecule population (energy-unconstrained elongation). We execute REGIR and DelaySSA across $N_\text{sim} = 10^4 / \sqrt{\beta}$ simulations and compare the root mean square deviation (RMSD) of normalized averaged populations for different values of the scale parameter $\beta$ (Figure~\ref{fig:hes1_model}B). As a baseline, we use the RMSD between two independent realizations of DelaySSA simulations. The number of simulations is scaled by $1/\sqrt{\beta}$ to compensate for the increased stochasticity in smaller systems.

The scale parameter $\beta$ represents a proxy for system size and, by design, is proportional to the total number of processes in the system \((N = N_N + N_M + N_P)\). Empirically, we observe that $N(t) \sim 25 \beta$ on average, where $N(t)$ denotes the total number of reactants at a given time. In the low-process regime $(\beta < 1)$, we find that the error scales inversely with $\beta$ (Figure~\ref{fig:hes1_model}B), consistent with theoretical expectations from Eq.~\ref{Eq:REGIR_final_error}, assuming a constant derivative rate $\langle \lambda' \rangle$:

\begin{equation}
    \mathbb{E}[\text{error}] \lesssim \frac{1}{N} \propto \frac{1}{\beta}.
\end{equation}

As \(\beta\) increases, the error decreases accordingly. Within our framework, we did not observe an RMSD significantly higher than the baseline for $\beta > 1$, roughly corresponding to $N=25$ RNAs and proteins in the system on average, even after averaging 10k simulations, showcasing REGIR's capabilities to model systems of moderate to large sizes. For smaller systems ($\beta < 1$), errors can be mitigated by increasing $\lambda_\text{max}$, albeit at the cost of higher computational demands (see Supplementary Section~\ref{arbitrary_accuracy}). While the standard implementation of REGIR sets $\lambda_\text{max} \geq \lambda_0$, as defined in Eq.\ref{Eq:lambda_max}, stricter constraints on $\lambda_\text{max}$ can substantially reduce errors. For example, setting the condition $N \lambda_\text{max} \geq 20\lambda_0$ ensured smaller time steps during simulations when the number of processes was low (below 20), resulting in an RMSD within the baseline for all scale parameters (Figure~\ref{fig:hes1_model}B, Supplementary Section~\ref{arbitrary_accuracy}). 

In Figure~\ref{fig:hes1_model}C \& D, we present the time-dependent dynamics of nascent RNA, mature mRNA, and proteins for $\beta = 20$, averaged over 1,000 simulations. Oscillations are evident in both the energy-limited ($\gamma = 0.1$) and energy-unconstrained ($\gamma = 0$) scenarios, with the delay between population maxima approximately doubling in the energy-limited case. The increased delay ratio ($\Delta t_2 / \Delta t_1$) between minima (low population, faster elongation) and maxima (high population, slower elongation) underscores REGIR’s capacity to model the effects of elongation time on transcription and translation. This capability is particularly relevant for investigating biological systems in which the availability of energetic resources modulates gene expression, such as during cellular stress or metabolic changes~\cite{smith2018metabolic, gwinn2008ampk}. 

Next, we assess the computational time required to simulate the system for 12 hours using various scale parameters $\beta$. In Figures~\ref{fig:hes1_model}E\&F, we highlight how the computational complexity scales with different scale parameters. We observe that nMGA scales as O($N^2$), while both REGIR and SG scale as O($N$). This is expected, as the parameter $\beta$ is proportional to the number of reactants in the system ($N \sim 25 \beta$). Initially, DelaySSA's running time appears similar to that of REGIR; however, when examining the time per successful event, DelaySSA's iteration time gradually increases, while REGIR's remains constant. This corresponds to the increase in computational expense associated with sorting delayed reactions in DelaySSA, resulting in an additional cost of $O(\log(N))$. Figure~\ref{fig:hes1_model}E\&F also shows that SG and REGIR's running times are proportional by a factor of $r$,  equal to 1.6 in our model. This value may vary depending on the rate values of each channel and the choice of the delay's distribution. In the $\beta < 1$ regime, the $N\lambda_\text{max} \leq 20\lambda_0$ implementation of REGIR retains a constant computational cost as its time steps are fixed at a fixed, minimum, making it less efficient than DelaySSA in this context. Nevertheless, the $\beta < 1$ regime represents a biologically unrealistic scenario where RNA transcription activity is below one molecule per hour. Under typical conditions, genes produce dozens to thousands of RNA molecules per hour~\cite{sidaway2014direct, churchman2011nascent, larsson2019genomic, kim2013inferring}, making such low transcription rates unrepresentative of most biological systems. Thus, while REGIR is effective for moderate to large systems, its performance in the $\beta < 1$ regime is of limited practical significance in this system.

Overall, REGIR demonstrates clear advantages in modeling transcription and translation dynamics with its ability to continuously adjust the reaction rates of individual reactants while maintaining competitive computational efficiency.

%For instance, an exponential distribution has a ratio of 0 (since $\hat\lambda =  \lambda_\text{max} = \lambda_i \; \forall i$, no reaction is rejected). On the other side of the spectrum, the Weibull instantaneous rate increases polynomially with time (SI, Section 2), so the maximum rate $\lambda_\text{max}$ increases quickly with $\alpha$, thus increasing the number of rejections. Other longer-tailed distributions with reaction rates increasing sub-linearly with time, e.g. the gamma distribution, will be less affected by changes in their respective shape parameter (Figure~\ref{fig:REGIR}D, SI Section 2). In general, simulating a distribution with smaller variance will increase the maximum rate and as a result also increase the computational cost. This is intuitively clear from Figure~\ref{fig:REGIR}D, where the rejected over accepted reaction ratio ($R/A$) monotonically increases with the shape parameter of the gamma and Weibull distributions, which inversely correlate to the variance of their respective distribution. On the other hand, as both $R$ and $A$ are proportional to the rate $\lambda_0$, the ratio $\nicefrac{R}{A}$ is independent of the mean IED $\nicefrac{1}{\lambda_0}$ and thus scale invariant.

%\newpage

\subsection{Application II: Stochastic processes with individual properties}
\label{IP_benchmark}
\noindent We now highlight the advantages of REGIR in modeling non-Markovian systems with individual properties. The Gillespie algorithm was originally developed for simulating stochastic chemical systems under mass-action kinetics—where all particles in the same reaction channel are identical. However, many biological processes involve reactants with distinct, evolving properties.
%
%\sout{Although the Gillespie algorithm was originally developed for simulating stochastic chemical systems with mass-action kinetics, it has since been adapted to a wide variety of applications. Many of these systems feature reactants with unique properties, which may be predefined at initialization or evolve dynamically through interactions}. 
%
These individual properties may represent parameters that influence reaction rates or the shape of inter-event distributions (IEDs) and may evolve dynamically as the system progresses. Accurately capturing these properties is essential for simulating complex systems, including epidemiological models with varying susceptibility~\cite{grossmann2020efficient}, social networks with individual sociability patterns~\cite{le2023modeling}, and biochemical interactions with evolving affinity~\cite{thomas2019probabilistic, pelissier2020computational}. These dynamics require stochastic frameworks that dynamically represent individual properties and their impact on system behavior.

Traditional algorithms like SG 
%handle varying properties within a channel by discretizing them into subchannels of particles with similar values. However, this significantly increases the number of processes to monitor, resulting in high computational costs and inaccuracies caused by the discretization process. \sout{
struggle with systems where reactions involve multiple reactant types with individual properties, as the number of processes and reaction channels scales rapidly with population size. For example, for two populations, A and B, with $N_\text{A}$ and $N_\text{B}$ having individual properties, SG requires handling  $N = N_\text{A} \cdot N_\text{B}$ distinct reaction channels independently.
%If each reactant pair has a unique rate influenced by individual properties, the number of processes scales as $N = N_\text{A} \cdot N_\text{B}$. 
Simulating such a system with SG  requires significant computational resources or approximations, such as discretizing reactants into subpopulations with similar properties~\cite{thomas2019probabilistic}. 
These approximations, however, come at the cost of accuracy, as they do not represent the full range of variability of the considered properties. 
In contrast, REGIR can handle such systems by directly determining the instantaneous reaction rate for each process $\mathcal{P}_j$, i.e. $\lambda_j(\mathcal{P}_j)$. Reactions are then accepted with a probability:
\begin{equation}
    p_\text{accept} = \frac{\lambda_j(\mathcal{P}_j)}{\lambda_\text{max}} \, ,
\end{equation}
where 
\begin{equation}
    \lambda_\text{max} = \max_{j \leq N} \left\{\lambda_j(\mathcal{P}_j)\right\}.
\end{equation}
This formulation allows REGIR to bypass the need for coarse approximations, preserving the full variability of individual reactant rates. Moreover, REGIR achieves computational efficiency by reducing the complexity of each simulation step to $O(1)$, compared to the $O(N_\text{A} \cdot N_\text{B})$ scaling of SG. By directly accounting for individual reaction rates without requiring major algorithmic modifications, REGIR emerges as a highly advantageous approach for modeling systems with a large number of dynamic processes (Table~\ref{table:algorithm_comparison_IP}).

\subsubsection*{Modeling B-cell maturation during an immune response}

\begin{figure}[h!t]
    \centering
    \captionsetup{width=1\linewidth}
    \includegraphics[width=0.83\linewidth]{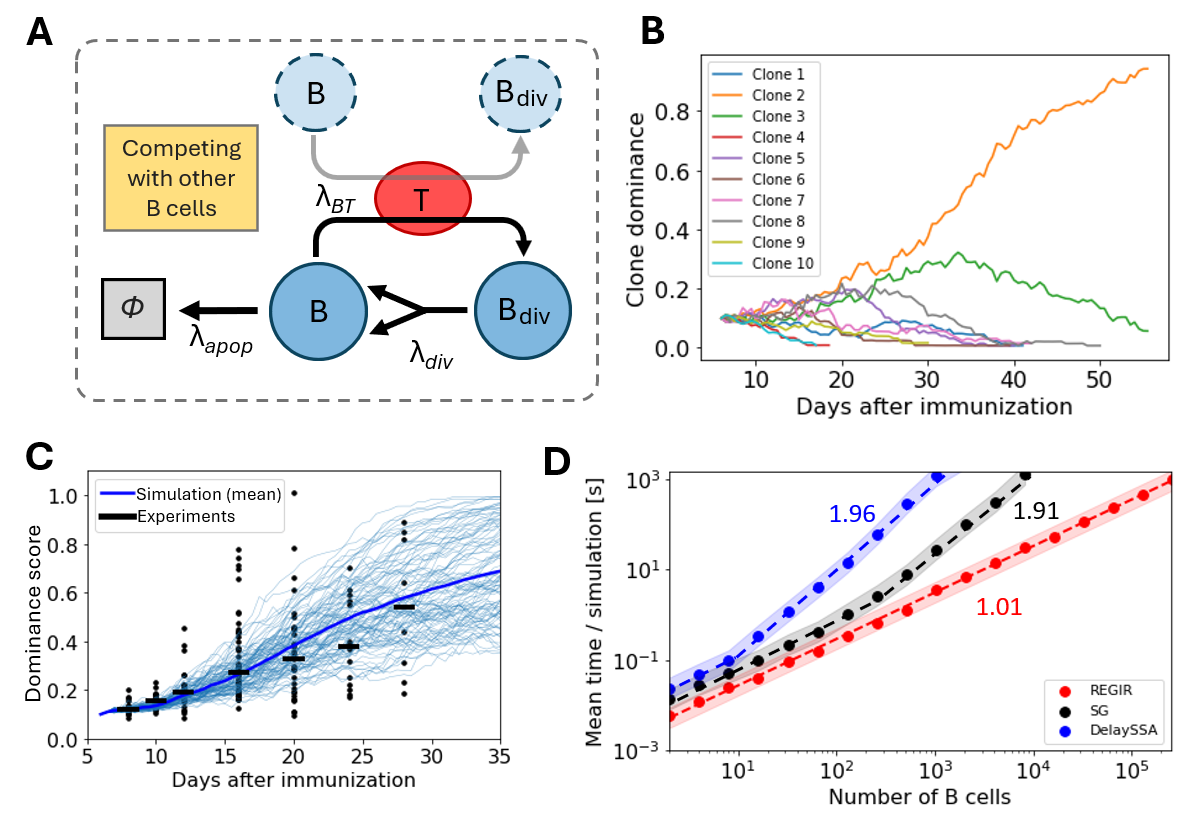}
    \caption{.3
    (A) Cartoon depicting B-cell affinity maturation in response to a foreign pathogen during an immune response. A B~cell with an antigen-binding receptor competes with other B cells (shown in light blue) for survival signals from a T~cell or undergoes apoptosis. This T-cell-mediated selection favors B cells with the highest-affinity receptors, promoting their proliferation and maturation, leading to the emergence of high-affinity clones. (B) Evolution of 10 distinct clonal families over time. Family 2 becomes dominant, leading to the disappearance of the other families. (C) Progression of the dominance score, defined as the fraction of the most expanded clone, throughout the immune reaction. Thin blue lines represent individual simulations, while the thick blue line shows the average of over 1000 simulations.  Experimental results~\cite{tas2016visualizing} are represented by a scatter plot, with individual points representing single measurements and a bar indicating the median. (D) Computational complexity is quantified as mean running time per simulation as the number of B~cells increases while the number of T~cells is kept constant. Running times are averaged over 100 runs and slopes are obtained with piecewise linear fits, with separate fits applied to different segments of the curve. Shaded areas around each fit indicate the standard deviation of simulation times across repeated runs.}
    \label{fig:immune_response}
\end{figure}

\noindent To demonstrate the advantages of REGIR in non-Markovian systems with individual properties, we consider the dynamics of B-cell maturation during an immune response where each cell follows distinct evolutionary paths leading to different phenotypic fate~\cite{martinez_quantitative_2012}. Namely, each B~cell expresses a different B-cell receptor (BCR) with varying affinity for a target antigen. B cells with higher-affinity BCRs have a competitive advantage, as they are more likely to receive survival signals from T cells and proliferate.
(Figure~\ref{fig:immune_response}A). This stochastic system has been described in detail in~\cite{thomas2019probabilistic,pelissier2020computational}. Here, we present a simplified system to illustrate REGIR's modeling capabilities. To study the impact of individual properties without introducing additional complexity, we assume all reactions follow Markovian dynamics, as this approximation has been shown to effectively capture key aspects of B~cell evolution.

\begin{itemize}
    \item \textbf{B cell binding to T~cells}: B + T $\xrightarrow{\lambda_\text{BT}}$ $[$BT$]$. A B~cell initiates an interaction with a T~cell. Here, we assume that T~cells are all identical and thus recognize the same antigen.
   
    \item \textbf{B-cell competition for T-cell help}: $[$B$_1$T$]$ + B$_2$ $\xrightarrow{(IP)}$ $[$B$_2$T$]$ + B$_1$. Even if a cell B$_1$  is already interacting with a T cell and receiving survival signals  (B$_1$T), it can be displaced by a new cell B$_2$ with a higher-affinity receptor, i.e.  $\text{affinity}(B_2) > \text{affinity}(B_1)$.
    %A new B~cell (B$_2$) exchanges positions with another B~cell (B$_1$) that was engaged with a T~cell. 
    %This reaction occurs only if B$_2$'s receptor can bind the antigen with higher affinity compared to that of B$_1$. 
    This competition is central to driving affinity maturation in B~cells.  (IP) in the reaction equation denotes that the process incorporates the individual properties (IP) of each B cell, accounting for variations in receptor affinity, and resulting in different propensities, or rates, for each existing (B, T) pair.
    
    \item \textbf{B-cell apoptosis}: B $\xrightarrow{\lambda_\text{apop}} \varnothing$. B~cells undergo apoptosis if not rescued by T~cells. This mechanism eliminates non-competitive B~cells from the system.
    
    \item \textbf{B-cell spontaneous unbinding from T~cells}: $[$BT$]$ $\xrightarrow{\lambda_\text{unbind}}$ B$_\text{div}$ + T. After receiving sufficient survival signals, B~cells detach from their T~cell and prepare for further divisions.
    
    \item \textbf{B-cell division}: B$_\text{div}$ $\xrightarrow{\lambda_\text{div}}$ B + B. Selected B cells undergo division, producing two daughter cells. Let $B_p$ and $B_d$ represent a parent and a daughter B~cell, respectively. After each division event, the affinity change of the daughter cell  $B_d$  is computed as follows:
    \begin{equation}
        \label{affinity_change}
        \Delta_\text{aff} = 
        \text{affinity}(B_d) - \text{affinity}(B_p) =
        \frac{u - \text{affinity}(B_p)}{\beta},
    \end{equation}
    where $u$ is a uniformly distributed random variable in the range $[0, 1]$, and $\beta$ is a scale parameter that determines the magnitude of the affinity change. This formulation makes it more challenging for cells with higher affinity values to further improve their affinity (Supplementary Figure~\ref{fig:GC_affinity}A) 
    %\maria{you might want to plot in the supplementary a plot showing the average affinity increase/decrease as a function of the parent initial affinity. This could be companion subplot of fig. S5}. 
    %Using the computed affinity change, the affinity of each daughter cell is then updated according to:
%
%    \begin{equation}
%        \label{affinity_function}
%        \text{affinity}(B_d) = \text{affinity}(B_p) + \Delta_\text{aff}.
%    \end{equation}
%    
    In our simulations, B-cell affinities start at 0 and evolve dynamically through successive divisions. After 50 days, affinity values reach approximately $0.7 \pm 0.05$ (Supplementary Figure~\ref{fig:GC_affinity}B), reflecting the gradual accumulation of affinity-enhancing mutations over time.
    %\begin{equation}
    %    \label{affinity_function}
    %    \text{affinity}(B_d) = \text{affinity}(B_p) + \frac{u-\text{affinity}(B_p)}{M},
    %\end{equation}

\end{itemize}

%These mechanisms can easily be implemented in REGIR, where we can reproduce the affinity maturation process and the diversity metrics of the repertoire.

Next, we use our model to investigate clonal dominance, where affinity-based competition drives the selective expansion of higher-affinity B-cell clones, often leading to the elimination of lower-affinity clones~\cite{tas2016visualizing}. Here, we define a \textit{clonal family} as the progeny derived from a single founder B cell. Following our prior work~\cite{pelissier2020computational}, we initialize the system with $N_B = 1000$ B~cells  and $N_T = 10$ T~cells. We run the simulation over a period of 50 days using the parameters $\beta = 10$, $\lambda_\text{BT} = \SI{0.146}{h^{-1}}$, $\lambda_\text{unbind} = \SI{2}{h^{-1}}$, $\lambda_\text{apop} = \SI{0.084}{h^{-1}}$ and $\lambda_\text{div} = \SI{0.134}{h^{-1}}$~\cite{pelissier2020computational}. Figure~\ref{fig:immune_response}B\&C illustrates the clonal competition among B-cell families.
Figure~\ref{fig:immune_response}B shows the progression of the dominance (the relative frequency of a given clone in the repertoire) of the 10 most expanded B-cell clones in a representative simulation, where the second clonal family becomes dominant after several days. 
Our simulations (Figure~\ref{fig:immune_response}C) align with experimental findings~\cite{tas2016visualizing}, showing a monotonic increase in the dominance of the most expanded clone, as it gradually overtakes the population, consistent with affinity-driven selection dynamics.

To evaluate the computational complexity of the standard Gillespie algorithm, DelaySSA, and REGIR, we ran 100 simulations for each method, progressively increasing the number of B cells ($N_B$) while keeping the number of T cells ($N_T$) constant (Figure~\ref{fig:immune_response}D). An important aspect to consider is that here, as the reaction channel involved with individual properties deals with pairs of reactants, the number of distinct processes scales up to $N = N_B \cdot N_T$, so that updating the propensities for each B-[BT] pair after a reaction becomes computationally expensive for some methods. Importantly, since each T cell is bound to a B cell, it can be regarded as inheriting individual properties from the B cell to which it is attached.
%\maria{I'm confused. T cells are identical in the simulation, i.e. they have no individual properties, correct? If so, increasing  $N_B$ should result in the number of processes scaling as $N_B^2$, as it is a competition between 2 B cells. Also, you should mention somewhere that, in this simplified model, we assume T cells are identical, i.e. they recognize the same antigen}. 
Hence, SG must account for $N = N_B \cdot N_T$ reaction channels instead of a single channel for indistinguishable B-[BT] pairs. This results in a computational complexity per time step of $O(N)$ time per step (Table~\ref{table:algorithm_comparison_IP}).

Methods have been proposed to address this computational inefficiency. For instance,  the tree-based Gillespie algorithm~\cite{gibson2000efficient} organizes reaction propensities in a sorted binary tree structure, reducing the time required to select the next reaction to $O(\log N)$. However, updating the tree after each event incurs an additional cost of $O(k \log N)$, where $k$ is the number of propensities that need to be updated, making the overall cost comparable to that of DelaySSA (Table \ref{table:algorithm_comparison_IP}). For methods like DelaySSA and tree-based Gillespie, the update process becomes particularly expensive in dynamic systems with frequently changing individual reaction rates, as seen in this study.
%
%Specifically, these methods require evaluating the affinity of all B~cells after each reaction, leading to a linear scaling of the update cost with the number of B~cells ($k \sim O(N_B)$). 
%In contrast, REGIR  handles each process individually resulting in a $O(rN)$ complexity, making it  more efficient in systems with many distinct processes. 
In contrast, REGIR does not require propensity updates at each iteration, resulting in a $O(1)$ complexity per time step.

\begin{table}[htbp]
\centering
\captionsetup{width=1\linewidth}
\setstretch{1.3} % Set the spacing for the table content
\resizebox{\textwidth}{!}{
\begin{tabular}{lcccll}
\toprule
\textbf{Algorithm}         & \textbf{Individual} & \textbf{Arbitrary inter-event } & \textbf{Accuracy}             & \textbf{Complexity/} & \textbf{Complexity /} \\
                           &                       \textbf{properties }                    &                                       \textbf{time distribution}            &                                & \textbf{simulation}   & \textbf{time step}        \\
\midrule
Standard Gillespie         & Yes                                        & No                                                & Exact                          & $O(N^2)$             & $O(N)$               \\
Tree-based Gillespie~\cite{gibson2000efficient}        & Yes                                        & No                                                & Exact                          & $O(N k \log N)$      & $O(k \log N)$        \\
DelaySSA~\cite{fu2022delayssatoolkit}                   & Yes                                        & Yes                                               & Exact                          & $O(N k \log N)$      & $O(k \log N)$        \\
REGIR               & Yes                                        & Yes                                               & \setstretch{0.8} \begin{tabular}[c]{@{}c@{}}Arbitrary high\\ (exact for Markovian)\end{tabular}    & $O(rN)$              & $O(1)$               \\
\bottomrule
\end{tabular}}
\caption{Comparison of existing algorithms and REGIR for the simulation of stochastic systems with individual properties. Here, we assume that executing the reaction and updating the population count incurs an $O(1)$ cost. $k$ refers to the number of propensity updates per iterations, $r$ is the ratio of attempted over accepted event. Details on the complexity derivation can be found in Supplementary Section~\ref{SI_complexity}.}
\label{table:algorithm_comparison_IP}
\end{table}
\renewcommand{\arraystretch}{1}

%\maria{Tables 1, 2 and 3 are almost identical. Can we merge them? Clearly, not all methods will be evaluated over all properties (e.g. DelaySSA will not be evaluated for Adaptive node rate adjustment, but we can still have a single unified table with different properties }

Interestingly, our polynomial fits of average running times per simulation (Figure~\ref{fig:immune_response}D) show that the computational costs do not universally scale as $O(N^\alpha)$ but instead exhibit different regimes. This observation can be explained by considering the following:  
For a small number of B~cells, all methods scale as $O(N_B)$. This scaling occurs because the time spent on propensity updates is small compared to other aspects of the simulation, such as updating the population counts and affinities, which scale as $O(1)$ per iteration. However, as the number of B~cells increases, propensity updates begin to dominate, shifting the scaling behavior upwards.
Beyond a certain number of B~cells (10 for DelaySSA and 100 for the standard Gillespie), the complexity of Gillespie and DelaySSA scales as $O(N_B^{1.91})$ and $O(N_B^{1.96})$, respectively, not as $O(N_B^2)$, as we previously reasoned. This happens because, as $N_B$  grows while keeping $N_T$ constant, an increasingly smaller fraction of B-[BT] pairs actually interact. As a result, the total reaction rate (propensity) decreases, leading to larger time steps in the simulation (as time steps are inversely proportional to the total propensity). This may explain the slight discrepancy observed in our estimated scaling behavior.

%\maria{i didn't fully understand the explanation you wrote, so I'm proposing this rewriting}   
%\sout{This outcome stems from the fact that, as $N_B$ increases, a smaller fraction of B-[BT] exhibits non-zero interaction rates. Consequently, the total propensity (the sum of all reaction rates) decreases, which in turn leads to larger time steps in the simulation, as the time step is inversely proportional to the total propensity. Specifically, with a fixed number of T~cells, there are roughly $N_B^* \cdot N_T$ pairs with non-zero rate, where $N_B^*$ represents the expected number of B~cells with higher affinity than those currently bound to T~cells. The value of $N_B^*$ is driven by the choice of the affinity function (Eq.~\ref{affinity_function}), and does not scale strictly as $O(N_B)$}.

In summary, REGIR offers significant advantages for modeling stochastic systems with individual properties, and particularly, it efficiently simulates large-scale systems with dynamic rates and complex pairwise interactions,  while maintaining a competitive computational cost.

\subsection{Application III: Non-Markovian temporal networks}

\noindent \noindent Temporal networks are network models in which the links between nodes, representing entities or reactants, are not static but change over time (Figure~\ref{fig:social_interactions}A). These evolving connections capture interactions between nodes at different times and of different durations, adding a temporal dimension essential for representing dynamical patterns that static network models cannot capture~\cite{holme2012temporal}. Until now, our discussion has focused on non-Markovian reaction channels involving a single reactant, where the inter-event time is simply defined as the waiting time for an individual reactant before the event occurs. The only exception has been a multi-reactant reaction involving competition for T cells ($[$B$_1$T$]$ + B$_2$ $\rightarrow$ $[$B$_2$T$]$ + B$_1$). However, this process is still governed by an exponential distribution and a binary rejection/acceptance framework based on individual node properties, eliminating the need for explicit tracking of the inter-event time distribution for each reactant pair. In general, in systems where multiple entities—such as chemical reactants or network nodes—interact, the concept of "inter-event time distribution" can refer to different aspects of the system's dynamics. For example, we might focus on the activity patterns of individual entities, such as $A_i$ or $B_j$, where the inter-event time distribution, $\psi_A(t)$ or $\psi_B(t)$, describes how often each entity is involved in an event. Alternatively, we might study the interactions between specific pairs, such as $(A_i, B_j)$, where the inter-event time distribution, $\psi_{AB}(t)$, describes the time intervals between the interaction of a given pair. In temporal networks, the activity of individual nodes is often the primary focus~\cite{ubaldi2017burstiness, sheng2023constructing}, but understanding pairwise interactions is also essential for capturing relationship-specific dynamics. Importantly, the IED of individual nodes, $\psi_A(t)$,  differs fundamentally from the IED of pairwise interactions, $\psi_{AB}(t)$. Specifically, $\psi_{AB}(t)$ can be expressed as a weighted sum of convolutions of $\psi_A(t)$, where the weights are determined by the probabilities of pairing events that either involve or exclude each possible node pair (Supplementary Section~\ref{distribution_relationship}). This relationship is inherently complex, and explicit derivation is only straightforward in the case of exponential distributions. Within the context of temporal networks,  this complexity highlights the importance of deciding whether to consider edge-associated or node-associated IEDs.

%heterogeneous activationpatterns and self-exciting processe

In reaction channels involving two reactants, each pair of interacting nodes represents a process, allowing methods such as REGIR and DelaySSA to accurately account for the inter-event time distribution associated with each connection (i.e., edges linking node pairs). However, in non-Markovian temporal networks, the primary focus often shifts to the activity patterns of individual nodes rather than their pairwise interactions~\cite{holme2012temporal}. This distinction renders DelaySSA and the version of REGIR previously introduced unsuitable for such scenarios, as these methods cannot independently disentangle the inter-event times of individual nodes from those of their pairwise connections.

%connectivity-driven model~\cite{erdHos1960evolution, watts1998collective, barabasi1999emergence} offer little flexibility on the choice of rules.

\subsubsection*{REGIR for node-driven non-Markovian temporal networks (REGIR-TN)}

\label{Section: REGIR-TN}

\noindent Previously, the standard Gillespie’s method has been used to simulate time-varying networks through a straightforward \textbf{A + A} $\to \texttt{interaction}$ kinetic channel~\cite{vestergaard2015temporal, unicomb2021dynamics}. A key advantage of this approach is that the time step is inherently defined by the modeling framework, allowing for a single event per step. This contrasts with activity-driven modeling, where multiple events are aggregated within a single time step, and varying levels of aggregation can lead to different resulting networks~\cite{ribeiro2013quantifying}. However, its application is restricted to Markovian systems, i.e. exponential distributions. Here we extend the REGIR framework to allow it to specify the inter-event time distribution of individual nodes (node interduration distribution) to an arbitrary IED. 

Let us consider a reaction channel \textbf{A + A} $\to$ \textbf{X}. We denote by $N_A$ the number of nodes. As each pair of nodes can interact, we have $N_A (N_A-1)/2$ processes. For each individual node $\text{A}_i$  $(1 \leq i \leq N_A)$ and $\text{A}_j$  $(1 \leq j \leq N_A)$, we consider $t_i$ and $t_j$ the time elapsed since the last interaction involving node $A_i$ and $A_j$, respectively. At each iteration, two nodes $A_i$ and $A_j$ are drawn, and their interaction propensity is determined by a time-dependent pairwise interaction rate $\Lambda_{ij}(t_i, t_j) > 0$, which accounts for the individual times of both nodes. Importantly, $\Lambda_{ij}$ can also incorporate pair-specific properties or interaction history of the node pair $(i, j)$, enabling the modeling of \textit{temporal neighborhood effects} — that is, interaction dynamics shaped by the past interactions or unique relationship between the two nodes. The REGIR algorithm for node-driven temporal networks (REGIR-TN) involves four steps:
\begin{enumerate}

    \item Set $\Lambda_\text{max}$, the maximum reaction rate over all processes, such that:
    
    \begin{equation}
        \Lambda_\text{max} \geq \ \max_{\{(i,j) \in [1,N_A], \ i \neq j\}} \Lambda_{ij}(t_i, t_j).
    \end{equation}

    \item Compute the time increment to the next event using $\Lambda_\text{max}$. Namely, a random variable  is uniformly drawn from the interval $[0, 1]$, i.e. $u \in \mathcal{U}^{[0,1]}$, and the time increment is computed as:  
    
    \begin{equation}
        \Delta t = \frac{2 \ln (1/u)}{N_A (N_A-1) \cdot \Lambda_\text{max}}.
        \label{deltat_TN}
    \end{equation}

    \item Select the two reactants $A_i$ and $B_j$ for the next event. All reactants have an equal probability of being drawn, and therefore, the probability of selecting $A_i$ and $A_j$ are:
    
    \begin{equation}
        p_i = \frac{1}{N_A} \ \text{and} \ p_j = \frac{1}{N_A - 1},  \ \ \text{respectively}
    \end{equation}

    \item Accept the process with probability $p_\text{accept}$, given by:
    \begin{equation}
        \label{Eq:REGIR_compute_pairwise_rate}
        p_\text{accept} = \frac{\Lambda_{ij} (t_i, t_j)}{\Lambda_\text{max}}, 
    \end{equation}
    and update the reactants' population accordingly. If the process is rejected, the next event is set to an empty event, i.e. the reactant populations remain unchanged.

\end{enumerate}
Here, the definition of the pairwise rate $\Lambda_{ij}$ is crucial, as it directly governs both the interaction rate between node pairs and the resulting inter-event time distribution for individual nodes. We assume that, each node possesses an \textit{intrinsic} instantaneous rate, denoted as $\lambda_{i}(t_i)$ and $\lambda_{j}(t_j)$ for nodes $A_i$ and $A_j$, respectively. These rates, akin to activity-driven models, quantify the node's propensity to interact and can be conceptualized as a broadcast signal indicating its willingness to engage in interactions. We also introduce a scale free rate  $\tilde{\lambda}(t) = \lambda(t) / \lambda_0$ and an arbitrary scaling constant $\Lambda_0$ for clarity.

For a given pair of nodes, we can consider different interaction scenarios. In one scenario, interactions are dominated by the most active node, reflecting an independent additive process. Here, the pairwise rate can be expressed as $\Lambda_{ij} \propto \Lambda_0 \left[\tilde{\lambda}_i(t_i) + \tilde{\lambda}_j(t_j)\right]$, making it well-suited for modeling directed communication networks, such as epidemic spreading, email exchanges or message transmissions. Alternatively, in a scenario where interactions require synchronous availability between the two nodes, a multiplication emerges as a natural choice, $\Lambda_{ij} \propto \Lambda_0 \left[\tilde{\lambda}_i(t_i) \cdot \tilde{\lambda}_j(t_j)\right]$, accounting for the fact that the interaction rate is limited by the node with the lower availability, as neither can sustain the interaction independently. This type of temporal system, requiring synchronization or simultaneous availability of interacting agents, is prevalent across various fields. In neuroscience, the precise timing of signals between neurons is critical for effective communication, as synaptic transmission often relies on the synchronized release of neurotransmitters and receptor activation~\cite{kelly2012framework}. Similarly, in social interactions, effective engagement typically requires both participants to be active simultaneously~\cite{sheng2023constructing}. In molecular biology, ligand-receptor binding events also depend on synchronization, as successful interactions occur only when both molecules are available at the same time~\cite{guevara2014use}.\\

%\maria{this does not work regarding dimensions. if you look at eq. 17, $\Lambda_{ij}$ has the units of $1/t$, and hence, the case of $\Lambda_{ij} \propto [\lambda_i(t_i) + \lambda_j(t_j)]$ is correct as we expect $\lambda_j(t_j)$ would also have $1/t$ dimension. however, then, $\Lambda_{ij} \propto [\lambda_i(t_i) \times \lambda_j(t_j)]$ would lead to $\Lambda_{ij} \sim 1/t^2$ ... You might want to define a dimensionless $\tilde{\lambda}_i(t_i) = \frac{\lambda_i(t_i)}{\lambda_0} $, where $\lambda_0$ is a typical time scale of the system. Then, in the additive case, $\Lambda_{ij} = \Lambda_0 (\lambda_0) \left[ \tilde{\lambda}_i(t_i) + \tilde{\lambda}_j(t_j) \right]$. In the multiplicative case, $\Lambda_{ij} = \Lambda_0 (\lambda_0) \, \left[ \tilde{\lambda}_i(t_i) \cdot \tilde{\lambda}_j(t_j) \right] .$} 

%\maria{this requires more explanations. first, can you give some intuitive explanation of why the additive case leads to higher interaction rates? maybe because these rates a typically small and hence, addition leads to a larger probability of interaction than multiplication? would this  still be true if the rates were not so small? Also, why does this larger probability of interaction "restrict flexibility in shaping the inter-event time distribution of individual nodes "? }
In Supplementary Section~\ref{REGIR_paired_rate}, we explore different formulations of the pairwise interaction rate $\Lambda_{ij}$ and their influence on the inter-event time distribution of individual nodes. With the additive rate, nodes can still engage in interactions even when their intrinsic rate is zero, as they may be randomly paired with a highly active node. As a result, there remains a nonzero probability of events occurring at $t_i \approx 0$, limiting the flexibility in shaping the inter-event time distribution. This constraint is particularly relevant for distributions such as Gamma or Weibull ($\alpha > 1$), which inherently assign zero probability to events occurring at $t=0$. On the other hand, we prove in Supplementary Section~\ref{REGIR_paired_rate_AA} that defining a multiplicative pairwise rate $\Lambda_{ij}$ with
\begin{equation} 
    \label{Eq:paired_rate}
    \Lambda_{ij} (t_i, t_j) = \frac{N}{N-1} \cdot \frac{\lambda_i(t_i) \cdot \lambda_j(t_j)}{\sum_{k=1}^{N} \lambda_k(t_k)}
\end{equation}
allows for the inter-event time distribution of node $A_j$ to converge to the PDF of its intrinsic rate as $N \to \infty$, through the relation:
\begin{equation}
    \psi_A{_j}(t) \underset{N \to \infty}{=} \lambda_j(t) \times \exp \left({-\int^{t}_{0} \lambda_j(\tau) \ d\tau}\right).
    \label{eq_PDF_TN}
\end{equation}
The formulation of the multiplicative rate $\Lambda_{ij}$, emerges naturally from considering the minimum event times of $N$ independent processes, governed by a total rate of $\sum_{k=1}^{N} \lambda_k(t_k)$ (Supplementary Section~\ref{pairwise_rate}). It reflects the combined likelihood of both nodes being available, weighted by their respective contributions to the total interaction rate. The additional term $N/(N-1)$ is a correction factor to account for the fact that a node cannot interact with itself, as derived in Supplementary Section~\ref{REGIR_paired_rate_AA}.

In summary, REGIR-TN addresses key limitations of AD models by enabling the simulation of arbitrary node IEDs and dynamically optimizing time steps through a rigorous mathematical framework (Figure~\ref{fig:social_interactions}A). Unlike the spanning-tree method, which constructs temporal networks to reproduce specific IEDs but may lack flexibility in handling complex interaction patterns~\cite{sheng2023constructing}, REGIR-TN offers enhanced adaptability. This flexibility is crucial for simulating intricate real-world systems, including temporal neighborhood effect~\cite{le2024flow, bail2025generalizing}, and higher-order relationships~\cite{han2024probabilistic, lin2024higher} that extend beyond pairwise connections by considering interactions where nodes preferentially engage with neighbors of their neighbors~\cite{majhi2022dynamics, le2023modeling, iacopini2024temporal}.

\subsubsection*{Modeling a face-to-face social interaction network}

\begin{figure}[h!t]
    \centering
    \captionsetup{width=1\linewidth}
    \includegraphics[width=1\linewidth]{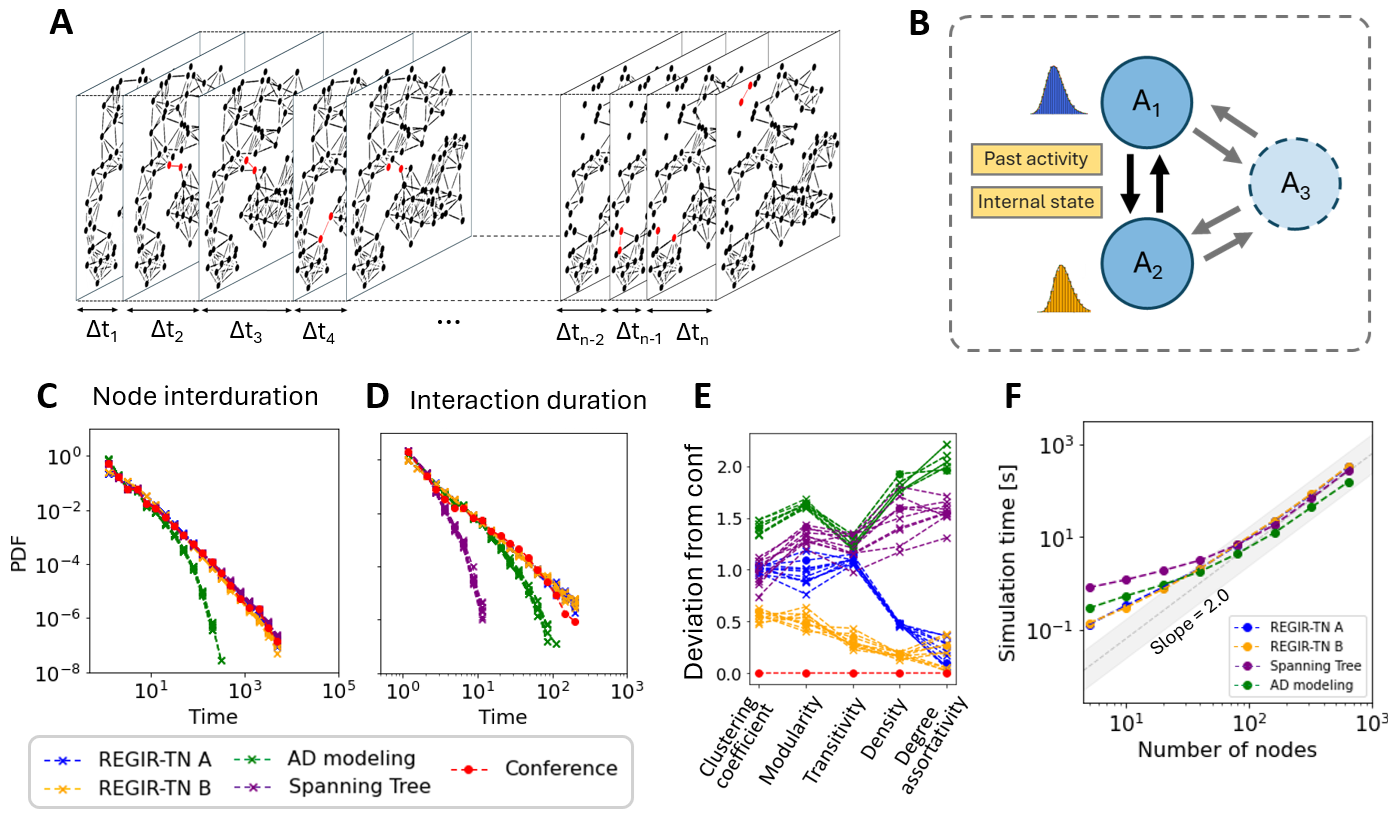}
    \caption{(A) Schematic illustration of a temporal network simulated using REGIR-TN.At each time increment, at most one link between nodes is updated, with the changing links and involved nodes highlighted in red. No links are modified when a step is rejected. The time step $\Delta t$ is variable and determined by REGIR-TN according to Eq.~\ref{deltat_TN}. (B) Cartoon describing our model of agents' face-to-face interactions. The probability of interaction for each pair of agents is influenced by their prior interactions and their individual activity rates. Each node tracks its own IED, depicted by the blue and brown distributions. (C) PDF of the time between two interactions for a given node, shown for both the conference data~\cite{genois2022combining} and simulated networks, with 10 reproductions for each model. (D) PDF of the duration of interactions between two agents, comparing the conference data with simulated networks. (E) Absolute deviations of network metrics from conference data, derived from the aggregated social network. Each deviation is normalized by the corresponding standard deviation across all models to ensure consistency. Results are shown for 10 independent simulations per model. (F) Computational runtime for simulating temporal networks with increasing numbers of nodes ($N_A$). For $N_A > 100$, all models exhibit a scaling behavior of $O(N_A^2)$, highlighted by the shaded region.}
    \label{fig:social_interactions}
\end{figure}

\noindent To highlight the potential of REGIR-TN for modeling non-Markovian temporal networks, we consider a social network where nodes correspond to individuals and edges represent face-to-face interactions (Figure~\ref{fig:social_interactions}B). As our reference dataset, we leverage interaction data collected during the International Conference on Computational Social Science, which took place from July 10 to 13, 2017~\cite{genois2022combining}. Conferences, unlike structured environments such as classrooms or workplaces, offer a setting where interactions are less constrained, allowing attendees to freely engage with others~\cite{le2023modeling}. In most social settings, interactions have been observed to follow a bursty pattern~\cite{genois2022combining, le2023modeling}, where episodes of rapid node activity are interspersed with periods of minimal or no activity, a regime effectively captured by Pareto or power law distributions, $\psi(t) \sim t^{-\alpha}$ (Figure~\ref{fig:social_interactions}C \& D). Therefore, we model our temporal social network with a straightforward system of two kinetic reactions, with the inter-event time of each individual agent following a Pareto distribution $\mathcal{P}$ (Eq.\ref{pareto-equation}):

%\begin{equation}
%Pareto(t, \lambda_0, \alpha ) = \begin{cases} \frac{\alpha}{t^{\alpha+1}} \cdot \frac{1}{2 \lambda_0^\alpha} & t \geq \frac{1}{\lambda_0} 2^{-1/\alpha} \\ 0 & t < \frac{1}{\lambda_0} 2^{-1/\alpha} \end{cases}
%\end{equation}

%which we simulate with multiplicative rates as outlined in Section~\ref{AA-multiplicative}:
%
\begin{itemize}
    \item \textbf{A$_i$ + A$_j$ $\xrightarrow{\mathcal{P}(\lambda_\text{A}, \ \alpha_\text{A})}$ A$_i$ + A$_j$ + $[$A$_i$A$_j]$:} Two agents encounter each other and begin an interaction. They remain available for interactions with other agents, as simultaneous interactions are possible. This translates into adding an edge in the temporal network. To maintain consistency with prior studies on temporal networks, we refer to this as the \textit{node interduration}, corresponding to the time between two initiated interactions for a given agent. A Pareto fit to the observed node interduration distribution in the conference data yields a shape parameter of $\alpha_A = 0.76$ and rate parameter $\lambda_A = 0.4$.

    %a minimum interaction time of unit $t_\text{min} = 1$, from which we derive the rate parameter $\lambda_A = 2^{-1/\alpha_A}/t_\text{min} = 0.40$ (see Eq.~\ref{Eq.Pareto_mu}) \maria{you can see here that the  time scale of the system emerges naturally. i would rename $t_\text{min} = 1 \rightarrow \lambda_0 = 1$ }. It is worth noting that the choice of 1 as the time unit is arbitrary and does not affect the generality of the results.
    
    \item \textbf{$[$A$_i$A$_j]$ $\xrightarrow{\mathcal{P}(\lambda_\varnothing, \ \alpha_\varnothing)} \varnothing$:} The two agents end their interaction after a certain duration, corresponding to the removal of an edge in the temporal network. This is simulated with the normal REGIR. A fit to the \textit{interaction duration} of conference data yields $\alpha_\varnothing = 1.4$ and $\lambda_\varnothing = 0.61$. 
\end{itemize}
Furthermore, empirical studies indicate that agents are more inclined to interact with those they have previously engaged with~\cite{le2023modeling}. This means that not all pairs of agents are equally likely to interact, even in similar circumstances. These considerations can be easily implemented with REGIR-TN by keeping track of the nodes' interaction history and adjusting the interaction rate of each pair accordingly (Methods~\ref{method:REGIR_temporal_network}). Namely, we consider two different versions of REGIR-TN for the reaction A$_i$ + A$_j$ $\to$ \texttt{interaction}:
\begin{itemize}

    \item \textbf{(REGIR-TN~A)} In this model, each node follows its own internal activity dynamics but does not retain the memory of past interactions with specific agents. As a result, every pair of agents is selected with equal probability at each iteration, meaning that interactions are not influenced by prior connections—there is no temporal neighborhood effect.  

    \item \textbf{(REGIR-TN~B)} In contrast, this model incorporates a first-order temporal neighborhood effect. After selecting the first agent $A_i$, the second agent $A_j$ is $w$ times more likely to have previously interacted with $A_i$. The parameter $w$ thus controls the strength of this temporal reinforcement, allowing past interactions to influence future connections.

    %\item \textbf{(REGIR-TN~A)} Every pair of agents is equally likely to be selected at each iteration. It corresponds to having ech node activity independent from other nodes
    %\item \textbf{(REGIR-TN~B)} Upon selecting the first agent $A_i$, the second agent $A_j$ is $w$ times more likely to have previously interacted with $A_i$. Here, the parameter $w$ captures the influence of first-order relationships in the dynamics. This is an introduction of the simplest mporal neighborhoods relationship.
\end{itemize}
To assess the fidelity of REGIR simulations to the conference network and to fit the models' parameters, we simulate the system with $N = 274$ nodes throughout $T = 7249$ units, consistent with the conference data. We evaluate REGIR-TN~A and REGIR-TN~B  using several metrics derived from the aggregated network, where each edge represents the total interaction duration between node pairs over the entire simulation (Methods~\ref{louvain_metrics}). Namely:
\begin{itemize}
    \item \textbf{Clustering coefficient}~\cite{watts1998collective}: Measures the tendency of nodes to form tightly connected groups.
    \item \textbf{Modularity}~\cite{newman2006modularity}: Reflects how strongly the network is divided into modules or communities, with higher values indicating more distinct community structures.
    \item \textbf{Transitivity}~\cite{newman2003structure}: Captures the global tendency of the network to form triangles.
    \item \textbf{Network density}~\cite{posfai2016network}: Measures the overall interconnectedness of the nodes. 
    \item \textbf{Degree assortativity}~\cite{newman2002assortative}: Quantifies nodes' likelihood of connecting with others of similar degree.
\end{itemize}
%
%\maria{These metrics are described in further detail in Section XX}. 
Using the aggregated conference network as a case study, i.e. the static network obtained by merging interactions across all time points into a single network, we fit the parameter $w$ of REGIR-TN~B to minimize the deviation from the conference network averaged over the five metrics (Methods~\ref{param_optimization}), where each value of $w$ are evaluated over 10 replicates of the simulation. This prioritization of first-order interactions significantly improves REGIR-TN~A in four out of the five metrics, with the only one not showing improvement being degree assortativity because it was already close to conference data in REGIR-TN~A (Figure~\ref{fig:social_interactions}E). This highlights the importance of considering first-order relationships in dynamic social interactions.

As a baseline for comparison, we first consider the Spanning Tree method, which provides a mathematical guarantee on the node interduration distribution, $\mathcal{P}(\lambda_\text{A}, \ \alpha_\text{A})$. While this approach effectively captures the desired distribution (Figure~\ref{fig:social_interactions}C), it lacks the flexibility needed to properly handle interaction durations between node pairs. As a result, its ability to fit real-world network data remains limited, as illustrated in Figure~\ref{fig:social_interactions}D.
Next, we adopt AD modeling as a second baseline, given its widespread use in temporal network modeling and its inherent adaptability to social interactions. This method allows for extensive customization of interactions between agent pairs. We specifically employ the implementation by Le Bail \textit{et al.}\cite{le2023modeling}, which was previously fine-tuned for the conference data through a genetic parameter search, leveraging an underlying unobserved social bond network to guide interactions. AD modeling primarily relies on an empirical approach, as neither the node interdurations nor the interaction durations can be explicitly controlled. Instead, obtaining a good fit typically requires exhaustive parameter tuning~\cite{le2023modeling}. In contrast, REGIR-TN directly constructs networks with a predefined IED, providing greater precision and control over system dynamics.

To evaluate the performance of REGIR-TN against our two baselines, we generate 10 replicates for each model and compare their ability to reproduce the metric values inferred from the conference network data. As shown in Figure~\ref{fig:social_interactions}E, both REGIR-TN and Spanning Tree successfully replicate the node IED. In contrast, AD modeling performs poorly at longer timescales, as its characteristic exponential tail fails to capture the heavy-tailed nature of the observed power-law IED in the conference data (Figure~\ref{fig:social_interactions}C).
Regarding interaction durations, REGIR-TN closely aligns with the reference data, while Spanning Tree systematically underestimates interaction lengths due to its lack of explicit parameterization. Similarly, AD modeling struggles to reproduce longer interactions, once again constrained by its exponential tail (Figure~\ref{fig:social_interactions}F). Still, since AD modeling is typically used to study the emergence of node IEDs rather than to construct them explicitly, as is done in REGIR-TN and spanning tree models, it is more appropriate to evaluate how this greater control over IEDs affects other network observables.

When examining the aggregated temporal network, REGIR-TN~B outperforms both AD modeling and Spanning Tree across all evaluated metrics, underscoring its ability to accurately capture the dynamics of social interactions (Figure~\ref{fig:social_interactions}E).
Finally, we assess the computational complexity of each model by simulating networks with an increasing number of nodes. As expected, all models exhibit a scaling behavior proportional to $N_A^2$ for sufficiently large node counts ($N_A > 100$) (Table~\ref{table:algorithm_comparison_network}, Figure~\ref{fig:social_interactions}F). No significant runtime differences are observed between REGIR-TN~A and B, whereas AD modeling demonstrates faster execution—approximately twice as fast for $N_A > 100$—due to its use of fixed steps of 1 time unit and its ability to process multiple nodes simultaneously at each step.
\begin{table*}[ht]
\renewcommand{\arraystretch}{1.3}
\setstretch{1.3} % Set the spacing for the table content
\centering
\captionsetup{width=1\linewidth}
    \resizebox{1\textwidth}{!}{
    \begin{tabular}{lccccc}
    \toprule
    \textbf{Algorithm} & \setstretch{0.8} \textbf{\begin{tabular}[c]{@{}c@{}}Arbitrary IED of\\individual nodes\end{tabular}} & \setstretch{0.8} \textbf{\begin{tabular}[c]{@{}c@{}}Adaptive node\\rate adjustment\end{tabular}} & \textbf{Accuracy} & \textbf{Complexity} & \setstretch{0.8} \textbf{\begin{tabular}[c]{@{}c@{}}Complexity\\ per time step\end{tabular}} \\
    \midrule
    Standard Gillespie         & No & Yes    & Exact & $O(N^4)$ & $O(N^2)$ \\
    %DelaySSA~\cite{fu2022delayssatoolkit}          & No & No    & x & x & x \\
    Spanning Tree~\cite{sheng2023constructing}       & Yes & No   & Exact & $O(dN^2)$ & $O(N^2)$ \\
    AD modeling~\cite{perra2012activity}       & No   & Yes & Approximate & $O(dN^2)^\dagger$ & $O(N^2)^\dagger$ \\
    REGIR-TN      & Yes & Yes & \setstretch{0.7} \begin{tabular}[c]{@{}c@{}}Approximate\\ (exact for $N \to \infty$)\end{tabular}  & $O(rN^2)$ & $O(1)$ \\
    \bottomrule
    \end{tabular}}
\caption{Comparison of existing algorithms and REGIR-TN for modeling temporal networks with $N$ nodes. Here, $r$ represents the ratio of rejected to accepted events, and $d$ denotes the number of time steps in Spanning Tree and AD modeling. The parameter $d$ is user-defined and typically set based on the resolution of the experimental data. In contrast, in Gillespie-style algorithms, the number of time steps is determined inherently by the algorithm. $^\dagger$For AD modeling, the complexity depends on the rules used to find interacting pairs of nodes at each iteration, which may vary on the chosen distribution of individual agents. For some very simple networks that choose interactions randomly, the complexity can be reduced to $O(dN)$ per simulation, or $O(N)$ per time step. However, more complex networks can scale as $O(N^2)$ and higher per time step. Details on the complexity derivation can be found in Supplementary Section~\ref{SI_complexity}.
}
\label{table:algorithm_comparison_network}
\end{table*}
\renewcommand{\arraystretch}{1}

Overall, our findings underscore REGIR-TN’s ability to accurately capture complex temporal network dynamics while maintaining computational efficiency, establishing it as a powerful tool for modeling dynamic social interactions. Notably, an important distinction between REGIR-TN and AD modeling lies in their treatment of IEDs. AD modeling primarily aims to study how node IEDs emerge organically rather than explicitly constructing them, whereas, in REGIR-TN, the IEDs serve as a structural assumption. Specifically, REGIR-TN assumes in our example that each node follows an intrinsic activity pattern of $\sim 1/t_i$, a behavior that naturally arises from priority-driven decision processes—where the likelihood of engaging in an event decreases over the time since the last interaction~\cite{barabasi2005origin, jo2012time}. By directly constructing the system's IEDs, REGIR-TN simplifies one aspect of the system, allowing a more focused investigation into how other factors shape network observables, such as first-order temporal neighborhood dynamics (REGIR-TN B). This structured approach provides a solid foundation for further refinement and exploration of model design for social interactions.

\section{Discussion}

%Non-Markovian stochastic systems are prevalent in various domains, such as social, biological, and cellular processes, where memory effects arise due to non-exponential waiting times, hidden states, feedback loops, and reaction time delays.
% Accurately describing such systems is notoriously challenging and requires specialized analytical or simulation frameworks. While the classical chemical master equation does not impose specific assumptions on waiting time distributions, its standard Markovian formulation inherently assumes memoryless transitions. To incorporate non-exponential waiting time distributions, extensions have been developed that relax this Markovian assumption~\cite{aquino_chemical_2017, zhang2019markovian}.

\noindent Non-Markovian stochastic systems are prevalent in many systems, such as in social, biological, and cellular processes, where non-exponential waiting times, hidden states, feedback loops, and reaction time delays introduce memory effects.  
A correct description is notoriously difficult and requires the use of non-Markovian analytical or simulation frameworks. Analytically, extensions of the standard Markovian formulation of the chemical master equation that allows for non-exponential waiting time distributions have been developed~\cite{aquino_chemical_2017, zhang2019markovian}. %\maria{is this correct? the classical master equation does not make any assumption regarding distributions or waiting times. maybe you are referring to a Markovian formulation?}.
While, in principle, its solution yields the system's probability density vector at any time, in practice, analytical solutions exist only for the simplest cases, and numerical approaches become computationally expensive or even infeasible as the number of reachable states becomes large or even infinite. 
At the stochastic simulation level, non-Markovian frameworks face similar challenges and often require strong simplifying assumptions for computational feasibility. These include pre-annotating fixed reaction times~\cite{barrio2006oscillatory, cai2007exact, anderson2007modified, ramaswamy2011partial}, modeling inter-event times with a mixture of exponentials~\cite{masuda2018gillespie} or with exponential variables whose rates are allowed to change with time~\cite{boguna2014simulating}, to become computationally feasible. Even with these assumptions, non-Markovian frameworks remain computationally more demanding than the standard Gillespie algorithm. Due to these complexities, most stochastic systems are analyzed using the Markovian assumption, which assumes that the system is memoryless, despite this being an unrealistic description of many real-world systems. Some generalizations have been proposed, such as Hidden Markov Models, which increase modeling flexibility by introducing hidden intermediate states, although this comes at the cost of a much larger number of parameters, which makes inference considerably harder.  

To address these challenges, we have introduced REGIR, a practical and scalable algorithm to efficiently simulate complex non-Markovian stochastic systems. REGIR enables flexible and accurate modeling of stochastic processes with commonly used inter-event time distributions, such as the exponential, gamma, normal, log-normal, Pareto, and Weibull distributions. REGIR can simulate complex systems with many different reaction channels, while exploiting a rejection sampling approach to maintain computational cost. Concretely, rejection sampling enabled us to (i) reduce the computational complexity of non-Markovian frameworks from $O(N^2)$ (as in the  nMGA~\cite{boguna2014simulating}) or $O(k N \log N)$ (as in DelaySSA~\cite{fu2022delayssatoolkit} and Laplace Gillespie~\cite{masuda2018gillespie}) to $O(N)$, thus reaching the same asymptotic complexity of the original Gillespie algorithm, and (ii) achieve an arbitrary user-define accuracy by introducing additional rejections, which further reduce the time step $\Delta t$ and, consequently, the approximation error. Moreover, since REGIR treats each process individually, integrating variable rates for each process is straightforward and does not introduce extra computational costs. 
This enabled the design of computationally efficient hybrid simulations that connect the internal dynamics of individual processes to coarse-grained stochastic dynamics at the population level.

We have demonstrated REGIR's capabilities across three different classes of non-Markovian systems, namely delayed reaction channels, stochastic processes with individual reactant properties, and activity-driven temporal networks. Firstly, REGIR effectively modeled the Hes1 RNA transcription and protein synthesis system with a negative feedback loop. It accurately simulated energy-limited elongation and performed competitively with DelaySSA in scenarios where delays were independent of reactant populations, becoming more efficient than maintaining a sorted array of reactants once the system exceeded $\sim 2000$ reactants in the system ($\beta \sim 100$). 
The advantages of REGIR became particularly evident in our second example, which modeled B-cell maturation during the immune responses and where reactants possessed individual properties. 
Here, each B~cell was endowed with a unique receptor that responded distinctively to external stimuli delivered by T cells and antigen-presenting cells~\cite{thomas2019probabilistic,pelissier2020computational}. 
Finally, we demonstrated REGIR's capability in simulating node-driven temporal networks by modeling face-to-face interactions during a scientific conference.  In this application, we introduced REGIR-TN, which defines pairwise interaction rates based on the intrinsic activity rates of individual nodes. 
This approach enabled the construction of a dynamic temporal network in which individual nodes could follow any desired IEDs while dynamically adjusting the time step based on the system’s total propensity. REGIR-TN combines key strengths from both activity-driven modeling~\cite{perra2012activity} —- including the ability to adapt node activities while accounting for temporal neighborhood relationships —- and Spanning Tree methods~\cite{sheng2023constructing}, which offer mathematical guarantees. By combining these advantages, REGIR-TN establishes a robust and versatile framework for modeling temporal networks.

%This overcomes critical challenges in AD models, where the choice of time step significantly impacts the aggregated network structure~\cite{ribeiro2013quantifying}, and where mathematical guarantees for inter-event time distributions are lacking. REGIR-TN combines the strengths of both AD modeling—such as the ability to dynamically adapt node activities—and Spanning Tree methods~\cite{sheng2023constructing}, which provide mathematical guarantees, creating a robust framework that unifies these advantages. Specifically, REGIR-TN ensures mathematical guarantees at any given time step, allowing for the real-time adjustments of both individual node activities and pairwise interaction rates while maintaining the same asymptotic complexity. For instance, it is particularly well-suited for modeling higher-order interaction dynamics~\cite{han2024probabilistic}, where nodes preferentially interact with those connected to their neighbors—an essential feature for accurately capturing real-world social networks~\cite{majhi2022dynamics, le2023modeling, iacopini2024temporal}.

A key factor influencing REGIR's computational efficiency is the ratio of attempted to accepted reactions. If this ratio becomes too high, the increase in rejections leads to a proportional rise in computational cost, potentially reducing efficiency. Therefore, it is essential to carefully consider this factor during system design. 
Firstly, while REGIR allows for arbitrary accuracy, increasing $\lambda_\text{max}$ to reduce the time step $\Delta t$ will directly increase the computational complexity. In this study, we imposed the condition $\lambda_\text{max} \geq \lambda_0$ to prevent large fluctuations in $\Delta t$ when many processes have nearly zero rates. However, as we have seen in the case of Hes1 transcription, this condition might not be sufficient for systems with a low number of processes ($N$), in which case further raising $\lambda_\text{max}$ might be necessary to achieve a desired accuracy (Supplementary Section~\ref{arbitrary_accuracy}). Our simulations show that this factor only significantly affects simulation accuracy when $N < 20$. In this regime,  and assuming there are no delays that co-evolve with the system, DelaySSA~\cite{fu2022delayssatoolkit} may provide a better alternative for simulating systems with a small number of reactants. 
On the other hand, REGIR's benefits are more evident for systems of moderate to large size, both in terms of accuracy and computational cost.

Secondly, while REGIR's rejection approach is advantageous when the system requires a costly update of propensities at each iteration, it can slow down the simulation when managing a wide array of individual rates, as processes with low rates often get rejected~\cite{thanh2014efficient}. This issue is notably significant in processes involving pairs of reactants with individual properties. In such cases, a small number of pairs with high interaction rates can disproportionately slow down the system, with many zero or near-zero propensity pairs still undergoing selection and subsequent rejection. When these high-propensity interacting pairs can be easily identified, adopting a hybrid version of REGIR that still uses rejection but applies varying weights to the selection of each process is a desirable strategy (Supplementary Section~\ref{REGIRbin}). However, wide variations in rates can sometimes be unavoidable, particularly with distributions characterized by a rapidly increasing rate like a Weibull distribution with a high shape parameter, where $\lambda(t) \propto t^\alpha$. Typically, simulating IEDs with low variance results in a higher rejected to accepted ratio, an empirical guideline being approximately $r \propto \text{mean/std}$ for most distributions (Supplementary Figure~\ref{fig:REGIR_ratio}).  Therefore, REGIR may not be the most suitable choice for simulating IEDs with a low standard deviation to mean ratio, as this could lead to an excessively high number of rejected reactions.

Nevertheless, all stochastic simulation frameworks can be slow when analyzing systems with a very high number of reactants and reaction channels. In such situations, viable options include using analytical simplifications to reduce the computational time, or running a few stochastic simulations, and training a neural network to perform a much larger number of simulations~\cite{jiang2021neural}. More importantly, both Markovian and non-Markovian frameworks struggle 
to efficiently simulate systems where the reaction propensities span several orders of magnitudes. In this regime, the vast majority of the time steps execute the reactions with the highest propensities, while reactions with lower propensities only occur rarely. This can result in a massive amount of computational time needed to reach steady-state, as has been observed in the numerical analysis of RNA transcription~\cite{herbach2017inferring}. Piece-wise-deterministic Markov process (PDMP) models provide a viable alternative, where low-frequency events are modeled with the Gillespie algorithm while high-frequency events are modeled by locally solving a system of ordinary differential equations. PDMP approaches were shown to considerably speed up simulations in several applications~\cite{herbach2017inferring,boxma2005off,pakdaman2010fluid, koshkin2024stochastic}.

%Finally,  the rejection sampling approach implemented in REGIR allows for very high customization of both stochastic models and simulations. Since each reactant in the population can follow its own stochastic process, it is possible to tag each reactant with \textit{individual properties}, such that its reaction rate changes continuously according to external variables or functions. This possibility was exploited to model the birth and maturation of B~cells, where each B~cell was endowed with an individual B-cell receptor that reacted differently to external cues~\cite{thomas2019probabilistic,pelissier2020computational}. This modeling flexibility opens the exciting possibility of designing computationally efficient hybrid simulations, where the internal dynamics underlying each process are coupled to macro-population stochastic dynamics. In biochemistry, this approach can be used to simulate stochastic interactions between a high number of cells while maintaining the individual cell \textit{identity} (i.e., DNA sequence, mRNA expression)~\cite{pelissier2020computational, sun2020stochastic}. 
%

To conclude, we described in this article a simple yet powerful framework that can simulate complex non-Markovian stochastic systems, with the hope that it becomes a new benchmark for the study of stochastic dynamic systems in various applications. Although this article focused on biochemical and social applications, other research areas can greatly benefit from such stochastic simulation frameworks, such as finance, epidemiology, and the study of propagation mechanisms in temporal networks. Indeed, the design of non-Markovian systems with individual properties can better capture the internal complexity and diversity of individuals, avoiding dependence on large-scale simulations based on overly simplistic assumptions~\cite{filatova2016regime}. REGIR facilitates the efficient simulation of complex non-Markovian systems, enabling the investigation of how the distribution of inter-event times and various individual properties impact population dynamics. These characteristics, often challenging to measure directly in experiments, are critical to understanding the stochastic processes underlying system dynamics. In fact, REGIR's ability to describe the IED for each process with two or more parameters, the rate $~\lambda_0$ and the shape parameter $\alpha$, offers greater flexibility to model realistic population dynamics than Markovian models, which only admit one parameter per reaction channel.

\section{Materials and Methods}

\label{Methods}

\subsection{REGIR implementation}
\label{Methods_REGIR}
\noindent The implementation focuses on two aspects to optimize the computational cost of the algorithm. The first is about the time since the last reaction of each process occurred ($t_j$). Rather than updating all of them at each time step, we only keep track of the timestamp the last reaction occurred $t_{0j}$ for the process $j$, and then obtain $t_j$ when needed as $t_j = t - t_{0j}$. The second avoids the computation of $\lambda_\text{max}$ at each time step. Consider four possible scenarios:

\begin{itemize}
    \item For distributions with a bounded instantaneous rate, such as the Pareto, log-normal, Cauchy, or delayed exponential distribution, $\lambda_\text{max}$ can be kept constant as the upper bound.
    
    \item For distributions with an unbounded rate that increases monotonically with time, such as normal, Weibull ($\alpha > 1$), or gamma ($\alpha > 1$), the maximum rate is known by keeping track of the reactant that did not react for the longest time. $\lambda_\text{max} = \lambda(t_\text{max})$. As such, the maximum rate does not need to be updated until that reactant with maximum waiting time does not react.
    
    \item For distributions with an unbounded rate that decreases monotonically with time, such as Weibull ($\alpha < 1$), gamma ($\alpha < 1$), or power laws, the maximum rate is known by keeping track of the reactant that did not react for the shortest time. As this can result in a very high maximum rate in some situations, we do not recommend using REGIR for these long-tailed distributions, since they can be simulated more efficiently with the Laplace Gillespie algorithm~\cite{masuda2018gillespie}.
    
    \item For distributions with arbitrary rates, such as those obtained numerically from experimental data, the user should precompute some bounds on $\lambda(t)$ for specific intervals, so that it can return to one of the three cases described above for each interval.
\end{itemize}

\subsection{REGIR for node-driven temporal network}
\label{method:REGIR_temporal_network}

\noindent In the main text, we model the IED activity of the node as following a Pareto distribution $\mathcal{P}(\lambda_\text{A}, \alpha_\text{A})$, and simulate it using REGIR-TN. To reduce computational cost and avoid reevaluating node rates at each iteration (see Eq.~\ref{Eq:REGIR_compute_pairwise_rate} and Eq.~\ref{Eq:paired_rate}), we rely on two key observations. First, given the properties of the Pareto distribution, the instantaneous rates of individual nodes $\lambda_i(t)$ are bounded above by $\lambda_\text{max} = \alpha_\text{A} / t_\text{min}$, where $t_\text{min} = \frac{1}{\lambda_\text{A}} 2^{-1/\alpha_\text{A}}$. Applying Eq.~\ref{Eq:paired_rate}, we then obtain the bound on the pairwise rate $\Lambda_\text{max} = \lambda_\text{max} / (N_A - 1)$, which we hold fixed throughout the simulation. Then, we approximate $\left\langle \sum_{k=1}^{N_A} \lambda_k(t_k) \right\rangle \approx N_A \lambda_A$. Here, we recall that $\lambda_A$ serves as both the inverse of the mean node IED (since it directly parametrizes the IED) and the expected observed interaction rate (see Supplementary Section~\ref{mean_observed_rate}). Based on this simplification, we compute the pairwise rate function as:
\begin{equation} 
    \Lambda_{ij} (t_i, t_j) = \frac{\lambda_i(t_i) \cdot \lambda_j(t_j)}{(N_A-1) \cdot \lambda_A}.
\end{equation}

Regarding REGIR-TN~B, we simulate the process by drawing nodes non-uniformly (weight $w$) rather than with equal probabilities followed by rejection. While this deviates from the original REGIR framework, as nodes are not sampled with equal likelihood, it remains valid in the case of Weibull and Pareto distributions, as detailed in Supplementary Section~\ref{REGIRbin}. This approach allows for the direct sampling of more probable interactions, thereby preventing a sharp increase in the proportion of rejected reactions, which would otherwise increase by approximately tenfold for $w=9.4$ with the standard REGIR-TN.

\subsection{AD modeling temporal network}
\noindent For AD modeling temporal networks, we utilized the V9 network from Le Bail \textit{et al.}~\cite{le2023modeling}, identified as the highest-performing model. Optimized simulation parameters for the conference dataset, along with the code for implementing AD modeling, were sourced from \url{https://github.com/DidierLeBail/Temporal-networks-PhD-code/tree/aurelien}. The conference dataset is labeled as \texttt{conf17} within this repository.

\subsection{Spanning Tree temporal network}
\noindent Spanning Tree temporal networks were generated by assigning node activity rates based on a Pareto distribution fitted to the conference data. The repository \url{https://github.com/anzhisheng/Temporal-networks-by-spanning-trees} was used for this purpose. Before generating the spanning trees, a Barabási-Albert graph~\cite{albert2002statistical} was constructed with $m = N/2$ neighbors per node, reflecting the observed conference data trend where nodes interact on average with half of the population at least once.

\subsection{Louvain communities in aggregated temporal networks}
\label{louvain_metrics}
\noindent Upon completing a simulation, the network interactions were aggregated by summing the interaction durations for each pair of agents. The resulting values were log-transformed, normalized by their maximum value, and defined as an \textit{interaction score}. A graph was then constructed by creating edges between pairs of individuals with an interaction score exceeding 0.4. To identify groups within the network, the Louvain community detection algorithm~\cite{blondel2008fast} was applied with a resolution parameter of 1. Key network metrics were computed using the \texttt{networkx} library as follows: the clustering coefficient: \texttt{nx.average\_clustering(G)}, modularity: \texttt{nx.algorithms.community.quality.modularity(G, communities)}, transitivity: \texttt{nx.transitivity(G)}, network density: \texttt{nx.density(G)}, degree assortativity: \texttt{nx.degree\_pearson\_correlation\_coefficient(G)}.

\subsection{Parameter optimization}
\label{param_optimization}

\noindent For each set of parameters used to run a REGIR simulation, we quantify its \textit{accuracy} by comparing the simulated to the experimentally measured metrics of interest. These metrics can be RNA and protein populations, the dominance scores of B-cells, or some properties of the temporal networks in the context of face-to-face interactions. We define the \textit{score} of the model as the root mean square deviations (RMSD) between the measured and simulated metrics averaged over all timepoints and reactants. All metrics are normalized by their maximum in order to assign the same weight to each metric. Then, we minimize this score function with the \texttt{maxLIPO} algorithm~\cite{davis2017optimization}, which is both parameter-free and provably better than a random search. It is a good alternative to Bayesian optimization methods~\cite{snoek2012practical}, that typically require the definition of prior assumptions about the function being optimized and thus require domain knowledge.

\subsection*{Code availability}
\noindent The REGIR and DelaySSA implementations, along with the data and the code to reproduce all figures presented in this article, are publicly available on GitHub at \url{https://github.com/AI-SysBio/REGIR}.

\subsection*{Acknowledgement}
\noindent The authors thank Jonathan Karr, Farshid Jafarpour, Srividya Iyer-Biswas, and Peter Ashcroft for their valuable suggestions. This research was supported by the COSMIC European Training Network, funded by the European Union’s Horizon 2020 research and innovation program under grant agreement No 765158.

\subsection*{Competing interests}
\noindent The authors declare no competing interest

\newpage

\onecolumn 

\appendix
\setcounter{figure}{0}
\renewcommand{\thefigure}{S\arabic{figure}}

\section{Probability density functions and instantaneous rates in the REGIR framework}

\subsection{Parametrization of inter-event time distributions}

\label{REGIR_distributions}

\noindent For consistency with other works in the literature of stochastic simulations, we parameterize processes using their mean $\frac{1}{\lambda_0}$, where $\lambda_0$ would be the instantaneous rate if the distribution was exponential. Additionally, some distributions require a second parameter to describe their \textit{shape}. For the normal and log-normal distribution, we use $\gamma$, a scale-free standard deviation defined as the standard deviation over the mean. For the gamma and Weibull distributions we use $\alpha$, the shape parameter in their standard parametrization. 
Thus, all IEDs in this article are parametrized with two variables, ($\lambda_0, \alpha$) or ($\lambda_0, \gamma$).

\paragraph{Normal distribution.} The normal distribution is typically parametrized by the mean $\mu$  and  the standard deviation $\sigma$, as follows:
\begin{equation}
    normal(t; \mu,\sigma)  = \frac{1}{\sigma \sqrt{2 \pi}} \exp \left({-\frac{1}{2}\left(\frac{t-\mu}{\sigma}\right)^{2}}\right) . 
\end{equation}

\noindent The parameters $\mu$ and $\sigma$ can be chosen such as the mean inter-event time distribution $1/\lambda_0$ with:
$$\mu = \frac{1}{\lambda_0} \ \text{and} \ \sigma = \frac{\gamma}{\lambda_0}$$
Where $\gamma$ corresponds to the ratio between the standard deviation and the mean, representing a scale-free standard deviation. The alternative parametrization of the distribution for REGIR is then given by

\begin{equation}
\boxed{
    normal(t; \lambda_0,\gamma)  = \frac{\lambda_0}{\gamma \sqrt{2 \pi}} \exp \left({-\frac{1}{2}\left(\frac{\lambda_0 t-1}{\gamma}\right)^{2}}\right). 
}
\end{equation}

\paragraph{Log-normal distribution.} The log-normal distribution is typically parametrized by $\mu$ and $\sigma$, as follows:
\begin{equation}
    \textit{log-normal}(t; \mu,\sigma) = \frac{1}{t \sigma \sqrt{2 \pi}} \exp \left(-\frac{\left(\log (t)-\mu\right)^{2}}{2 \sigma^{2}}\right),
\end{equation}
The mean and variance are given by:
\begin{align*}
mean =  & \exp \left(\mu+\frac{\sigma^{2}}{2}\right) ,  \\
variance = & \left[\exp \left(\sigma^{2}\right)-1\right] \exp \left(2 \mu+\sigma^{2}\right) .
\end{align*}
\noindent If we want the mean inter-event time distribution to be $1/\lambda_0$, we chose $\mu$ and $\sigma$ as follows:
%    sigma0 = alpha/rate
%    mu0 = \frac{1}{\lambda_0}
\begin{equation*}
\mu = \log \left(\frac{1}{ \lambda_0 \sqrt{1 + \gamma^2}}\right) \ \text{and} \ \sigma = \sqrt{\log \left(1+ \gamma^2\right)} \, . 
\end{equation*}
Where, as with the normal distribution, $\gamma$ corresponds to the ratio between the standard deviation and the mean, and thus, it is scale invariant. The alternative parametrization of the distribution for REGIR is then given by
\begin{equation}
\boxed{
    \textit{log-normal}(t; \lambda_0,\gamma) = \frac{1}{t \sqrt{2 \pi \log (1+ \gamma^2)}} \exp \left(-\frac{\left(\log (t)+\log \left( \lambda_0 \sqrt{1 + \gamma^2}\right)\right)^{2}}{2 \log (1+ \gamma^2)}\right),
}
\end{equation}

\paragraph{Gamma distribution.} The gamma distribution admits 2 constants, $\alpha$ and $\beta$:
\begin{equation}
    \label{Eq:gamma-param}
    gamma(t; \alpha,\beta) = \frac{\beta^{\alpha}}{\Gamma(\alpha)} t^{\alpha-1} e^{-\beta t} . 
\end{equation}
Mean and variance can be computed to be:
\begin{align*}
mean  = & \frac{\alpha}{\beta} \, , \\
variance  = & \frac{\alpha}{\beta^2} \, .
\end{align*}
The parameter $\beta$ can be chosen such as the mean inter-event time distribution is $1/\lambda_0$:
$$\beta= \alpha \lambda_0 \, . $$ 
Thus, we parametrize the gamma distribution as
\begin{equation}
\boxed{
    gamma(t; \lambda_0,\alpha) = \frac{(\alpha \lambda_0)^{\alpha}}{\Gamma(\alpha)} t^{\alpha-1} e^{-\alpha \lambda_0 t} . 
}
\end{equation}
We note that in this case the ratio between the standard deviation and the mean is related the the shape parameter $\alpha$ with:
\begin{equation*}
    \gamma = \frac{1}{\sqrt{\alpha}} \, , 
\end{equation*}

\paragraph{Weibull distribution.}

\label{weibull_appendix}
The Weibull distribution~\cite{jiang2011study} can be parametrized  with constants $\lambda$ and $\alpha$ as follows:
\begin{equation}
    Weibull(t; \lambda,\alpha) =  \frac{\alpha}{\lambda} \left( \frac{t}{\lambda} \right)^{\alpha-1} e^{-(t/\lambda)^\alpha} . 
\end{equation}
The mean and variance are:
\begin{align*}
mean = & \lambda \ \Gamma \left(1 + \frac{1}{\alpha} \right) \, , \\
variance = & \lambda^{2}\left[\Gamma\left(1+\frac{2}{\alpha}\right)-\Gamma^{2}\left(1+\frac{1}{\alpha}\right)\right] \, .
\end{align*}
If $t$ represents a "time-to-failure", the Weibull distribution gives a distribution for which the failure rate is proportional to a power of time. 
%A value of $k<1$ indicates that the failure rate decreases, while $k>1$ indicates that the failure rate increases over time. \mrm{is this supposed to be evident from Eq. 13? I cannot see it. If this is important, please, explain it further, otherwise, remove it.}
An alternative parametrisation often found in text books is  $\lambda = \left( \frac{\alpha}{\beta} \right) ^{\frac{1}{\alpha}}$, under which the PDF of the Weibull distribution becomes:

\begin{equation}
\label{Weibull_param}
    Weibull \left( t; \beta, \alpha \right) =  \beta t^{\alpha-1} \times \exp \left(- \frac{\beta t^{\alpha}}{\alpha}\right) . 
\end{equation}
To make the mean inter event time distribution equal to $1/\lambda_0$, $\beta$ has to be chosen as follows:
\begin{equation*}
\beta=\alpha\left[\lambda_{0} \Gamma\left(\frac{\alpha+1}{\alpha}\right)\right]^{\alpha}.
\end{equation*}
Thus, we parametrize the gamma distribution as
\begin{equation}
\boxed{
    Weibull \left( t; \lambda_0, \alpha \right) =  \alpha\left[\lambda_{0} \Gamma\left(\frac{\alpha+1}{\alpha}\right)\right]^{\alpha} t^{\alpha-1} \times \exp \left(- \frac{\alpha\left[\lambda_{0} \Gamma\left(\frac{\alpha+1}{\alpha}\right)\right]^{\alpha} t^{\alpha}}{\alpha}\right) . 
}
\end{equation}
We note that with this choice, the ratio between the standard deviation and the mean is related to $\alpha$ with:
\begin{equation*}
\gamma = \frac{\Gamma\left(\frac{\alpha+2}{\alpha}\right)}{\Gamma^{2}\left(\frac{\alpha+1}{\alpha}\right)}-1 \, .
\end{equation*}

\paragraph{Cauchy distribution.} The Cauchy distribution admits two parameters,  $\mu$ and $\sigma$:
\begin{equation}
    Cauchy(t; \mu,\sigma) = \frac{1}{\pi \sigma\left[1+\left(\frac{t-\mu}{\sigma}\right)^{2}\right]} \, . 
\end{equation}
The Cauchy distribution represents the distribution of the ratio of two independent and normally distributed random variables with mean zero. The mean and variance are 	undefined, as the integrals necessary to compute these values do not exist\footnote{ \small The integral associated with the mean, $\int_{-\infty}^\infty x f(x)\,dx $, 
does not exist. This can be proven, for instance, by noticing  that  	$\lim _{a\to \infty }\int _{-a}^{a} x f(x)\,dx $ and $\lim _{a\to \infty }\int _{-2a}^{a} x f(x)\,dx $ converge to different values. Similar arguments can be used to show that the variance does not exist either.}.
Intuitively, this happens because extremely large number can be drawn with non-zero probability. Nevertheless, we can choose the parameter $\mu$ as the inverse of the median inter event time $1/\lambda_0$:
$$\mu = \frac{1}{\lambda_0}\, . $$
Similarly, we can define $\gamma$ as the \textit{analogue} of ratio between the standard deviation and the mean, which is scale invariant:
$$\sigma = \frac{\gamma}{\lambda_0} \, .$$
Thus, we parametrize the Cauchy distribution as
\begin{equation}
\boxed{
    Cauchy(t; \lambda_0,\gamma) = \frac{1}{\pi \frac{\gamma}{\lambda_0}\left[1+\left(\frac{\lambda_0 t-1}{\gamma}\right)^{2}\right]} \, .
}
\end{equation}

\paragraph{Pareto distribution.}

The Pareto distribution admits two parameters,  $\mu$ and $\sigma$:
\begin{equation}
Pareto(t, \mu, \alpha ) = \begin{cases} \frac{\alpha \mu^\alpha}{t^{\alpha+1}} & t \geq \mu \\ 0 & t < \mu \end{cases}
\end{equation}
As with the Cauchy distribution, the Pareto distribution does not always have a finite mean, so we choose the parameter $\mu$ such as its median equals $1/\lambda_0$:
\begin{equation}
\label{Eq.Pareto_mu}
    \mu = \frac{1}{\lambda_0} 2^{-1/\alpha}\, . 
\end{equation}
Thus, we parametrize the Pareto distribution as
\begin{equation}
\boxed{
    Pareto(t, \lambda_0, \alpha ) = \begin{cases} \frac{\alpha}{t^{\alpha+1}} \cdot \frac{1}{2 \lambda_0^\alpha} & t \geq \frac{1}{\lambda_0} 2^{-1/\alpha} \\ 0 & t < \frac{1}{\lambda_0} 2^{-1/\alpha} \end{cases}
}
\label{pareto-equation}
\end{equation}
 
\subsection{Relationship between probability density functions and instantaneous rates}
\label{rate_proof}

%\subsection*{Relationship} 
\noindent We consider the survival distribution function (SDF) $\Psi(t) $ of a renewal process:
\begin{equation}
\Psi(t) =\int_{t}^{\infty} \psi\left(\tau\right) \mathrm{d} \tau 
\end{equation}
and its relationship with the probability distribution function (PDF) $\psi(t)= - \frac{d \Psi(t)}{d t} $. Using the definition of the instantaneous rate function:
\begin{equation}
    \lambda(t) = \frac{\psi(t)}{\Psi(t)} \, ,
\end{equation}
we can describe the time evolution of $\Psi(t)$ as a first order homogeneous differential equation:
\begin{equation}
    \lambda(t) \, \Psi(t) + \Psi'(t)   = 0  \, . 
\end{equation}
The general solution is easily written as~\cite{boyce2017elementary}:
\begin{equation}
    \label{Psi}
    \Psi\left(t\right)= K \exp \left({-\int^{t}_{0} \lambda(\tau) d\tau}\right) ,  
\end{equation}
where $K \in \mathbb{R}$ is an  integration constant. However, since by definition $\Psi(0) = 1$, we conclude that $K = 1$.
The PDF of that process can now by computed as follows: 
\begin{equation}
    \label{psi}
    \psi(t)
    = - \frac{d \Psi}{d t} (t)
    = \lambda(t) \, \exp \left({-\int^{t}_{0} \lambda(\tau) d\tau}\right)
\end{equation}

% Since the SDF at $t=0$ is 1 by definition:
% \begin{equation}
% \begin{aligned}
%    &\Psi(0) = 1\\
%     \Leftrightarrow & \ K \exp \left({-\int^{0}_{0} \lambda(\tau) d\tau}\right) = 1\\
%      \Leftrightarrow & \ K = 1
% \end{aligned}
% \end{equation}

Note that $\Psi(t)$ verifies the additional normalization condition $\Psi(\infty) = 0$. This implies:
\begin{equation}
\begin{aligned}
    \Psi(\infty)  =  & \ \exp \left({-\int^{\infty}_{0} \lambda(\tau) d\tau}\right) = 0\\
    \Rightarrow & \ \int^{\infty}_{0} \lambda(\tau) d\tau = \infty\\  \, .
\end{aligned}
\end{equation}
This means that $\lambda(t)$ has to be chosen such as its definite integral from 0 to $\infty$ is infinite, otherwise $\psi(t)$ and $\Psi(t)$ do not represent a renewal process.

\subsection*{Examples} 
\noindent In general, the instantaneous rate for any distribution can be computed as $\lambda(t) = \frac{PDF(t)}{SDF(t)}$, where the survival distribution function (SDF) is related to the cumulative distribution function (CDF) according to $SDF = 1 - CDF$. 

We provide here a few examples of instantaneous rate functions and their associated PDFs:
\begin{itemize}

    \item $\lambda(t) = a_0$ leads to $PDF =a_0 \times \exp \left( - a_0 t \right)$ and $SDF = \exp \left( - a_0 t \right)$, which represent an exponential distribution.

    \item $\lambda(t) = \beta t^{\alpha-1}$ leads to $PDF =\beta t^{\alpha-1} \times \exp \left( - \frac{ \beta t^{\alpha}}{\alpha} \right)$ and $SDF =  \exp \left( - \frac{ \beta t^{\alpha}}{\alpha} \right)$, associated with the Weibull distribution~\cite{jiang2011study}.
    
    \item $\lambda(t) = \dfrac{c^2 t}{1+ct}$ leads to $ PDF = c^2 t \times \exp \Bigl(-ct\Bigr)$ and $SDF = (1+ct) \times \exp \Bigl(-ct\Bigr)$.
    
    \item Many important distributions do not have a simple analytic form for the instantaneous rate. For instance, the normal distribution, $PDF = \frac{1}{\sigma \sqrt{2 \pi}} e^{-\frac{1}{2}\left(\frac{t-\mu}{\sigma}\right)^{2}}$ and $SDF = \frac{1}{2}\left[1-\operatorname{erf}\left(\frac{t-\mu}{\sigma \sqrt{2}}\right)\right]$, with \textit{erf} being the error function \cite{gautschi1972error}, results in an instantaneous rate that cannot be expressed in terms of basic functions. An approximation is however possible at large times, where the instantaneous rate asymptotically approximates a linear function $\lambda(t) \approx \frac{t-\mu}{\sigma^{2}}$.

\end{itemize}

\begin{figure*}[h!t]
    \centering
    \includegraphics[width=0.85\linewidth]{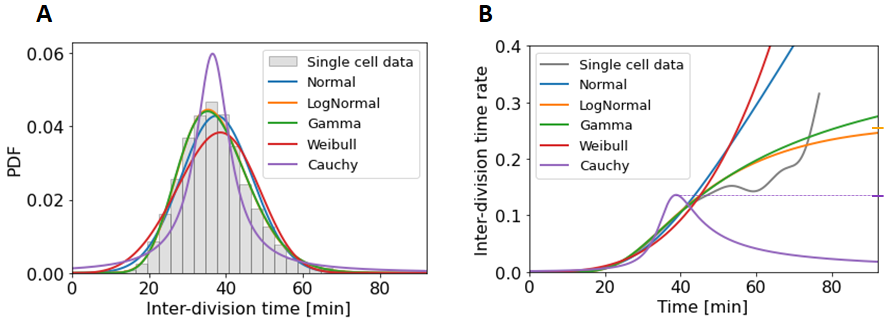}
    \caption{\small (A) PDF of several  distributions typically used to represent biochemical waiting times, where the parameters have been chosen to fit the measured inter-division time of \textit{Bacillus subtilis}~\cite{sauls2019control} at constant temperature. (B) Although the PDFs are relatively similar, the instantaneous rates show markedly different behaviors as a function of time. The single-cell instantaneous rate was estimated with a Gaussian kernel density estimator. When finite, the maximum instantaneous rates are displayed on the right of the figure.}
    \label{fig:distributions}
\end{figure*}

In Figure~\ref{fig:distributions}B, we show that the normal, Weibull ($\alpha \geq 1$) and gamma ($\alpha \geq 1$) distributions have monotonically increasing rates, while the Cauchy and log-normal distributions exhibit a maximum.

\newpage

\section{The Rejection Gillespie algorithm for non-Markovian Reactions}

\label{REGIR_proof}

\noindent We consider $N$ Poisson processes running in parallel, each with their respective reaction rate $\lambda_j$ $(1 \leq j \leq N)$, and denote by $a_0 = \sum \lambda_j$ the sum of the individual rates. The standard Gillespie (SG) algorithm is a popular stochastic simulation framework that can generate statistically correct trajectories of a stochastic equation system. The algorithm assumes that the reaction rates are known and constant, under which assumption, a trajectory can be obtained using the following iterative rules (see~\cite{masuda2018gillespie} for the full derivation):

\renewcommand{\labelenumi}{\theenumi}
\renewcommand{\theenumi}{(\roman{enumi})}

\begin{enumerate}
    \item Draw $u \in \mathcal{U}^{[0,1]}$ a random variate from the uniform density on the interval $[0, 1]$, and compute the time increment to the next event with
    \begin{equation}
        \Delta t = \frac{\ln (1/u)}{a_0},
        \label{eq:deltat2}
    \end{equation}
    
    \item Draw the process $j$ that has produced the event with probability
    \begin{equation}
        P_j = \frac{\lambda_j}{a_0},
        \label{eq:deltat3}
    \end{equation}
\end{enumerate}

In the main text, we introduced the Rejection Gillespie for non-Markovian Reactions (REGIR) that is statistically exact in the limit of $\Delta t \rightarrow 0$ (Equation~\ref{eq_PDF} of the main text). Below, we provide the proof, which we decompose into two independent subparts. First, we show that the introduction of rejected steps in the Gillespie model is mathematically equivalent to the standard Gillespie (SG) algorithm. Second, we show that locally considering each process as Poissonian during the time step $\Delta t$ at each Gillespie iteration yields statistically exact results for non-Markovian simulations when $\Delta t \rightarrow 0$. We note that both of these aspects were already proven separately in reference \cite{thanh2014efficient} and \cite{boguna2014simulating}, respectively. Here we put them together for the reader's convenience. As the rejection framework allows for the arbitrary reduction of $\Delta t$ by increasing $\lambda_\text{max}$, REGIR is exact in the limit  $\lambda_\text{max} \rightarrow \infty$ or $N \rightarrow \infty$.

\subsection{Proof for the Rejection-based Gillespie Algorithm}
\label{rejection_proof}
\noindent The rejection-based Gillespie algorithm  follows the same distribution as the Standard Gillespie (SG) if and only if (iff): 

\begin{itemize}[label={}]
    \item (i) The reaction $R_j$ occurs with probability $P = {\lambda_j}/{a_0}$.
    \item (ii) The $\Delta t$ time increment follows the same exponential distribution as in SG, i.e. $f_{\Delta t}(x) = a_0 \cdot \exp({-a_0 x})$.
\end{itemize}

\subsection*{(i) Reaction $R_j$ accepted with the same probability as in SG} 
\noindent We define $p_\text{accept}(R_j)$ as the joint probability of $R_j$ being first selected and then accepted, $\lambda_{\text{max}}$ as the upper propensity bound for all reactions, and $a_{0, \text{max}} = N \lambda_{\text{max}}$. We can write:
$$
p_\text{accept}(R_j) = \frac{\lambda_{\text{max}}}{a_{0, \text{max}}} \times \frac{\lambda_j}{\lambda_\text{max}} = \frac{\lambda_j}{a_{0, \text{max}}}
$$
We then denote by $p_\text{accept}(R)$ the probability of any reaction being accepted:
$$p_\text{accept}(R) = \frac{a_0}{a_{0, \text{max}}}.
$$
Finally, we write the conditional probability of reaction $R_j$ being accepted given that some reaction had been accepted as:
$$
p_{\text{accept}}(R_j \mid R) = \frac{p_{\text{accept}}(R_j)}{p_{\text{accept}}(R)} = \left(\frac{\lambda_j}{a_{0, \text{max}}}\right) / \left(\frac{a_0}{a_{0, \text{max}}}\right) = \frac{\lambda_j}{a_0} \, , 
$$ 
\noindent in agreement with SG framework. 

\subsection*{(ii) The time increment $\Delta t$ follows an exponential PDF}
\noindent We denote by $k$ the number of trials until the reaction is accepted (thus, there are $k-1$ rejections until success), with time being advanced by an increment of $\Delta t = - \ln{(u)} / a_{0,\text{max}}$ after each attempt. It follows that after $k$ attempts, the time increment is:
\begin{equation}
  \label{eqSI:t_erlang}
  \Delta t = - \frac{1}{a_{0, \text{max}}}\ln \left(\prod_{i=1}^{k}{u_i}\right)
\end{equation}
which corresponds to an Erlang distribution with parameters $k$ and $a_{0, \text{max}}$. This distribution represents the time elapsed until the $k$th event of a Poisson process with rate  $a_{0, \text{max}}$. In addition, $k$ is geometrically distributed with probability $p_\text{accept}(R)$, i.e. $$P(X=k) = (1 - p_\text{accept}(R))^{k-1} \times p_\text{accept}(R).$$

\noindent As the PDF for $\Delta t$ can be expressed as the derivative of its CDF, we can write
\begin{align*}
   f_{\Delta t}(x) &= \frac{d}{dx}F_{\Delta t}(x)\\
                   &= \frac{d}{dx}P(\Delta t \leq x),\\
   \intertext{ where $P(\Delta t \leq x)$ can be partitioned for values of k:}
                   &= \frac{d}{dx}\sum_{k=1}^{\infty}P(\Delta t \leq x \mid X=k)P(X=k)\\
                   &= \frac{d}{dx}\sum_{k=1}^{\infty}P(\Delta t \leq x \mid X=k)\left( 1 - \frac{a_0}{a_{0, \text{max}}} \right)^{k-1}\frac{a_0}{a_{0, \text{max}}}\\
   \intertext{ and as shown in Eq.~\ref{eqSI:t_erlang}, the distribution of $\Delta t$ parametrized by $k$ follows an Erlang distribution:}
                   &= \sum_{k=1}^{\infty}\frac{d}{dx}F_{\text{Erlang}(k, \lambda_{0,\text{max}})}\left( 1 - \frac{a_0}{a_{0, \text{max}}} \right)^{k-1}\frac{a_0}{a_{0, \text{max}}}\\
                   &= \sum_{k=1}^{\infty}f_{\text{Erlang}(k, \lambda_{0,\text{max}})}\left( 1 - \frac{a_0}{a_{0, \text{max}}} \right)^{k-1}\frac{a_0}{a_{0, \text{max}}}\\
                   &= \sum_{k=1}^{\infty}\frac{a_{0,\text{max}}^k \cdot x^{k-1} \cdot \exp({-a_{0,\text{max}} x}) }{(k-1)!} \cdot \left( \frac{a_{0, \text{max}} - a_0}{a_{0, \text{max}}} \right)^{k-1} \cdot \frac{a_0}{a_{0, \text{max}}}\\
                   &= a_0 \exp({-a_{0, \text{max}}x}) \sum_{k=1}^{\infty}\frac{({a_{0, \text{max}} - a_0})^{k-1} \cdot x^{k - 1}}{(k-1)!}\\
                   &= a_0 \exp({-a_{0, \text{max}}x}) \cdot \exp(x \cdot (a_{0, \text{max}} - a_0))\\
                   &= a_0 \cdot \exp({-a_0x}).
\end{align*}
Hence, in Rejection Gillespie, $\Delta t$ follows the same exponential distribution as in the SG.

\subsection{Proof for the Non-Markovian Gillespie Algorithm (nMGA)}
\label{nMGA}
\noindent In this section, we summarize the proof given by Boguna~\cite{boguna2014simulating}, and consider a second-order approximation of their algorithm. 
We consider $N$ renewal processes running in parallel, and denote by $t_{j}$ the time elapsed since the last event of the $j$th process $(1 \leq j \leq N)$. We denote by $\psi_{j}(t_j)$ the probability density function of inter-event times for the $j$th process, and by
\begin{equation}
    \Psi_{j}\left(t_{j}\right)=\int_{t_{j}}^{\infty} \psi_{j}\left(\tau\right) \mathrm{d} \tau  \, , 
\end{equation}
the survival function of the $j$th process, i.e., the probability that the inter-event time is larger than $t_{j}$. We also set
\begin{equation}
    \label{eq:REGIR_prob_next_event}
    \Phi\left(\Delta t \mid\left\{t_{j}\right\}\right)=\prod_{j=1}^{N} \frac{\Psi_{j}\left(t_{j}+\Delta t\right)}{\Psi_{j}\left(t_{j}\right)}  \, 
\end{equation}
which is the probability that no process generates an event for time $\Delta t$~\cite{masuda2018gillespie}. Then in the non-Markovian Gillespie algorithm (nMGA), the time until the next event, $\Delta t$, is computed by solving $\Phi\left(\Delta t \mid\left\{t_{j}\right\}\right) = u$, where $u \in \mathcal{U}^{[0,1]}$ is a random variate drawn from the uniform density on the interval $[0, 1]$. This can be time-consuming for some distributions~\cite{boguna2014simulating}. In the limit of a large number of processes $N\to\infty$, we can simplify the numerical computation of the time $\Delta t$ needed in the algorithm. We start by rewriting the function $\Phi(\Delta t |{tj})$ as:
\begin{equation}
\label{eqSI:nMGA_approx}
    \Phi\left(\Delta t \mid\left\{t_{j}\right\}\right)=\exp \left[-\sum_{j=1}^{N} \ln \left(\frac{\Psi_{j}\left(t_{j}\right)}{\Psi_{j}\left(t_{j}+\Delta t\right)}\right)\right]
\end{equation}
The sum within the exponential function is a sum of $N$ monotonously increasing functions of $\Delta t$. Therefore, when $N\to\infty$, the survival probability $\Phi\left(\Delta t \mid\left\{t_{j}\right\}\right)$ is close to zero everywhere except when $\Delta t \sim 0$. Hence, we only need to consider $\Phi\left(\Delta t \mid\left\{t_{j}\right\}\right)$ around $\Delta t=0$. In this neighborhood, we can perform a Taylor expansion for small $\Delta t$, namely,  $\Psi_{j}\left(t_{j}+\Delta t\right) \approx \Psi_{j}\left(t_{j}\right)-\psi_{j}\left(t_{j}\right) \Delta t+O\left(\Delta t^{2}\right)$. Plugging this expression into Eq. \ref{eqSI:nMGA_approx}, using the approximation $\ln(1+x) \approx x + O\left(x^{2}\right)$ and $1/(1-x) \approx 1 + x + O\left(x^{2}\right)$ when $x \rightarrow 0$, we can write:
\begin{equation}
    \begin{aligned}
    \Phi\left(\Delta t \mid\left\{t_{j}\right\}\right) &=\exp \left[-\sum_{j=1}^{N} \ln \frac{\Psi_{j}\left(t_{j}\right)}{\Psi_{j}\left(t_{j}+\Delta t\right)}\right] \\
    & \approx \exp \left[-\sum_{j=1}^{N} \ln \frac{\Psi_{j}\left(t_{j}\right)}{\Psi_{j}\left(t_{j}\right)-\psi_{j}\left(t_{j}\right) \Delta t+O\left(\Delta t^{2}\right)}\right]\\
    &\approx \exp \left[-\Delta t\left(\sum_{j=1}^{N} \lambda_j(t_{j})\right)  +O\left(\Delta t^{2}\right) \right] \, ,
    \end{aligned}
    \label{eqSI:Taylor_appr}
\end{equation}
where the instantaneous rate $\lambda_j$ is defined as:
$$\lambda_j(t_j) = \frac{\psi_{j}(t_j)}{\Psi_{j}(t_j)}.$$
With this approximation, the time until the next event is determined by 

\begin{equation}
    \Phi\left(\Delta t \mid\left\{t_{j}\right\}\right) \approx \exp \left[-\Delta t\left(\sum_{j=1}^{N} \lambda_j\left(t_{j}\right)\right)\right] . 
    \label{eq:38}
\end{equation}
Denoting by $ u $  a uniform random variable sampled from $[0,1]$
%, i.e. $u \in \mathcal{U}^{[0,1]}$, 
to  represent $\Phi\left(\Delta t \mid\left\{t_{j}\right\}\right)$, we can   approximate $\Delta t$ as follows:
%
%Denoting by $u$ a random variable sampled according to $\Phi\left(\Delta t \mid\left\{t_{j}\right\}\right)$, we can approximate Eq.~\ref{eq:38} as follows:
%
\begin{equation}
    \Delta t \approx \frac{\ln (1/u)}{\sum_{j=1}^{N} \lambda_j(t_{j})}.
    \label{eqSI:nMGA}
\end{equation}
We note that when  $\lambda_j(t_{j})$ depends on time,  the uniform random variable $u$ may not directly correspond to the distribution $\Phi\left(\Delta t \mid\left\{t_{j}\right\}\right)$. However, the approximation remains valid within a small neighborhood around $\Delta t = 0$ and when $\lambda_j(t_{j})$ changes slowly with time.
By eliminating the time dependency, i.e. setting $\lambda_{j}\left(t_{j}\right)=\lambda_{j}$, we recover the SG algorithm, where  $\Delta t$ is exponentially distributed. In the limit where $\lambda_j(t_{j})$ changes slowly with time, $\Delta t$ is still approximately exponentially distributed. nMGA exploits this approximation and locally treats each process as Poissonian during the time step $\Delta t$.

\subsection*{Second-order approximation}
\noindent We can also consider a quadratic expansion of Eq.~\ref{eqSI:nMGA_approx}, which results in a second-order approximation of  the time interval $\Delta t$. To do so, we first expand up to second order $\Psi_{j}\left(t_{j}+\Delta t\right)$:
\begin{equation}
    \Psi_{j}\left(t_{j}+\Delta t\right)
    \approx \Psi_{j}\left(t_{j}\right)-\psi_{j}\left(t_{j}\right) \Delta t - \psi_{j}'\left(t_{j}\right) \Delta t^{2}/2 +O\left(\Delta t^{3}\right), 
\end{equation}
\noindent where we remind the reader that $\Psi_j' = - \psi_j$. Additionally, the instantaneous rate $\lambda_j = \psi_j / \Psi_j$ is related to its derivative $\lambda'_j$ through the relation:

\begin{equation}
    \lambda'_j = \left(\frac{\psi_j }{\Psi_j}\right)' = \frac{\psi_j' \Psi_j - \psi_j \Psi_j'}{\Psi_j^2} = \frac{\psi'_j }{\Psi_j} + \left(\frac{\psi_j }{\Psi_j}\right)^2 = \frac{\psi'_j }{\Psi_j} + \lambda_j^2.
\end{equation}

\noindent Plugging this expression into Eq.~\ref{eqSI:nMGA_approx}, using the approximation $1/(1-x) = 1 + x + x^2 +  O\left(x^{3}\right)$ and $\ln(1+x) = x - x^{2}/2 + O\left(x^{3}\right)$ for $x \rightarrow 0$, we obtain:
\begin{equation}
    \label{Eq:REGIR_second_order}
    \begin{aligned}
    \Phi\left(\Delta t \mid\left\{t_{j}\right\}\right) &
    \approx \exp \left[-\sum_{j=1}^{N} \ln \frac{1}{1-\lambda_{j}\left(t_{j}\right) \Delta t  - \left(\lambda_j'(t_{j}) - \lambda_j^2(t_{j}) \right) \dfrac{\Delta t ^2}{2} + O\left(\Delta t^{3}\right)}\right]\\
    & \approx \exp \left[-\sum_{j=1}^{N} \ln \left( 1+\lambda_{j}\left(t_{j}\right) \Delta t  + \left(\lambda_j'(t_{j}) + \lambda_j^2(t_{j})\right) \frac{\Delta t ^2}{2} + O\left(\Delta t^{3}\right) \right)\right]\\
    & \approx \exp \left[-\sum_{j=1}^{N} \left(  \lambda_{j}\left(t_{j}\right) \Delta t  + \lambda_j'(t_{j})\frac{\Delta t ^2}{2} + O\left(\Delta t^{3}\right) \right) \right]\\
    \end{aligned}
\end{equation}

%\mrm{The derivation is correct, but difficult to follow as you have introduced the new functions $\lambda_j(t_j) $ and $\lambda'_j(t_j) $. it is fine to leave it as it is, but if you want to make it easier to follow, I would make the derivation in terms of $\psi_{j}$ and $\Psi_{j}$, and only introduce the new functions at the very end, when you solve $\Delta t$}

\noindent Solving $\Phi\left(\Delta t \mid\left\{t_{j}\right\}\right) = u$ to determine the next time increment $\Delta t$, we get a quadratic equation for which we take the positive solution:
\begin{equation}
\label{eqSI:dt_quadratic}
    \Delta t \approx \frac{- \sum_{j=1}^{N} \lambda_j (t_j) + \sqrt{\left(\sum_{j=1}^{N} \lambda_j(t_j)\right)^2 + 2 \left(\sum_{j=1}^{N} \lambda_j^{'}(t_j) \right) \cdot \ln(1/u)}}{\left(\sum_{j=1}^{N} \lambda_j^{'}(t_j) \right)} \, . 
\end{equation}
Once again,  Eq.~\ref{eqSI:dt_quadratic} makes explicit that $\Delta t$ is no longer exponentially distributed when $\lambda_j^{'}(t_j) \neq 0$, and hence, the rejection framework described in the first part cannot be applied with the second-order approximation of nMGA.

\subsection{Quantifying the errors of nMGA and REGIR }

\label{quant_error_REGIR}

\noindent Let us derive the error for both nMGA and REGIR. As a reminder, at each time step, both nMGA and REGIR utilize the first-order approximation of the survival distribution function to generate the next event with
\begin{equation}
    \Psi_{j}\left(t_{j}+\Delta t\right)=\Psi_{j}\left(t_{j}\right)-\psi_{j}\left(t_{j}\right) \Delta t+O\left(\psi_j'(t_j) \Delta t^{2}\right),
    \label{eq:Taylor_approx}
\end{equation}
from which the probability that no process generates an event for time $\Delta t$ with (Eq.~\ref{eq:REGIR_prob_next_event}) is computed from
\begin{equation}
    \Phi\left(\Delta t \mid\left\{t_{j}\right\}\right)=\prod_{j=1}^{N} \frac{\Psi_{j}\left(t_{j}+\Delta t\right)}{\Psi_{j}\left(t_{j}\right)},
\end{equation}
where $N$ is the total number of processes being considered. The error in this approximation is given by the difference between the actual value $\Phi\left(\Delta t \mid\left\{t_{j}\right\}\right)$ and the approximated value derived from the first term of the Taylor expansion. According to Taylor's theorem, and looking at the first and second order expression of Eq.\ref{Eq:REGIR_second_order}, we write the error term for the first-order approximation of $\ln \left[ \Phi\left(\Delta t \mid\left\{t_{j}\right\}\right) \right]$ as the second-order remainder~\cite{taylor1717methodus}:
\begin{equation}
    R_2^{\ln \Phi} = \sum_{j=1}^{N} \lambda_j'(t_j) \frac{\Delta t^2}{2} + O(\Delta t^3),
\end{equation}
where $\lambda_j'(t_j)$ is the derivative of the instantaneous rate function at time $t_j$.

\paragraph{Error of nMNGA.} Plugging in the value of $\Delta t$ of nMGA (Eq.\ref{eqSI:nMGA}), we estimate the error $R^{\ln \Phi}_2$ in nMGA per time step as
\begin{equation}
   \left( R^{\ln \Phi}_2 \right)_{ \mbox{\tiny nMGA}} \sim \sum_{j=1}^{N} \frac{\lambda_j'(t_j)}{2} \left( \frac{\ln (1/u)}{\sum_{i=1}^{N} \lambda_i(t_i)} \right)^2 \sim \frac{\ln^2(1/u)}{2 N} \cdot \frac{\langle \lambda' \rangle}{\langle \lambda \rangle^2},
    \label{eq:error_nmga}
\end{equation}
where we have introduced $\langle \lambda \rangle$ and $\langle \lambda' \rangle$, the average rate and the average derivative of the rates respectively, i.e.:
\begin{equation}
  \langle \lambda \rangle =  \frac{1}{N} \sum_{j=1}^{N}   \lambda_j(t_j) \ \ \text{and} \ \ \langle \lambda' \rangle =  \frac{1}{N} \sum_{j=1}^{N}   \lambda_j'(t_j) \, .
\end{equation}

\paragraph{Error of REGIR.} Using instead REGIR's time increment (Eq.\ref{eq:deltat}),  
\begin{equation}
    \Delta t = \frac{\ln(1/u)}{N \lambda_\text{max}} \, ,
\end{equation}
we obtain REGIR's error estimate: 
\begin{equation}
    \left( R^{\ln \Phi}_2 \right)_{ \mbox{\tiny REGIR}}
    \sim \frac{\ln^2(1/u)}{2 N} \cdot \frac{\langle \lambda' \rangle}{\lambda_\text{max}^2} \, .\\
    \label{eq:error_regir}
\end{equation}
We note that Eqs.~\ref{eq:error_nmga} and \ref{eq:error_regir} depend on $\ln^2(1/u)$, which does not have an upper bound so the error in a particular iteration can become arbitrarily large. However, we can write the expected error by substituting $\ln^2(1/u)$ with $\mathbb{E}[\ln^2(1/u)] = \int_0^{\infty} t^2 e^{-t} \, dt = 2$, where
\begin{equation}
    \label{eq:phi_error}
    \mathbb{E} \left[ \left( R^{\ln \Phi}_2 \right)_{ \mbox{\tiny nMGA} } \right] 
    \sim \frac{\langle \lambda' \rangle}{N \cdot \langle \lambda \rangle^2} \ \ \ \ \text{and} \ \ \ \     \mathbb{E} \left[ \left( R^{\ln \Phi}_2 \right)_{ \mbox{\tiny REGIR} }\right] 
    \sim \frac{\langle \lambda' \rangle}{N \cdot \lambda_\text{max}^2} \, .
\end{equation}
Here, we make several noteworthy observations. First, in both nMGA and REGIR, the error is directly proportional to the first derivative of the rate function, so both methods yield exact results for the exponential distributions where $\lambda_j'(t_j) = 0$.  In contrast, distributions with rapidly changing rates, such as the Weibull distribution with a high shape parameter, will exhibit significantly higher errors compared to distributions with more gradually varying rates, such as long-tailed distributions like Cauchy or Pareto.

Then, the primary difference between the two methods lies in the denominators of the expected error, which dictate how errors scale. Specifically, the error in nMGA scales with $1/\langle \lambda \rangle^2$, while REGIR's scales with $1/\lambda_\text{max}^2$. Consequently, nMGA is particularly prone to large errors when agents are initialized with low instantaneous rates, as $\langle \lambda \rangle$ can be very small at the start of the simulation. This issue is not encountered in REGIR, as $\lambda_\text{max}$ can be chosen independently, providing greater flexibility and accuracy.

%Another observation we can make is that for distributions where event rates increase monotonically with time, such as gamma, normal, log-normal, and Weibull, incorporating a second-order term generally reduces the value of $\Phi\left(\Delta t \mid \{t_j\}\right)$ (as shown in Eq.~\ref{eq:46}). This reduction reflects that the actual probability of no event occurring is lower than estimated by the first-order approximation.

\paragraph{Error in the time increment $\mathbf{\Delta t}$.} The errors derived above reflect the uncertainty in estimating $\Phi(\Delta t)$ at a specific time step, the probability that no events occur within the interval $\Delta t$. To quantify the error in the time step increment itself, we consider the sensitivity of $\Delta t$ to variations in $\ln \Phi(\Delta t)$. This error propagation can be determined by applying the chain rule~\cite{ku1966notes}:
\begin{equation}
    \text{Error}[\Delta t] \sim \left| \frac{\partial \Delta t}{\partial \ln \Phi(\Delta t)} \right| \cdot \text{Error}[\ln \Phi(\Delta t)],
\end{equation}
where the term  $\frac{\partial \Delta t}{\partial \ln \Phi(\Delta t)}$ represents how sensitive the time increment is to changes in $\ln \Phi(\Delta t)$. Here, we have ignored the sign, focusing only on the magnitude of the error. To analytically derive the error of $\Delta t$ in REGIR, we differentiate
\begin{equation}
    \Phi_{\mbox{\tiny REGIR}}(\Delta t) \sim \exp \left[-\Delta t N \lambda_\text{max} \right] 
\end{equation}
and find that
\begin{equation}
    \left| \frac{\partial \ln \Phi_{\mbox{\tiny REGIR}}(\Delta t)}{\partial \Delta t} \right| \sim N \lambda_\text{max}.
\end{equation}
The error in $\Delta t$ can thus be propagated as
\begin{equation}
    \text{Error}[\Delta t_\text{REGIR}] \sim \frac{1}{N \lambda_\text{max}} \cdot \text{Error}[\ln \Phi_{\mbox{\tiny REGIR}}(\Delta t)],
\end{equation}
which we also write as
\begin{equation}
    \text{Error}[\Delta t_\text{REGIR}] \sim \frac{1}{N \lambda_\text{max}} \cdot \left( R^{\ln \Phi}_2 \right)_{ \mbox{\tiny REGIR}}.
\end{equation}
Applying the same approach for nMGA, we write
\begin{equation}
    \Phi_{ \mbox{\tiny nMGA}}\left(\Delta t\right) \sim \exp \left[-\Delta t\left(\sum_{j=1}^{N} \lambda_j\left(t_{j}\right)\right)\right] \sim \exp \left[-\Delta t N \langle \lambda \rangle \right] 
\end{equation}
and obtain
\begin{equation}
    \text{Error}[\Delta t_\text{nMGA}] \sim \frac{1}{N \langle \lambda \rangle} \cdot \left( R^{\ln \Phi}_2 \right)_{\mbox{\tiny nMGA}}.
\end{equation}

\paragraph{Accumulated error in the IED.} In a stochastic simulation, errors accumulate across multiple time steps, denoted here as $n$. The total error in the simulated IED can be expressed as the sum of errors over all time steps:
\begin{equation}
    \text{Error}[\text{IED}] = \sum_{i=0}^n \text{Error}[\Delta t_i] \sim n \cdot \mathbb{E}\Bigl[\text{Error}[\Delta t]\Bigr].
\end{equation}
In the case of the standard Gillespie algorithm and nMGA, the number of time steps required to simulate a system for a fixed time $T_\text{end}$ scales with the number of processes, $N$, and the mean reaction rate, $\langle \lambda \rangle$, such that $ n \sim N \langle \lambda \rangle $. Consequently, the accumulated error for nMGA can be approximated as:
\begin{equation}
    \text{Error}[\text{IED}_\text{nMGA}] \sim N \langle \lambda \rangle \cdot \text{Error}[\Delta t_\text{nMGA}] \sim \mathbb{E}\Bigl[\bigl(R^{\ln \Phi}_2\bigr)_{\mbox{\tiny nMGA}}\Bigr].
\end{equation}
In the case of REGIR, where some reactions are rejected, the number of time steps increases compared to nMGA, scaling as $\sim N \lambda_\text{max}$. Substituting this relation, the accumulated error for REGIR can be written as:
\begin{equation}
    \text{Error}[\text{IED}_\text{REGIR}] \sim N  \lambda_\text{max} \cdot \text{Error}[\Delta t_\text{REGIR}] \sim \mathbb{E}\Bigl[\bigl(R^{\ln \Phi}_2\bigr)_{\mbox{\tiny REGIR}}\Bigr].
\end{equation}
Here we observe that, while REGIR involves a greater number of time steps due to the rejection of some reactions, the total accumulated error over the course of the simulation remains comparable with nMGA, since they both equal $R_2^{\ln\Phi}$. This equivalence can be intuitively understood as a balance: the increase in the number of time steps in REGIR is effectively counteracted by the proportional reduction in error magnitude per time step. Substituting the expression for $R_2^\Phi$ derived in Eq.~\ref{eq:phi_error}, the final expressions for the errors are:
\begin{equation}
    \label{eq:REGIR_IED_error}
    \text{Error}[\text{IED}_\text{nMGA}] 
    \sim \frac{\langle \lambda' \rangle}{N \cdot \langle \lambda \rangle^2} \ \ \ \ \text{and} \ \ \ \ 
    \text{Error}[\text{IED}_\text{REGIR}]
    \sim \frac{\langle \lambda' \rangle}{N \cdot \lambda_\text{max}^2}.
\end{equation}\\
We point out that, for illustrative purposes, we assumed here that $\langle \lambda' \rangle$, $\langle \lambda \rangle$, and $\lambda_\text{max}$ remain approximately constant throughout the simulation, a reasonable assumption for systems that quickly reach a steady state, where these parameters do not fluctuate significantly. However, this framework can also be generalized to systems with time-varying parameters. In such cases, the IED error at any given time step can be interpreted as the cumulative error that would result if the conditions (i.e., the values of $\langle \lambda' \rangle$, $\langle \lambda \rangle$, and $\lambda_\text{max}$) at that specific time step were held constant and applied across the entire simulation.

\paragraph{Accumulated error in the population dynamics.} While we successfully derived the expected error for the simulated IED, most non-Markovian studies report observables in terms of population dynamics rather than the IED, as population dynamics are generally more accessible experimentally~\cite{stumpf2017stem, england2010global, zeisel2011coupled} due to the difficulty of directly measuring IED in experimental systems. Here, we describe how errors in the time step $\Delta t$ propagates into population count dynamics over the course of the simulation. We consider $N$ processes and define the total rate as $\lambda_{\text{total}} = N \lambda_\text{max}$. For a given iteration, we estimate that the error in the time step propagates to the population as 
\begin{equation}
    \text{Error}[P] \sim \text{Error}[\Delta t] \cdot \frac{\lambda_{\text{total}}}{N} \sim \text{Error}[\Delta t] \cdot \lambda_\text{max}.
\end{equation}
Here, we normalize the population count error by the number of processes because we are interested in the relative population rather than the absolute errors in the counts. Due to the accumulated error during the simulation, we write the deviation of the population at time $t$ from the ground truth as
\begin{equation}
    \Delta P (t) = \sqrt{t} \cdot \text{Error}[P],
\end{equation}
where we have assumed that the error propagates through time like a diffusing behavior. Then, we can write the total error of the simulation as a sum through the number of total events $T$ 
\begin{align}
    \text{Error}[\text{POP}_\text{REGIR}] &= \sum_{k=0}^T  \Delta P (t_k)\\
    &= \text{Error}[P] \left(\sum_{k=0}^T \sqrt{t_k} \right)\\
    &\sim \lambda_\text{max} T^{3/2} \cdot \text{Error}[\Delta t_\text{REGIR}] \cdot
\end{align}\\
where the series scales approximately as $T^{3/2}$ due to the summation of $\sqrt{t}$ term. Substituting $T \sim \lambda_\text{max} N$ and the time step error derived under the REGIR framework
\begin{equation}
    \text{Error}[\Delta t_\text{REGIR}] \sim \frac{\langle \lambda' \rangle}{N^2 \cdot \lambda_\text{max}^3},
\end{equation}
we obtain:
\begin{equation}
    \text{Error}[\text{POP}_\text{REGIR}] \sim \sqrt{\frac{1}{N \lambda_\text{max}}} \langle \lambda' \rangle.
\end{equation}

\newpage

\subsection{Rejection sampling allows for arbitrary approximation accuracy}

\label{arbitrary_accuracy}

\begin{figure*}[h!t]
    \centering
    \captionsetup{width=1\linewidth}
    \includegraphics[width=1\linewidth]{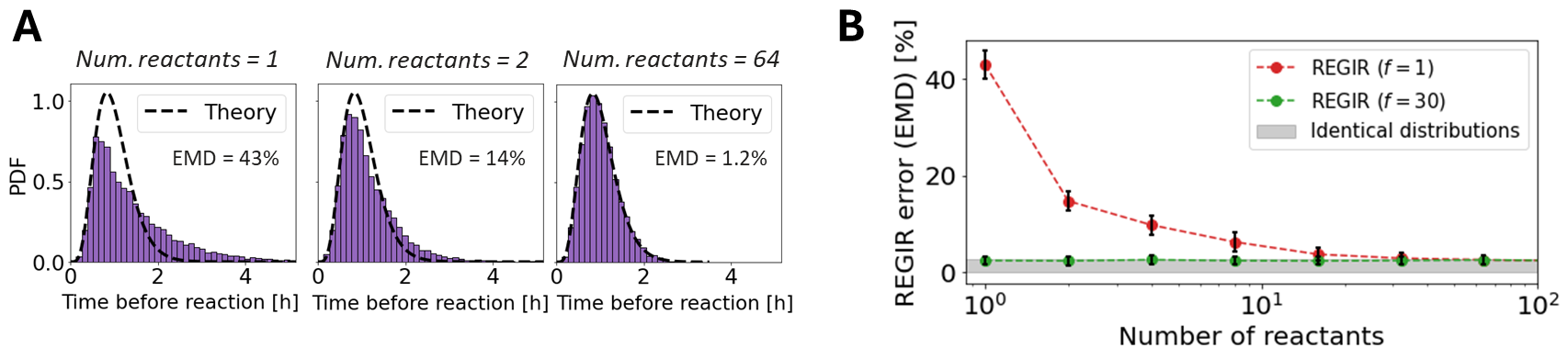}
    \caption{(A) REGIR ($f = 1$) approximation accuracy on a toy reaction $A \rightarrow \varnothing$ with a gamma inter-event distribution of shape parameter 6, visualized for different population sizes. The accuracy is computed using the earth mover distance (EMD) between the theoretical and simulated distributions, given in units of the distribution mean $1/\lambda_0$. In (B), we show how the EMD scales with the population size for two variations of REGIR, where the difference between the two lies in the additional parameter $f$ used to define $\lambda_\text{max}$.}
    \label{fig:REGIR_accuracy}
\end{figure*}

\noindent In this section, we discuss the differences between nMGA~\cite{boguna2014simulating} and REGIR in terms of simulation accuracy. We recall that both REGIR and nMGA are exact only when $\Delta t \rightarrow 0$, as they use a first-order Taylor approximation of the survival distribution function. Indeed, the survival distribution function can be Taylor approximated as follows:
\begin{equation}
    \Psi_{j}\left(t_{j}+\Delta t\right)=\Psi_{j}\left(t_{j}\right)-\psi_{j}\left(t_{j}\right) \Delta t+O\left(\psi_j'(t_j) \Delta t^{2}\right).
    \label{eq:Taylor_approx}
\end{equation}
The approximation breaks down when the inequality $\psi_j(t_j) \gg \psi_j'(t_j)\Delta t$ is no longer verified. The main difference between REGIR and nMGA lies in the introduction of the rejection step. In nMGA, the time increment until the next event is computed as:
\begin{equation}
    \Delta t\ \text{(nMGA)}= \frac{\ln (1/u)}{\sum_{j=1}^{N} \lambda_j(t_{j})} \,,
    \label{eq:nMGA_increment}
\end{equation}
while in REGIR, the expression becomes:
\begin{equation}
     \Delta t \ \text{(REGIR)} = \frac{\ln (1/u)}{N \cdot \lambda_\text{max}} \, .
     \label{eq:REGIR_increment}
\end{equation}
From Eq.\ref{eq:nMGA_increment}, it is clear that low rates are associated with large time increments $\Delta t$, and that in such a regime, the linear approximation nMGA used to compute $\Delta t$ might fail (Eq.\ref{eq:Taylor_approx}). Indeed, nMGA is only exact in the limit of an infinite number of processes ($N \rightarrow \infty$) where it can be assumed that $\sum_{j=1}^{N} \lambda_j(t_{j}) \rightarrow \infty$. For processes characterized by low rates at some time points (such as gamma distribution at $t=0$ for $\alpha > 1$), this approximation can be poor even in the limit of a large number of processes. REGIR circumvents this problem by setting $\lambda_\text{max}$ to an arbitrary large value (for example $\lambda_\text{max} \geq \lambda_0$), such that the time increment $\Delta t$ remains \textit{small enough} for the first order Taylor approximation (Eq.~\ref{eq:Taylor_approx}) to hold during the entire simulation. Still, this condition may not be sufficient when the number of processes $N$ is too low (Supplementary Figure~\ref{fig:REGIR_accuracy}A). To handle these cases, we can set $\lambda_\text{max}$ at each iteration such that:
\begin{equation}
    \label{eq:f_formula}
    \lambda_\text{max} \geq \lambda_0 \cdot \min \left\{ \frac{f}{N}, \ 1 \right\},
\end{equation}
where $\lambda_0$ is the inverse of the mean inter-event
time distribution, and $f$ a factor defined by the user to guarantee a desired upper bound for $\Delta t$, e.g. such as 
$\psi_j(t_j) \gg \psi_j'(t_j)\Delta t$ always holds:
\begin{equation}
    \Delta t \ \text{(REGIR)} \leq \frac{\ln (1/u)}{f \cdot \lambda_0}.
    \label{eq:REGIR_adjusted}
\end{equation}
An important consideration is that this definition will affect the system only when the systems contains a low number of processes. For systems with a large number of processes ($N \gg f$), the relationship
\begin{equation}
    \max_{\{j \in [1,N]\}} \lambda_j (t_j) \geq f \cdot \frac{\lambda_0}{N}
\end{equation}
is always verified with high probability, so that 
\begin{equation}
\lambda_\text{max} \geq \max_{\{j \in [1,N]\}} \lambda_j (t_j)
\end{equation}
regardless of the value of $f$. On the other hand, for systems with a low number of processes ($N < f$), increasing $f$ results in increased accuracy, but at the cost of additional computational time. More precisely, the computational cost of running REGIR will increase from $O(r N)$ to $O(r f)$, where $r$ is the attempted-over-accepted ratio of the system for $f=1$. Using the general equation for the REGIR error (Eq.~\ref{eq:REGIR_IED_error}), and substituting $\lambda_\text{max}$ from Eq.~\ref{eq:f_formula} we get that for $f \geq N$,
\begin{equation}
    \text{Error}[\text{IED}_\text{REGIR}]
    \leq \frac{N \cdot \langle \lambda' \rangle}{f^2 \cdot \lambda_0^2}.
\end{equation}\\
and otherwise ($f < N$)
\begin{equation}
    \text{Error}[\text{IED}_\text{REGIR}]
    \leq \frac{\langle \lambda' \rangle}{N \cdot \lambda_\text{0}^2}.
\end{equation}\\

\begin{figure}[h!t]
    \centering
    \includegraphics[width=0.8\linewidth]{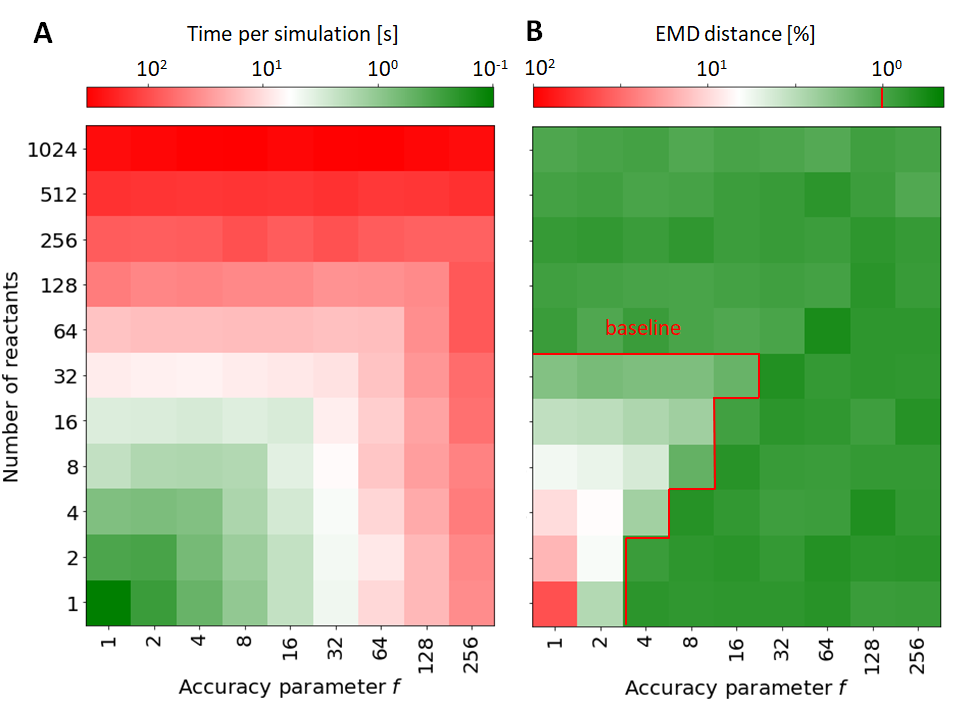}
    \caption{\small Trade-off between computational cost and accuracy. The toy reaction $A \rightarrow \varnothing$ is simulated with a gamma inter-event time distribution with parameters ($\lambda_0 = 1$, $\alpha = 6$). We display the heatmap of (A) the computational time for one simulation and (B) the EMD distance from the theoretical distribution. All values are averaged over 100 simulations, and EMD distances are given with a $\pm 10\%$ confidence interval. The red lines mark the border where the EMD becomes below the baseline EMD (defined as the EMD between two equal distributions with 10k sampling, equal here to 1\%).}
    \label{fig:EMD_tradeoff}
\end{figure}

Interestingly, our empirical investigation shows that such consideration only significantly affect the simulation accuracy when $N < 30$ (Supplementary Figure~\ref{fig:REGIR_accuracy}B \& Figure~\ref{fig:EMD_tradeoff}), which falls well below the number of reactants typically involved in most practical scenarios. In particular, the choice $f$ did not impact computational cost and accuracy the systems we discussed in this main article.

\newpage

\subsection{Rejection sampling reduces the computational complexity of REGIR to $\mathbf{\textit{O(N)}}$}
\label{proof_complexity}
\noindent We consider a simple reaction channel with $N$ reactants and an arbitrary IED over a duration $T$. Let us examine two different aspects of this simulation: (i) the computational complexity of a single time step and its scaling properties, and (ii) the number of  steps required to  simulate a fixed interval of length $T$. Let us first focus on the scaling properties of a single-time step. During a large simulation, two factors contribute to its computational cost, the calculation of the time step to the next reaction, and the update of the reactant populations~\cite{sanft2015constant}. For SG with one reaction channel, both of these steps are $O(1)$. In the case of nMGA however, these updates carry a computational cost of $O(N)$, as each reactant has its own reaction rate, which is equivalent to having its own reaction channel. In particular, the update of the instantaneous rates can be very expensive, as it requires recalculating them for each process in the entire population after every simulation iteration. On the other hand, REGIR reduces the computational complexity to $O(1)$ using a rejection base approach, where only the rate of the drawn reactant is computed at each step. The maximum rate $\lambda_\text{max}$ is either kept constant throughout the simulation or is updated using an ordered data structure for storing the $t_j$ values, thus also $O(1)$ (See Methods~\ref{Methods_REGIR}).

Regarding the number of steps required to simulate a fixed interval, SG, nMGA, Laplace, and DelaySSA scale linearly with the number of reactants $N$, because the time step becomes increasingly smaller as more channels are added to the simulation (Eq.\ref{eq:deltat}). However, the rejection approach of REGIR results in a larger number of time steps required to simulate a $ T_\text{end}$ interval, as a fraction of the time steps are rejected by the algorithm. The complexity then becomes $O(N) + O(R)$, where $R$ refers to the number of rejected steps. From Eq.\ref{eq:choice} and Eq.\ref{eq:rejection}, we can compute the probability of rejection for a given iteration as:
\begin{equation}
\begin{aligned}
  p_{reject} &  = 1 - p_{accept} \\
  & =  1 - \frac{\sum_{j=1}^{N} p_j \lambda_j(t_j) }{\lambda_\text{max} } \\
  & =  1 - \frac{1}{N \lambda_\text{max}}\sum_{j=1}^{N}  \lambda_j(t_j) \\
  & =  1 - \frac{\langle \lambda \rangle}{ \lambda_\text{max}}, 
\end{aligned}
\end{equation}

\noindent where we have introduced $\langle \lambda \rangle$, the average rate. Then, the expected number of rejections before the first accepted reaction ($A$) is given by the mean of the geometric distribution with success probability $p_\text{accept}$:

\begin{equation}
 \frac{R}{A}  = \frac{1}{p_{accept}} - 1 = \frac{(\lambda_\text{max} - \langle \lambda \rangle)}{ \langle \lambda \rangle}\\
\end{equation}

\noindent Thus we can conclude that the scaling of SG and REGIR running times, $T_{\text{R, SG} }$ and $T_{\text{R, REGIR}}$, are proportional according to the relation:

\begin{equation}
    \frac{T_{\text{R, REGIR}}}{T_{\text{R, SG}}} = \frac{A+R}{A} = \frac{\lambda_\text{max}}{\langle \lambda \rangle} 
\end{equation}

\begin{figure*}[h!t]
    \centering
    \captionsetup{width=0.8\linewidth}
    \includegraphics[width=0.5\linewidth]{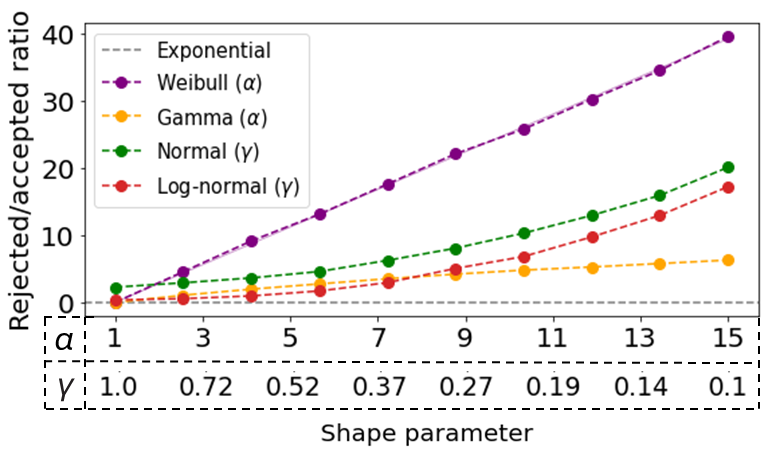}
    \caption{$R/A$ ratio for different shape parameters of different distributions, averaged over 100 simulations of on a toy reaction $A \rightarrow \varnothing$. A standard deviation of $\sim \pm 12\%$ was observed for all measurements. Note that a direct comparison of the ratio across distributions is not meaningful as they have different standard deviations. We also note that $R/A$ is scale-invariant and thus is independent of the distribution mean $1/\lambda_0$.}
    \label{fig:REGIR_ratio}
\end{figure*}

\noindent Supplementary Figure~\ref{fig:REGIR_ratio} shows the rejected over accepted reactions ratio ($\nicefrac{R}{A}$) for various distribution and shape parameters. We note that the ratio significantly varies with the choice of distribution and the parameters. For instance, an exponential distribution has a ratio of 0 (since $\hat\lambda =  \lambda_\text{max} = \lambda_i \; \forall i$, no reaction is rejected). On the other side of the spectrum, the Weibull instantaneous rate increases polynomially with time, so the maximum rate $\lambda_\text{max}$ increases quickly with $\alpha$, thus increasing the number of rejections. Other longer-tailed distributions with reaction rates increasing sub-linearly with time, e.g. the gamma distribution, will be less affected by changes in their respective shape parameter. In general, simulating a distribution with smaller variance will increase the maximum rate and as a result also increase the computational cost. This is intuitively clear from Supplementary Figure~\ref{fig:REGIR_ratio}, where the rejected over accepted reaction ratio ($R/A$) monotonically increases with the shape parameter of the gamma and Weibull distributions, which inversely correlate to the variance of their respective distribution. On the other hand, as both $R$ and $A$ are proportional to the rate $\lambda_0$, the ratio $\nicefrac{R}{A}$ is independent of the mean IED $\nicefrac{1}{\lambda_0}$ and thus scale invariant.

\newpage

\section{Relationship between the node and the pairwise interactions IEDs}

\label{distribution_relationship}

\noindent We consider two populations two type of reactants A and B, and a reaction channel \textbf{A + B} $\to$ \textbf{X}. When a reactant $A_i$ interacts at a given time $t$, the probability of pairing with a specific reactant $B_j$ is weighted by $w_{B_j}$, which represents $B_j$'s internal state and history of interactions. Specifically, the probability of $A_i$ to pair with $B_j$ at time $t$ is given by the ratio:
\begin{equation}
    \frac{w_{B_j}(t)}{\sum_{m} w_{B_m}(t)},
\end{equation}
where the denominator ensures normalization across all potential partners $B_m$. 

To compute the inter-event time distribution for the pair $(A_i, B_j)$, we must account for all possible scenarios in which $A_i$ does not pair with $B_j$ for $k$ consecutive interactions before finally pairing with $B_j$ on the $(k+1)$-th interaction, which we write as:
\begin{equation}
    P(\text{pairing with } B_j \text{ after } k \text{ steps}) =  \prod_{i=1}^k \left(1 - \frac{w_{B_j}(t_i)}{\sum_{m} w_{B_m}(t_i)}\right) \cdot \frac{w_{B_j}(t_{k+1})}{\sum_{m} w_{B_m}(t_{k+1})}.
\end{equation}
Here, the first term (the product) represents the probability that $A_i$ did not pair with $B_j$ during the first $k$ interactions, while the second term represents the probability of pairing with $B_j$ at the $(k+1)$-th interaction.

The inter-event time distribution for the pair $(A_i, B_j)$, denoted $\psi_{AB}(t)$, is influenced by the activity pattern of $A_i$, represented by $\psi_A(t)$, and the probabilities of pairing $(A_i, B_j)$ after $k$ prior interactions. Specifically, it is expressed as a weighted sum of convolutions of $\psi_A(t)$, where the weights are determined by the pairing probabilities:
\begin{equation}
    \psi_{AB}(t) = \sum_{k=0}^\infty \left[\prod_{i=1}^k \left(1 - \frac{w_{B_j}(t_i)}{\sum_{m} w_{B_m}(t_i)}\right) \cdot \frac{w_{B_j}(t_{k+1})}{\sum_{m} w_{B_m}(t_{k+1})}\right] \cdot \psi_A^{*(k+1)}(t),
\end{equation}
where $\psi_A^{*(k+1)}(t)$ denotes the $(k+1)$-fold convolution of $\psi_A(t)$. This formulation highlights how the inter-event time distribution of $(A_i, B_j)$ depends on both $A_i$'s activity and the probabilistic nature of its pairings with $B_j$.

Assuming $N_B$ a constant number of nodes over time, the term $\frac{w_{B_j}(t_i)}{\sum_m w_{B_m}(t_i)}$ can be approximated as $\frac{1}{N_B}$ on average. Under this assumption, the pairing probability simplifies, and the likelihood of $A_i$ pairing with $B_j$ after $k$ prior interactions follows a geometric distribution:
\begin{equation}
    P(\text{pairing with } B_j \text{ after } k \text{ steps}) = \frac{1}{N_B} \left(1 - \frac{1}{N_B}\right)^k.
\end{equation}
As a result, the inter-event time distribution for $(A_i, B_j)$ simplifies to:
\begin{equation}
    \psi_{AB}(t) = \sum_{k=0}^\infty \frac{1}{N_B} \left(1 - \frac{1}{N_B}\right)^k \cdot \psi_A^{*(k+1)}(t).
\end{equation}
As the mean of $k$ convoluted distribution is the sum of their respective mean, i.e. $\mathbb{E}\left[\psi_A^{*(k)}\right] = k \cdot \mathbb{E}\left[\psi_A\right]$, and recognizing the derivative of a geometric series, we conclude using the linearity of the expected value that the means of the distributions $\psi_A(t)$ and $\psi_{AB}(t)$ are related by:
\begin{equation}
    \mathbb{E}[\psi_{AB}] = N_B \cdot \mathbb{E}[\psi_A].
\end{equation}
For systems where the nodes' activity follows a Poisson process, i.e. $\psi_A(t) = \lambda_A \exp \left( - \lambda_A t  \right)$, we can also write
\begin{equation}
    \psi_{AB}(t) = \frac{\lambda_A}{N_B} \exp \left( - \frac{\lambda_A}{N_B} \cdot t \right).
\end{equation}

For non-Poissonian activity patterns, however, the combination of convolutions and geometric probabilities makes $\psi_{AB}(t)$ challenging to compute explicitly, often requiring numerical methods or approximations even if we assume a uniform probability of selected $B_j$. Nevertheless, we can still make general statements about $\psi_{AB}(t)$, particularly when $\psi_A(t)$ exhibits long-tailed behavior. If $\psi_A(t)$ follows a heavy-tailed distribution, such as a power law, the convolution $\psi_A^{*(k+1)}(t)$ retains this heavy-tailed nature, with a progressively slower decay as $k$ increases. This means that even without explicitly computing $\psi_{AB}(t)$, we know it will inherit the heavy-tailed characteristics of $\psi_A(t)$. Still, the link between the two distributions is not straightforward, and controlling one does not necessarily translate to control over the other. Given the complexity of the relationship between the inter-event time distribution for individual pairs and that of individual nodes in the network, it is crucial to clearly define and prioritize the specific aspect to focus on in the modeling process.

\section{Choice of the pairwise interaction rate function $\Lambda_{ij}$ in REGIR}
\label{REGIR_paired_rate}

\noindent Following the notation introduced in the main text (Section~\ref{Section: REGIR-TN}), we provide here the mathematical guarantees underlying REGIR-TN, as well as the rationale behind our choice of the pairwise rate formula. Here we consider a system in which each node, belonging to class $A$ or $B$, follows its own intrinsic rate function—$\lambda_A(\tau_A)$ and $\lambda_B(\tau_B)$, respectively—and that their pairwise interaction rate is given by $\Lambda(\tau_A, \tau_B)$. Here, $\tau_A$ and $\tau_B$ represent the internal times of reactants $A$ and $B$, respectively.\\

\noindent \textbf{Effective inter-event time distributions:} Let $\psi_A^\text{eff}$ and $\psi_B^\text{eff}$ denote the \textit{effective} PDF of inter-event time distribution of individual reactants $A$ and $B$, respectively. These PDFs are referred to as effective because they cannot be directly expressed as functions of the intrinsic rates $\lambda_A(t)$ and $\lambda_B(t)$, due to the coupling introduced by the interaction term $\Lambda$. These effective PDFs differ from the PDFs the reactants $A$ and $B$ in isolation $\psi_A (t) = \lambda_A (t) \cdot \exp(-\int_0^\infty \lambda_A (t))$ and $\psi_B = \lambda_B(t) \cdot \exp(-\int_0^\infty \lambda_B(t))$, which we refer to as the \textit{intrinsic} PDFs.

In the general case, the effective inter-event time distribution for reactant $A$ is determined by the minimum inter-event time over all possible interactions with $B$, which corresponds to the combined instantaneous rates of all such interactions, represented by the sum $\sum_{k=1}^{N_B} \Lambda(\tau_A, \tau_{Bk})$. Note that, while we focus here on the effective dynamics of reactant $A$, the same formulation can be applied to $B$ symmetrically without loss of generality. The effective SDF for $A$, denoted as $\Psi_A^\text{eff}(t)$, can thus be expressed as:
\begin{equation}
    \Psi_A^\text{eff}(t) = \exp\left(-\int_0^t \sum_{k=1}^{N_B} \Lambda(\tau_A, \tau_{Bk}) \, d\tau_A \right), \ \ \ \text{where} \ \ \ \tau_{Bk} \sim \psi_B^\text{obs} (\tau_B)
\end{equation}
represents the internal times of the $k$-th reactant $B$, sampled according to the \textit{observed} PDF $\psi_B^\text{obs}(\tau_B)$. Formally, $\psi_B^\text{obs} (\tau_B)$ is the PDF of the backward recurrence time associated with $\psi_B^\text{eff} (\tau_B)$, as we discuss in Supplementary Section~\ref{mean_observed_rate}.\\
%\begin{equation}
%    \tau_{Bk} \sim \psi_B^\text{obs}(\tau_B).
%\end{equation}

%\begin{equation}
%    \psi_A^\text{obs} (t) = \exp \left({- \mathlarger{\int}_{0}^{t} \sum_{k=1}^{N_B} \Lambda(\tau_A, \tau_{Bk}) \ d\tau_A}\right), \ \ \ \text{where} \ \ \ \tau_{Bk} \sim \psi_B^\text{obs} (\tau_B)
%\end{equation}

\noindent \textbf{Poisson processes.} The simplest case arises when the rates of both reactants, $\lambda_A$ and $\lambda_B$, are constant. In this scenario, for any pairwise rate function $\Lambda$, the effective survival function of $A$ is given by

    \begin{equation}
        \Psi_A^\text{eff} (t) = \exp \left({- N_B \mathlarger{\int}_{0}^{t} \Lambda \biggl(\lambda_A,  \lambda_B \biggr) \ d\tau}\right) = e ^ { - N_B \cdot \Lambda \bigl(\lambda_A,  \lambda_B \bigr) \cdot t}.
    \end{equation}
    If we account for individual properties within the population of $B$, where each individual reactant $B_k$ has its own rate $\lambda_{Bk}$, the observed inter-event time distribution for $A$ is influenced by the heterogeneity in $B$'s rates. In this case, the effective survival function for $A$ reflects the cumulative contributions of all interactions with $B_k$, and is given by
    \begin{equation}
        \Psi_A^\text{eff} (t) = e ^ { - \sum_{k=1}^{N_B} \Lambda \bigl(\lambda_A,  \lambda_{Bk} \bigr) \cdot t}.
    \end{equation}.

\noindent \textbf{Multiplicative rates:} For multiplicative rate, $\Lambda(\tau_A, \tau_B) = c \cdot \lambda(\tau_A) \cdot \lambda(\tau_B)$, we can write

    \begin{align}
        \Psi_A^\text{eff} (t) &= \exp \left({- c \mathlarger{\int}_{0}^{t} \sum_{k=1}^{N_B} \lambda(\tau_{Bk}) \cdot \lambda(\tau_A) \ d\tau_A}\right),\\
        \Psi_A^\text{eff} (t) &= \exp \left({- c \sum_{k=1}^{N_B} \lambda(\tau_{Bk}) \mathlarger{\int}_{0}^{t} \lambda(\tau_A) \ d\tau_A}\right),\\
    \end{align}
    Defining the mean observed rate $\langle \lambda_B \rangle = \frac{1}{N_B} \sum_{k=1}^{N_B} \lambda(\tau_{Bk})$, we can write
    \begin{align}
        \Psi_A^\text{eff} (t) &= \exp \left({- \mathlarger{\int}_{0}^{t} \lambda(\tau_A) \ d\tau_A}\right) ^ {c N_B  \cdot \langle \lambda_B \rangle }\\
       \Psi_A^\text{eff} (t) &= \biggl[\Psi_A (t)\biggr] ^ {c N_B  \cdot \langle \lambda_B \rangle }
    \end{align}
    Finally, we can conclude that, by setting $c = 1 / (N_B \cdot \langle \lambda_B \rangle)$, we always have $\Psi_A^\text{eff} (t) = \Psi_A (t)$.\\

\noindent \textbf{Additive rates:} For additive rate, $\Lambda(\tau_A, \tau_B) = c \cdot [\lambda(\tau_A) + \lambda(\tau_B)]$, we can write 

    \begin{align}
        \Psi_A^\text{eff} (t) &= \exp \left({- c \mathlarger{\mathlarger{\int}}_{0}^{t} \sum_{k=1}^{N_B} \lambda(\tau_A) + \lambda(\tau_{Bk}) \ d\tau_A}\right),\\
        \Psi_A^\text{eff} (t) &= \exp \left({- c N_B \mathlarger{\int}_{0}^{t} \lambda(\tau_A) \ d\tau_A}\right) \cdot \exp \left({- c N_B \mathlarger{\int}_{0}^{t} \langle \lambda_B \rangle \ d\tau_A}\right)\\
         \Psi_A^\text{eff} (t)  &\approx \biggl[ \Psi_A (t) \biggr]^{c N_B} \cdot \exp \biggl({- c N_B \langle \lambda_B \rangle \cdot t}\biggr)
    \end{align}
    Setting the constant $c = \nicefrac{1}{N_B}$, we obtain
    \begin{align}
         \Psi_A^\text{eff} (t)  &\approx \biggl[ \Psi_A (t) \biggr] \cdot e^ {- \langle \lambda_B \rangle t},
    \end{align}
    from which we derive the effective PDF as 
    \begin{equation}
        \psi_A^\text{eff} (t) \approx \big[ \Psi_A (t) \big] \cdot \langle \lambda_B \rangle e^{- \langle \lambda_B \rangle t}  +  \psi_A(t) \cdot e^{- \langle \lambda_B \rangle t}.
    \end{equation}
    Interestingly, this is the mixture of two PDFs. The first behave like an exponential distribution for low time, since $\Psi_A (0) = 1$, while the second is the intrinsic PDF but scaled with an additional exponential tail.

\subsection{The reaction A$_i$ + A$_j$ $\rightarrow$ X}
\label{REGIR_paired_rate_AA}

\noindent In the particular case of two reactants of the same type reacting with each others, the REGIR algorithm is slightly modified. The reaction $\text{A} + \text{A} \rightarrow X$ involve $\frac{1}{2} N_A (N_A - 1)$ processes, and the probabilities of choosing reactants $A_i$ and $A_j$ at each iteration are given by $\frac{1}{N_A}$ and $\frac{1}{N_A-1}$, respectively. In this case, the observed SDF is directly related to the observed PDF with
    \begin{equation}
        \Psi_A^\text{eff}(t) = \exp\left(-\mathlarger{\mathlarger{\int}}_0^t \sum_{k=1}^{N_A - 1} \Lambda(\tau_A, \tau_{Ak}) \, d\tau_A \right), \ \ \ \text{where} \ \ \ \tau_{Ak} \sim \psi_A^\text{obs} (\tau_A)
    \end{equation}
    That summation includes the reaction rates of $N_A - 1$ nodes, excluding the node of interest $A$. In the case of the multiplicative rate, $\Lambda(\tau_{A_i}, \tau_{A_j}) = c \cdot \lambda(\tau_{A_i}) \cdot \lambda(\tau_{A_j})$, we can write
    \begin{equation}
        \Psi_A^\text{eff} (t) = \exp \left({- c \sum_{k=1}^{N_A-1} \lambda(\tau_{Ak}) \mathlarger{\int}_{0}^{t} \lambda(\tau_A) \ d\tau_A}\right),
    \end{equation}
    Defining the mean observed rate $\langle \lambda_A \rangle = \frac{1}{N_A} \sum_{k=1}^{N_A}  \lambda(\tau_{Ak})$, we have
    \begin{equation}
        \sum_{k=1}^{N_A-1} \lambda(\tau_{Ak}) = \sum_{k=1}^{N_A} \lambda(\tau_{Ak}) - \lambda(\tau_{A}) = N_A \langle \lambda_A \rangle - \lambda(\tau_{Ak})
    \end{equation}
    and we can thus rewrite
    \begin{align}
        \Psi_A^\text{eff} (t) &= \exp \left({- \mathlarger{\int}_{0}^{t} \lambda(\tau_A) \ d\tau_A}\right) ^ {c \cdot [N_A \langle \lambda_A \rangle - \lambda(\tau_{Ak})]}.
    \end{align}
    Defining $c$ as
    \begin{align}
        c &= \frac{N_A}{N_A-1} \cdot \frac{1}{N_A \cdot \langle \lambda_A \rangle} \ ,
    \end{align}
    we get an effective SDF
    \begin{align}
        \Psi_A^\text{eff} (t) &= \left[\exp \left({- \mathlarger{\mathlarger{\int}}_{0}^{t} \lambda(\tau_A) \ d\tau_A}\right)\right] \text{{\LARGE \^{}}} \left(\frac{N_A - \frac{\lambda(\tau_A)}{\langle \lambda_A \rangle}}{N_A - 1}\right) \underset{N_A \to \infty}{=} \Psi_A (t),
    \end{align}
    which converge to the intrinsic density distribution for large $N_A$, as the factor involving self reaction becomes diluted by the number of processes and becomes negligible.

\subsection{Approximation of the mean observed rate}

\label{REGIR_total_approximation}

\noindent While computing the term $N_A \langle \lambda_A \rangle = \sum_{k=1}^{N_A} \lambda_k(t_k)$ may increase computational cost due to the need to iterate over all nodes at each step, it can be approximated by introducing the mean observed instantaneous rate, $\lambda_0 = \frac{1}{N_A} \left\langle \sum_{k=1}^{N_A} \lambda_k(t_k) \right\rangle$. Using this approximation, $\Lambda_{ij}$ can be expressed as:
\begin{equation} 
    \Lambda_{ij} (t_i, t_j) = \frac{N_A}{N_A-1} \cdot \frac{\lambda_i(t_i) \cdot \lambda_j(t_j)}{\left\langle \sum_{k=1}^{N_A} \lambda_k(t_k) \right\rangle} \approx \frac{\lambda_i(t_i) \cdot \lambda_j(t_j)}{(N_A-1) \cdot \lambda_0}.
\end{equation}
In Supplementary Section~\ref{mean_observed_rate}, we show that, in the steady state, the mean observed rate is given by $\lambda_0 = \frac{1}{N_A} \sum_{i=1}^{N_A} \lambda_{0_i}$, where $\lambda_{0_i} = 1/\mathbb{E}[\tau_i]$, and $\mathbb{E}[\tau_i]$ denotes the mean inter-event time of the intrinsic PDF for each individual node,
\begin{equation} 
    \mathbb{E}[\tau_i] = \int_{0}^{\infty} t \cdot \psi_A{_j}(t) \, dt.\\
\end{equation}

\newpage

\section{Derivation of the Pairwise Interaction Rate for Poisson Processes requiring Synchronous Availability}

~\label{pairwise_rate}

\noindent We consider $ N $ nodes, indexed by $ k = 1, 2, \ldots, N $, each emitting events according to independent Poisson processes with instantaneous rates $ \lambda_k $. Let $ T_k $ denote the time until the next event from node $ k $. The next pairwise interaction occurs at the minimum of these event times, $ T = \min(T_1, T_2, \ldots, T_N) $. 
The total rate of events occurring at time $ T $, $ \Lambda_{\text{total}} $, is derived from the property of the \textit{minimum of independent exponential random variables}. Specifically, if $ T_k \sim \text{Exp}(\lambda_k) $ for all $ k $, then $ T = \min(T_1, T_2, \ldots, T_N) $ is itself exponentially distributed with a rate equal to the sum of the individual rates:
\begin{equation}
    \Lambda_{\text{total}} = \sum_{k=1}^N \lambda_k.
\end{equation}
This result arises because the survival probability of the minimum is the product of the survival probabilities of the individual random variables:
\begin{equation}
    P(T > t) = \prod_{k=1}^N P(T_k > t) = \prod_{k=1}^N e^{-\lambda_k t} = e^{-\left(\sum_{k=1}^N \lambda_k\right)t}.
\end{equation}
For an interaction between two nodes $ i $ and $ j $ to occur, both nodes must contribute by aligning their events. The probability that node $ i $ contributes the next event is proportional to its rate, given by:
\begin{equation}
    P(T = T_i) = \frac{\lambda_i}{\sum_{k=1}^N \lambda_k},
\end{equation}
and similarly, the probability that node $ j $ contributes is:
\begin{equation}
    P(T = T_j) = \frac{\lambda_j}{\sum_{k=1}^N \lambda_k}.
\end{equation}
The pairwise interaction rate, $ \Lambda_{ij} $, is the product of the total rate of events, $ \Lambda_{\text{total}} $, and the joint probability of contributions from $ i $ and $ j $. This joint probability is proportional to the likelihood that both nodes emit events aligned in time:
\begin{equation}
    \Lambda_{ij} = \Lambda_{\text{total}} \cdot P(T = T_i) \cdot P(T = T_j).
\end{equation}
Substituting the expressions for $ \Lambda_{\text{total}} $, $ P(T = T_i) $, and $ P(T = T_j) $, we have:
\begin{equation}
    \Lambda_{ij} = \left( \sum_{k=1}^N \lambda_k \right) \cdot \frac{\lambda_i}{\sum_{k=1}^N \lambda_k} \cdot \frac{\lambda_j}{\sum_{k=1}^N \lambda_k}.
\end{equation}
Simplifying, this yields the effective pairwise interaction rate:
\begin{equation}
    \Lambda_{ij} = \frac{\lambda_i \cdot \lambda_j}{\sum_{k=1}^N \lambda_k}.
\end{equation}
%
%A similar reasoning can be applied to derive the instantaneous rate of interactions between pairs for arbitrary time-dependent instantaneous rates, resulting in:
%\begin{equation}
%    \Lambda_{ij}(t) = \frac{\lambda_i(t) \cdot \lambda_j(t)}{\sum_{k=1}^N \lambda_k(t)}.
%\end{equation}

 \newpage

\section{Observed backward recurrence time and instantaneous rates}

\label{mean_observed_rate}

\noindent  We consider a renewal process described by an arbitrary probability density function (PDF) $\psi_A(\tau)$, a survival density function $\Psi_A(\tau)$, and its instantaneous rate $\lambda_A(\tau) = \psi_A(\tau) / \Psi_A(\tau)$. When observing this process at a random point in time, the elapsed time $\tau$ since the last event corresponds to the \textit{backward recurrence time} in renewal theory~\cite{allison1985survival, zelen2004forward}, which represents the time elapsed since the most recent renewal event. The PDF of the backward recurrence time, $\psi_A^{\text{obs}}(\tau)$, is given by~\cite{allison1985survival}
\begin{equation}
    \psi_A^{\text{obs}}(\tau) = \frac{\Psi_A(\tau)}{\mathbb{E}[\tau]} = \lambda_0 \Psi_A(\tau),
\end{equation}
where we define $\lambda_0$ the reciprocal of the expected elapsed time $\mathbb{E}[\tau]$, defined as:
\begin{equation}
    \lambda_0 = \frac{1}{\mathbb{E}[\tau]} = \left(\int_{0}^{\infty} t \cdot \psi_A(\tau) \, d\tau\right)^{-1}.
\end{equation}
Specifically, the observed backward recurrence time does not follow the original PDF $\psi_A(t)$; instead, it is weighted by the survival of intervals, as longer intervals are more likely to be observed. This phenomenon, commonly referred to as the \textit{length-biased sampling effect}, is frequently discussed in studies related to disease screening~\cite{duffy2008correcting}.\\

In the context of REGIR, our focus is on the rate observed at each iteration, as it directly governs the probability of an event being accepted. The \textit{mean observed rate} can be computed by integrating the instantaneous rate $\lambda_A(\tau)$, weighted by the observed PDF $\psi_A^{\text{obs}}(\tau)$:
\begin{equation}
    \label{Eq:mean_observed_rate}
    \mathbb{E}_\text{obs}[\lambda_A(\tau)] = \int_{0}^{\infty} \lambda_A(\tau) \cdot \psi_A^{\text{obs}}(\tau) \, d\tau.
\end{equation}
Substituting $\psi_A^{\text{obs}}(\tau) = \lambda_0 \Psi_A(\tau)$, the expression becomes:
\begin{equation}
    \mathbb{E}_\text{obs}[\lambda_A(\tau)] = \lambda_0 \int_{0}^{\infty} \lambda_A(\tau) \cdot \Psi_A(\tau) \, d\tau.
\end{equation}
By noting that $\lambda_A(\tau) \cdot \Psi_A(\tau) = \psi_A(\tau)$, the integral simplifies to:
\begin{equation}
    \mathbb{E}_\text{obs}[\lambda_A(\tau)] = \lambda_0 \int_{0}^{\infty} \psi_A(\tau) \, d\tau.
\end{equation}
Since $\int_{0}^{\infty} \psi_A(\tau) \, d\tau = 1$, as $\psi_A(\tau)$ is a probability density function, the mean observed rate reduces to:
\begin{equation}
    \mathbb{E}_\text{obs}[\lambda_A(\tau)] = \lambda_0.
\end{equation}

Now, we extend this result to $N_A$ independent renewal processes, each characterized by its own PDF $\psi_{A_i}(\tau)$, survival function $\Psi_{A_i}(\tau)$, and mean rate $\lambda_{0_i} = 1 / \mathbb{E}[\tau_i]$. The observed PDF for randomly sampling a node and observing its rate is determined by the average contribution of all processes
\begin{equation}
    \psi_A^{\text{obs}}(\tau) =  \frac{1}{N_A} \sum_{i=1}^{N_A} \lambda_{0_i} \Psi_{A_i}(\tau).
\end{equation}
Substituting this expression into Eq.~\ref{Eq:mean_observed_rate} and following the same reasoning as in the single-process case, we find:
\begin{equation}
    \mathbb{E}_\text{obs}[\lambda_A(\tau)] = \frac{1}{N_A} \sum_{i=1}^{N_A} \lambda_{0_i}.
\end{equation}

\newpage

\section{Non uniform selection of reactants in REGIR}

\label{REGIRbin}

\noindent \textbf{Motivation.} The rejection approach in REGIR is particularly beneficial when individual rates are expensive to compute and fluctuate with each iteration. However, when handling a wide array of individual rates, this method can slow down the simulation, as processes with low rates are frequently rejected~\cite{thanh2014efficient}. This issue is notably significant in processes involving pairs of reactants. In such cases, a small number of pairs with high interaction rates can disproportionately slow down the system, while many pairs with zero propensity still undergo selection and subsequent rejection. If these zero-propensity pairs can be identified before each iteration without intensive computation, it would be more efficient to exclude them from selection altogether, rather than selecting and then rejecting them as REGIR does.

Here, we introduce a modified version of REGIR that adjusts the probability of selecting certain processes over others, thereby reducing the number of rejections up to several orders of magnitudes. This modification ensures that processes with highly disparate rates are handled more efficiently, and processes with zero rates are excluded from selection entirely.\\

\noindent \textbf{Algorithm.} We denote by $t_{j}$ the time elapsed since the last event of the $j$th process $(1 \leq j \leq N)$, and by $\lambda_{j}(t_j)$ the time-dependent reaction rate of the $j$th process. At each iteration, the modified REGIR performs 4 steps:

\begin{enumerate}

    \item Set $\lambda_\text{max}$, the maximum reaction rate over all processes, such that:
    
    \begin{equation}
        \lambda_\text{max} \geq \max_{\{j \in [1,N]\}} \lambda_j (t_j).
    \end{equation}

    \item Compute the time increment to the next event as in SG using $\lambda_\text{max}$. Namely, a random variable is uniformly drawn from the interval $[0, 1]$, i.e. $u \in \mathcal{U}^{[0,1]}$. The time increment is computed as:  
    
    \begin{equation}
        \Delta t = \frac{\ln (1/u)}{N \cdot \lambda_\text{max}}.
        \label{eq:deltat4}
    \end{equation}

    \item Assign a weight $w_j$ to each process, ensuring that the sum of all weights $\sum_{k=1}^N w_k = N$. Then, select the process $j$ that has triggered the event, with the probability of selecting each process being
    
    \begin{equation}
        p_j = \frac{w_j}{N}.
        \label{eq:choice}    
    \end{equation}

    \item Accept the process with probability $p_\text{accept}$, given by:
    \begin{equation}
        p_\text{accept} = \frac{\lambda_j (t_j)}{\lambda_\text{max}}, 
        \label{eq:rejection2}
    \end{equation}
    and update the reactants' population accordingly. If the process is rejected, the next event is set to an empty event, i.e. the reactant populations remain unchanged. 
    
\end{enumerate}

\noindent This algorithm's output resembles the original REGIR, with the difference that it scales the intrinsic rate of each process by $w_j$:
    \begin{equation}
        \lambda_j^\text{obs.}(t_j) = w_j \cdot \lambda_j(t_j),
        \label{eq:rejection3}
    \end{equation}

Importantly, This modification affects the computational cost of the simulation compared to using the original REGIR. Denoting by $M$ the number of unique weight $M = |set_{j \leq N}\{w_j\}|$, step (iii) now has a complexity of $O(M)$ instead of $O(1)$ as in the original REGIR framework. Consequently, the computational complexity of simulating the system becomes O($r' M N$),  with the key observation that the ratio of attempted to accepted reactions $r'$ is now lower than the ratio when all weights were equal $r$. Therefore, employing this modified version of REGIR is advantageous when the number of different weights to consider is less than the computational savings from reduced rejections, i.e. $M > r/r'$.

We note that scaling the rate by a weight at each iteration will affect the obtained distribution differently depending on the rate function. In the case of the exponential distribution, this is simply equivalent to scaling the inter-event time distribution of the $j$th process by $w_j$, i.e. $\lambda_{j0}^\text{obs.} = w_j \cdot \lambda_{j0}$. Similarly, if $\lambda_j (t_j)$ follows a Weibull distribution, then this is also equivalent to scaling the distribution by $w_j$, with its shape parameter $\alpha$ remaining unchanged. Finally, if $\lambda_j (t_j)$ adheres to a Pareto distribution, the obtained distribution will also be a Pareto but the weights will affect both the rate and the shape of the distribution. For most other rate functions however, the resulting distribution does not follow a simple form.\\

\noindent \textbf{Proof that modified REGIR follow the same distribution as Standard Gillespie.}  To prove each processes in the weighted REGIR follows $\lambda_j^\text{obs.}(t_j) = w_j \cdot \lambda_j$, we take a similar approach as we did in Supplementary Section~\ref{rejection_proof}. ie. we  prove that:

\begin{itemize}[label={}]
    \item \textbf{(i) The reaction $R_j$ occurs with probability  \bm{$p = {w_j\lambda_j}/\sum{w_j}\lambda_j$}}\\

\noindent We define $p_\text{accept}(R_j)$ as the joint probability of $R_j$ being first selected and then accepted, $\lambda_\text{max} \geq \max_{\{j \in [1,N]\}} \lambda_j$ as the upper propensity bound of all reactions , and $a_{0, \text{max}} = N \lambda_{\text{max}}$. We can write
$$
p_\text{accept}(R_j) = \frac{w_j}{N} \cdot \frac{\lambda_j}{\lambda_\text{max}} = \frac{w_j  \lambda_j}{a_{0, \text{max}}}
$$
We then denote by $p_\text{accept}(R)$ the probability of any reaction being accepted:
$$p_\text{accept}(R) = \frac{1}{N} \mathlarger{\mathlarger{\sum}} w_j \cdot \frac{\lambda_j}{\lambda_\text{max}} = \frac{\sum{w_j}\lambda_j}{a_{0, \text{max}}}.$$
The conditional probability $p_{\text{accept}}(R_j \mid R)$ can be exploited to show the probability of $R_j$ being accepted given that some reaction had been accepted:
$$p_{\text{accept}}(R_j \mid R) = \frac{p_{\text{accept}}(R_j)}{p_{\text{accept}}(R)} = \left(\frac{w_i \lambda_j}{a_{0, \text{max}}}\right) / \left(\frac{\sum{w_j}\lambda_j}{a_{0, \text{max}}}\right) = \frac{w_i\lambda_j}{\sum{w_j}\lambda_j}$$

\item \textbf{(ii) The $\Delta t$ increment in time follows the same exponential distribution as in SG, i.e.}

\textbf{\bm{$\ \ \ \ \ \ f_{\Delta t}(t) \sim \sum{w_j}\lambda_j \cdot \exp( - \sum{w_j}\lambda_j \cdot t)$}}\\

\noindent Here the derivation is exactly the same as in Supplementary Section~\ref{rejection_proof}. Given that $p_\text{accept}(R) = \frac{\sum{w_j}\lambda_j}{a_{0, \text{max}}}$, we obtain $\Delta t \sim (\sum{w_j}\lambda_j) \exp({-a_{0, \text{max}} \cdot t}) \cdot \exp(x \cdot (a_{0, \text{max}} - \sum{w_j}\lambda_j)) = (\sum{w_j}\lambda_j) \cdot \exp({-\sum{w_j}\lambda_j \cdot t}).$

\end{itemize}

         \section{Computational complexity of stochastic algorithms}

\label{SI_complexity}

\noindent In this section, we provide an analysis of the computational complexity of various stochastic simulation algorithms.

\begin{itemize}
    \item 	\textbf{Gillespie Algorithm:} The standard Gillespie algorithm has a known computational complexity of $O(M)$ per reaction step, where $M$ is the number of unique reaction rates. This complexity arises from the necessity of drawing a reaction based on propensity values at each iteration. Specifically, the reaction index $\mu$ is selected by searching for the smallest $\mu$ satisfying 
    \begin{equation}
        \sum_{j=1}^{\mu-1} a_j < r_2 \sum_{j=1}^{M} a_j, \leq \sum_{j=1}^{\mu} a_j.
    \end{equation}
    
    \item 	\textbf{Tree-Based Gillespie Algorithm}~\cite{gibson2000efficient}: By utilizing a binary tree structure, the selection process can be improved from $O(M)$ to $O(\log M)$. In this approach, reaction propensities are stored in a binary tree, where each node maintains a partial sum of the propensities of its subtree. Reaction selection is then performed using a binary search, reducing the selection time to $O(\log M)$. However, since the tree needs to be updated whenever reaction propensities change, its efficiency depends on the system. Specifically, each update requires modifying the relevant nodes along the path from the updated leaf to the root, leading to an update complexity of $O(k \log M)$, where $k$ is the number of updated reactions per step. For systems where propensities remain largely unchanged throughout the simulation, the tree-based approach can be highly efficient. Conversely, for systems with rapidly changing rates, maintaining the tree can be computationally expensive, sometimes outweighing the benefits over the standard Gillespie algorithm.\\
    
    \item 	\textbf{Laplace Gillespie Algorithm}~\cite{masuda2018gillespie}: The Laplace Gillespie algorithm extends the standard Gillespie algorithm to non-Markovian processes by incorporating non-exponential inter-event times. Instead of redrawing all $N$ rates at each step, it samples them initially from $ p(\lambda)$ and updates only the selected reaction’s rate thereafter. To efficiently handle selection and updates, the algorithm employs a tree-based structure where reaction propensities are stored hierarchically, enabling $O(\log N)$ selection via binary search and efficient updates. This structure maintains the algorithm’s efficiency while allowing for flexible non-Markovian dynamics.\\

    \item 	\textbf{nMGA (Non-Markovian Gillespie Algorithm)}~\cite{boguna2014simulating}: This method requires recomputing all $N$ reaction rates at each time step, making tree-based approaches inefficient. The overall complexity per iteration is thus $O(N)$, dominated by the full rate update.\\
    
    \item 	\textbf{Delayed Stochastic Simulation Algorithm (DelaySSA)}~\cite{fu2022delayssatoolkit}: This algorithm maintains a sorted list of scheduled events, allowing the next event to be extracted in $O(1)$ time. However, maintaining the sorted structure incurs an additional cost of $O(k \log N)$ per iteration, where $k$ is the number of events to be updated. New events are inserted at the correct position using binary search, followed by an efficient insertion operation in a balanced tree or linked structure~\cite{henriksen1983event}. Deletion is always performed at the front of the list in $O(1)$ time.

\end{itemize}

\noindent For stochastic simulations on temporal networks, where interactions occur among \( N \) nodes, there are \( N^2 \) potential interactions, or processes, to consider.

\begin{itemize}
    \item \textbf{Standard Gillespie Algorithm on Temporal Networks:} Since interactions take place between pairs of nodes, each with potentially different rates, the number of processes to track per time step is $O(N^2)$. Consequently, the total computational complexity for a full simulation scales as $O(N^4)$.\\
    
    \item 	\textbf{Spanning Tree Approach}~\cite{sheng2023constructing}: A spanning tree is constructed to model temporal interactions, ensuring that nodes and links follow predefined IED. The initial construction of the spanning tree incurs a computational cost of $O(N^2)$, given that each node interacts with a significant approximatly half of the other nodes at least once. During the simulation, interactions evolve over $d$ time steps, with each step requiring updates based on the IED distributions. In the worst case, all edges must be considered at each step, leading to a total computational complexity of $O(dN^2)$. \\
    
    \item 	\textbf{Activity-Driven (AD) Modeling}~\cite{perra2012activity}: At each time step, node pairs are assessed to determine potential interactions. Most activity-driven modeling studies consider complexities such as historical dependencies and heterogeneous activity levels when defining interactions~\cite{le2023modeling}, requiring iteration over all possible interactions. Given that the process unfolds over $d$ time steps, the overall computational complexity is $O(dN^2)$.
\end{itemize}

\section{Supplementary Figures}

\begin{figure*}[h!t]
    \centering
    \captionsetup{width=1\linewidth}
    \includegraphics[width=0.9\linewidth]{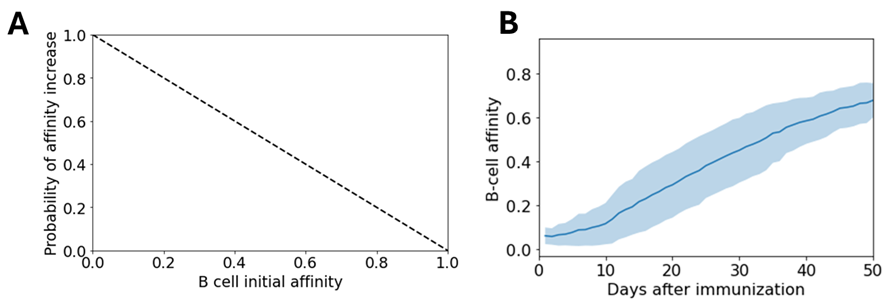}
    \caption{(A) Probability of the daughter cell increasing the affinity from the parent cell as a function of the initial affinity. (B) Affinity change of GC B cells during a germinal center simulation. The line is the average over all B~cells and the shaded area represent the mean plus one positive and one negative standard deviation.}
    \label{fig:GC_affinity}
\end{figure*}

%\printbibliography
\bibliographystyle{unsrt}
\bibliography{biblio}

\end{document}